\documentclass[12pt,aps,prd,preprint,onecolumn,nofootinbib,superscriptaddress]{revtex4-2}

\usepackage{amsmath,amssymb,mathtools}
\allowdisplaybreaks
\usepackage{slashed}
\usepackage{physics}
\usepackage{enumitem}
\usepackage{cancel}
\usepackage{graphicx}
   \usepackage{xcolor}
\usepackage[colorlinks=true, allcolors=black]{hyperref}
\usepackage{url}
 \usepackage{comment}

\usepackage{tikz}
\usetikzlibrary{decorations.pathmorphing}

\usepackage{microtype}
\usepackage{url}

\usepackage[colorlinks=true,allcolors=black]{hyperref}

\renewcommand{\Tr}{\operatorname{Tr}}
\newcommand{\SU}{\operatorname{SU}}
\newcommand{\U}{\operatorname{U}}
\newcommand{\Lag}{\mathcal{L}}

\usepackage[most]{tcolorbox}

\newtcolorbox{computation}[1][]{%
    enhanced,
    breakable,
    colback=gray!5,
    colframe=gray!50,
    fonttitle=\bfseries,
    title={Detailed Computation},
    #1
}

\begin{document}

\title{{\Large Strong {CP} and the QCD Axion} \\ Lecture Notes via Effective Field Theory}

\author{Francesco Sannino}
\affiliation{\mbox{$\hbar$QTC, Quantum Theory Center, \& Danish Institute for Advanced Study (Danish IAS)}, University of Southern Denmark, Campusvej 55, DK-5230 Odense M, Denmark}
\affiliation{\mbox{Dipt. di Fisica ``E. Pancini'', Universit\`a di Napoli Federico II, Via Cintia, 80126 Napoli, Italy}}


\begin{abstract}
These lecture notes provide a self-contained, graduate-level introduction to the strong $CP$ problem and QCD axion physics from an {effective field theory} (EFT) viewpoint.  We review the construction of the chiral EFT of QCD yielding a $\theta$-dependent potential, from which vacuum alignment, $\theta$ periodicity and branch structure follow. We further show how the framework leads to the Witten--Veneziano relation highlighting the role of the pure-glue topological susceptibility in organizing $\theta$-dependent hadronic observables. Using these tools, we show how to extract representative $CP$-odd mesonic and baryonic amplitudes, including the chiral estimate underlying the neutron EDM bound, and how to generalize the effective framework to confining $\SU(N)$ theories with fermions in arbitrary representations. We further show how to employ the Veneziano--Yankielowicz effective Lagrangian for $\mathcal{N}=1$ supersymmetric Yang--Mills theory to extract salient information on the $\theta$-dependent physics of one-flavour QCD via orientifold  planar equivalence. We also revisit a recent no strong $CP$ claim based on an ordering of limits in the sum over topological sectors and show, in the EFT language, that it amounts to introducing an extra non-propagating axion-like degree of freedom not required by QCD. We then present the standard dynamical resolution to the strong $CP$ problem, i.e. the Peccei--Quinn mechanism, the resulting axion potential and mass from chiral EFT and briefly review associated time-honored UV completions, and the axion quality problem from gravitational corrections.
\end{abstract}

\maketitle

\newpage 
\setcounter{tocdepth}{2}
\tableofcontents
\bigskip

\newpage 

\section{Introduction}

These lecture notes provide a self-contained, graduate-level introduction to the strong $CP$ problem and to QCD axion
physics from an {effective field theory} (EFT) viewpoint. The guiding idea is that the infrared
realization of gauge topology and the chiral dynamics of QCD can be organized most transparently in a low-energy
effective theory, where their dependence is encoded in an explicit potential and vacuum properties can be studied
systematically.

We start by recalling that the microscopic QCD Lagrangian featuring $N_f$ flavours may be written as
\begin{equation}
\Lag_{\rm QCD}
=
-\frac14\,G^a_{\mu\nu}G^{a\,\mu\nu}
-\theta\,\mathcal{Q}(x)
+\bar q\, i\slashed{D}\, q
-\Big(\bar q_L  \mathcal{M} q_R+\bar q_R  \mathcal{M}^\dagger q_L\Big),
\label{eq:intro_LQCD}
\end{equation}
where $ \mathcal{M}$ is the quark mass matrix. It further admits a renormalizable, gauge-invariant operator,
\begin{equation}
\Lag_{\theta}=-\theta\,\mathcal{Q}(x),
\qquad
\mathcal{Q}(x)\equiv \frac{g_s^2}{32\pi^2}\,G^a_{\mu\nu}\,\widetilde G^{a\,\mu\nu},
\qquad
\widetilde G^{a\,\mu\nu}\equiv \frac12\,\epsilon^{\mu\nu\rho\sigma}G^a_{\rho\sigma},
\label{eq:intro_Ltheta}
\end{equation}
where $\mathcal{Q}(x)$ is the topological charge density and $g_s$ is the strong coupling constant. Here the non-abelian field strength and covariant derivative are
\begin{equation}
G^a_{\mu\nu}=\partial_\mu A^a_\nu-\partial_\nu A^a_\mu+g_s f^{abc}A^b_\mu A^c_\nu,
\qquad
D_\mu=\partial_\mu-i g_s T^a A^a_\mu,
\label{eq:GandD_intro}
\end{equation}
with \(T^a\) the generators in the fundamental representation  of the gauge $\SU(3)$ group and $a=1,...,8$. One can write
$\mathcal{Q}(x)\propto \vec E^{\,a}\!\cdot\!\vec B^{\,a}$, so the $\theta$-term is odd under $P$ and hence under $CP$.
At the classical level $\mathcal{Q}(x)$ is a total derivative,
$G^a_{\mu\nu}\widetilde G^{a\,\mu\nu}=\partial_\mu K^\mu$,
but nonperturbative gauge topology makes $\theta$ physically meaningful in the quantum theory:
the configuration space splits into topological sectors and the vacuum is naturally described as a $\theta$-state
\cite{tHooft:1976snw,Jackiw:1976pf,Callan:1976je}.

\medskip
 Two structural facts are central for what follows.
First, in the chiral limit $\mathcal{M}\to 0$ the classical global symmetry contains a flavor-singlet axial factor $\U(1)_A$.
Second, this singlet axial symmetry is broken  at the quantum level by the Adler--Bell--Jackiw anomaly
\cite{Adler:1969gk,Bell:1969ts}. Equivalently, under a global singlet axial rotation
$q\to e^{i\beta\gamma_5}q$ the fermion path integral measure produces a Jacobian that shifts the vacuum
angle as \cite{Fujikawa:1979ay,Fujikawa:1980rc,Fujikawa:1980eg}
\begin{equation}
\theta \;\longrightarrow\; \theta-2N_f\,\beta.
\label{eq:intro_thetashift}
\end{equation}
At the same time, the same chiral rotation rephases the quark mass matrix,
$ \mathcal{M}\to e^{2i\beta} \mathcal{M}$, so that $\arg\det  \mathcal{M}\to \arg\det  \mathcal{M}+2N_f\beta$.
As a result, only a particular combination of $\theta$ and quark mass phases can be physically meaningful.
After establishing this point carefully within the low energy effective theory, we define the standard invariant parameter
\begin{equation}
\bar\theta \;\equiv\; \theta+\arg\det  \mathcal{M},
\label{eq:intro_thetabar}
\end{equation}
and explain why experimental constraints on hadronic $CP$ violation, most notably the neutron electric dipole
moment, require $|\bar\theta|$ to be extremely small (the precise translation depends on hadronic matching inputs)
\cite{Pospelov:2005pr,Abel:2020pzs}. The strong $CP$ problem is the absence, within the minimal Standard Model, of a
compelling principle that enforces this smallness.

\medskip
The EFT developments in these notes are organized in layers.
We begin with the discrete symmetry properties of the $\theta$-term and the topological input that makes $\theta$
observable. We then construct the chiral effective theory for the pseudoscalar mesons, first in the fictitious limit
where the $U(1)_A$ anomaly is absent, and subsequently including the anomaly in a convenient low-energy form.
A particularly economical implementation is to ``integrate in'' an auxiliary pseudoscalar background field
$\mathcal{Q}(x)\sim G\widetilde G$ so that $\mathcal{Q}(x)$ appears algebraically in the effective Lagrangian.
Integrating out $\mathcal{Q}(x)$ then generates an explicit $\theta$-dependent effective potential with controlled
expansion, closely connected to large-$N$ considerations and classic anomaly-effective constructions
\cite{DiVecchia:1980yfw,Witten:1979vv,Veneziano:1979ec}. This framework leads directly to the Witten--Veneziano
relation between the $\eta'$ mass and the pure-glue topological susceptibility.

\medskip
Once a $\theta$-dependent effective potential is available, vacuum alignment follows from stationarity conditions
(often referred to as Dashen equations \cite{Dashen:1970et} in this context). These equations determine the vacuum phases as functions of
$\theta$ and the quark masses, ensure $2\pi$ periodicity, and make manifest the multibranch structure characteristic of
confining gauge theories. A special role is played by $\theta=\pi$, where the vacuum need not respect $CP$. Instead,
the theory can develop two degenerate vacua exchanged by $CP$, corresponding to spontaneous $CP$ breaking (the Dashen
phenomenon) \cite{Dashen:1970et}. We treat this as a concrete illustration of how $\theta$-dependence and anomaly
dynamics constrain vacuum structure.

\medskip
With the vacuum fixed, the same EFT yields $CP$-odd interactions and allows one to organize $CP$-violating observables
systematically. We extract representative \(CP\)-odd mesonic interactions \cite{DiVecchia:1980yfw,DiVecchia:2013swa}, discuss how they feed into baryonic observables,
and review the standard chiral estimate connecting \(\bar\theta\) to the neutron EDM bound
\cite{Pospelov:2005pr,Abel:2020pzs}.
Where possible, we keep the discussion at the level of parametric control and clear matching logic. We also generalize the construction to confining $\SU(N)$ theories with fermions
in arbitrary representations, where the anomaly coefficient depends on the representation. This generalization is
useful both for composite extensions of the Standard Model and for planar-equivalent limits that interpolate between
QCD-like and supersymmetric dynamics \cite{Sannino:2004qp,Dietrich:2006cm,DiVecchia:2013swa,Armoni:2003gp,Armoni:2004uu, Sannino:2003xe,Sannino:2024xwj}.

\medskip
We use supersymmetric Yang--Mills effective theory and planar equivalence as controlled laboratories for the
\(\theta\)-vacuum structure, emphasizing what can and cannot be transported back to QCD.
This perspective provides an explicit bridge between exact results, large-\(N\) reasoning, and QCD-like dynamics
\cite{Veneziano:1982ah,Novikov:1983uc,Shifman:1986zi,Shifman:1987ia}.  An aspect of these lectures not present in \cite{Sannino:2003xe} is an explicit determination of the $\theta$-dependence of the one-flavour QCD
singlet pseudoscalar mass (the $\eta'$ analogue) via planar-equivalent orientifold arguments \cite{Armoni:2003fb,Armoni:2003gp,Sannino:2003xe,Armoni:2004uu,Sannino:2005sk} to connect  one-flavour QCD to super Yang--Mills, where the $\theta$-vacuum branch structure and effective potential are analytically controlled. Extensions and related interesting applications can be found in \cite{Armoni:2007jt,Armoni:2007kd,Armoni:2008gg,Armoni:2008nq,Armoni:2014ywa,Sannino:2004qp,Sannino:2024xwj,DellaMorte:2023ylq,Martins:2023kcj}. 

\medskip
Two later sections motivate additional material beyond a standard introduction. Section~\ref{sec:extra_phase_xi} examines a recent claim
that strong $CP$ violation is absent once the infinite-volume limit is taken in a particular order relative to the sum
over topological sectors \cite{Ai:2020ptm}. We reproduce the relevant effective description and
analyze its logical content from the EFT perspective. The conclusion reached in these notes is that removing strong
$CP$ in this manner corresponds, in the low-energy language, to imposing an additional stationarity condition
equivalent to introducing a non-propagating axion-like variable; this step is not implied by QCD alone, and therefore
cannot be taken as a resolution of the strong $CP$ problem.

Section~\ref{sec:PQ_axion} develops the QCD axion as the standard dynamical mechanism: the Peccei--Quinn idea promotes the vacuum angle
to a field-dependent quantity and introduces an approximate shift symmetry so that the vacuum relaxes dynamically to
the $CP$-conserving point \cite{Peccei:1977hh,Peccei:1977ur,Weinberg:1977ma,Wilczek:1977pj}. We emphasize how this
appears directly within the same chiral EFT used throughout, how the axion mass and couplings are fixed in terms of QCD
inputs \cite{GrillidiCortona:2015jxo}, and how these inputs have been recently refined in
\cite{DiLuzio:2020wdo}. Beyond the minimal construction, we also use this section to place the axion in a broader
model-building and phenomenological context: we summarize the role of UV completions \cite{Kim:1979if,Shifman:1979if,Zhitnitsky:1980tq,Dine:1981rt},
the origin and implications of the domain-wall number and model-dependent couplings, and the cosmological and
experimental logic underlying the standard search program. We then discuss the axion quality problem ~\cite{Kamionkowski:1992mf,Holman:1992us,Barr:1992qq,Ghigna:1992iv}, the fact
that generic Planck-suppressed violations of the PQ symmetry (including semiclassical gravitational effects) can
misalign the axion vacuum and reintroduce an effective $\bar\theta$ unless additional UV structure protects the shift
symmetry. The section closes with brief remarks on closely related topological angles and alternative directions,
including the possibility of a weak axion associated with electroweak topology \cite{Dvali:2024hsb,Cacciapaglia:2025xmr,Dvali:2025pcx,Cacciapaglia:2025dme,Davoudiasl:2025qqv}.

\medskip
The notes assume familiarity with basic quantum field theory and non-abelian gauge theory. They aim to be
self-contained at the level of effective Lagrangians, vacuum alignment conditions, and the main matching relations,
with explicit conventions chosen so that the reader can translate directly between the EFT discussion, modern lattice
inputs, and phenomenological constraints.
\section{Why you should care about \texorpdfstring{$\theta$}{theta}}
 
 At the quantum level \(\Lag_\theta\) is physical because the
space of gauge configurations has nontrivial topology (instantons, large gauge transformations, \(\theta\)-vacua)
\cite{tHooft:1976snw,Jackiw:1976pf,Callan:1976je}. This single term is therefore the cleanest portal from
nonperturbative gauge topology to observable hadron physics, and it lies at the core of the strong \(CP\) problem
(see e.g.\ \cite{Crewther:1977ce,DiVecchia:1980yfw,DiVecchia:2013swa}).

\medskip
A symmetry check already shows why \(\theta\) matters. In a fixed Lorentz frame define
\begin{equation}
E_i^a\equiv G^a_{0i},
\qquad
B_i^a\equiv \frac12\,\epsilon_{ijk}G^{a\,jk}
\qquad\Rightarrow\qquad
G^{a\,ij}=\epsilon^{ijk}B_k^a,
\label{eq:EBdefs_intro}
\end{equation}
so that
\begin{equation}
G^a_{\mu\nu}G^{a\,\mu\nu}=2\left(\vec B^{\,a\,2}-\vec E^{\,a\,2}\right),
\qquad
G^a_{\mu\nu}\tilde G^{a\,\mu\nu}=-4\,\vec E^{\,a}\cdot \vec B^{\,a}.
\label{eq:GGdualEB_intro}
\end{equation}
Under parity \(P:(t,\vec x)\mapsto(t,-\vec x)\) one has
\begin{equation}
A_0(t,\vec x)\xrightarrow{P}A_0(t,-\vec x),
\qquad
\vec A(t,\vec x)\xrightarrow{P}-\vec A(t,-\vec x),
\label{eq:Aparity_intro}
\end{equation}
so \(\vec E\) is a polar vector while \(\vec B\) is an axial vector:
\begin{equation}
\vec E^{\,a}\xrightarrow{P}-\vec E^{\,a},
\qquad
\vec B^{\,a}\xrightarrow{P}\phantom{-}\vec B^{\,a},
\qquad\Rightarrow\qquad
\vec E^{\,a}\cdot\vec B^{\,a}\xrightarrow{P}-\,\vec E^{\,a}\cdot\vec B^{\,a}.
\label{eq:EBparity_intro}
\end{equation}
For non-abelian gauge fields, charge conjugation acts by \(A_\mu=A_\mu^a T^a\to -A_\mu^a (T^a)^T\), and the
color-singlet contraction \(G^a_{\mu\nu}\tilde G^{a\,\mu\nu}\) is \(C\)-even. Hence the sign under \(CP\) is
inherited from parity,
\begin{equation}
G^a_{\mu\nu}\tilde G^{a\,\mu\nu}\xrightarrow{CP}-G^a_{\mu\nu}\tilde G^{a\,\mu\nu}
\qquad\Rightarrow\qquad
\Lag_\theta\xrightarrow{CP}-\Lag_\theta,
\end{equation}
so \(\theta\neq 0\) produces \emph{strong} CP violation.

\medskip

It is useful to split the Dirac quark fields  \(q=(q_1,\dots,q_{N_f})^T\) into their chiral components \(q_{L,R}=\frac12(1\mp\gamma_5)q\) linked by the quark mass matrix \( \mathcal{M}\). 
 In the chiral limit \(\mathcal{M}\to 0\), the classical global symmetry group enlarges to
\begin{equation}
\left[\SU(3) \right] \times \SU(N_f)_L\times \SU(N_f)_R\times \U(1)_V\times \U(1)_A,
\label{eq:globalsym_intro}
\end{equation}
where \(\U(1)_V\) is baryon number and \(\U(1)_A\) is the flavor-singlet axial rotation. 
For later reference, the transformation properties of the fundamental fields are summarized in
Table~\ref{tab:chiraltransf}. The group in square brackets is the gauge one. 

\begin{table}[t]
\centering
\renewcommand{\arraystretch}{1.25}
\begin{tabular}{lccccc}
\hline
 & $[\SU(3)]$ & $\SU(N_f)_L$ & $\SU(N_f)_R$ & $\U(1)_V$ & $\U(1)_A$ \\
\hline
$A_\mu^a$ & Adj & $\mathbf{1}$ & $\mathbf{1}$ & $0$ & $0$ \\
$q_L$     & Fund & Fund & $\mathbf{1}$ & $+1$ & $+1$ \\
$q_R$     & Fund & $\mathbf{1}$ & Fund & $+1$ & $-1$ \\
\hline
\end{tabular}
\caption{Transformation properties of QCD fields under the classical global symmetries in the chiral limit. In the non-Abelian columns, the entries indicate the representation of each field (with $\mathbf{1}$ denoting the singlet). In the Abelian columns, the entries give the corresponding $\U(1)$ charges. ``Adj'' and ``Fund'' denote the adjoint and fundamental representations, respectively.}
\label{tab:chiraltransf}
\end{table}

The associated Noether currents are
\begin{align}
J^{\mu,a}_V &= \bar q \gamma^\mu T^a q,
&
J^{\mu,a}_A &= \bar q \gamma^\mu \gamma_5 T^a q,
\label{eq:nonSingletCurrents_intro}
\\
J^\mu_V &= \bar q \gamma^\mu q,
&
J^\mu_5 &= \bar q \gamma^\mu \gamma_5 q.
\label{eq:singletCurrents_intro}
\end{align}
The central point for strong $CP$ is that the singlet axial symmetry is anomalous. Quantum mechanically, the
flavor-singlet axial current satisfies the exact operator identity \cite{Adler:1969gk,Bell:1969ts}
\begin{equation}
\partial_\mu J_5^\mu
=
2N_f\,\mathcal{Q}(x)
+2i\,\bar q\,  \mathcal{M} \gamma_5 q
-2i\,\bar q\, \gamma_5  \mathcal{M}^\dagger q \ .
\label{eq:ABJfull_intro}
\end{equation}
 Equivalently, as mentioned in the introduction, under the global axial rotation
\begin{equation}
q \to q' = e^{i\beta\gamma_5}q,
\qquad
\bar q \to \bar q' = \bar q\,e^{i\beta\gamma_5},
\label{eq:axialRot_intro}
\end{equation}
the fermion measure produces a (Fujikawa) Jacobian that shifts the \(\theta\) angle as
\cite{Fujikawa:1979ay,Fujikawa:1980rc}
\begin{equation}
\theta \to \theta'=\theta-2\beta N_f.
\label{eq:thetashiftAnom_intro}
\end{equation}
Because chiral rotations also rephase the quark mass matrix, \(\theta\) by itself is not invariant. Classically,
\eqref{eq:axialRot_intro} implies the spurion transformation for $\mathcal{M}$
\begin{equation}
q_L\to e^{i\beta}q_L,
\qquad
q_R\to e^{-i\beta}q_R,
\qquad\Rightarrow\qquad
 \mathcal{M} \to  \mathcal{M}' = e^{2i\beta} \mathcal{M},
\label{eq:MspurionShift_intro}
\end{equation}
so
\begin{equation}
\det  \mathcal{M} \to \det \mathcal{M}' = e^{+2i\beta N_f}\det  \mathcal{M},
\qquad\Rightarrow\qquad
\arg\det  \mathcal{M} \to \arg\det  \mathcal{M} + 2\beta N_f.
\label{eq:argdetshift_intro}
\end{equation}
Combining \eqref{eq:thetashiftAnom_intro} and \eqref{eq:argdetshift_intro}, the invariant, physically meaningful
\(CP\)-violating parameter is (see e.g.\ \cite{Crewther:1977ce,DiVecchia:1980yfw})
\begin{equation}
\boxed{\;\bar\theta\equiv \theta +\arg\det  \mathcal{M}\;}
\label{eq:thetaBar_intro}
\end{equation}
and it is \(\bar\theta\) (not \(\theta\) alone) that is constrained by experiment.

A nonzero \(\bar\theta\) induces \(CP\)-odd hadronic observables, in particular a neutron electric dipole moment
(see e.g.\ \cite{Pospelov:2005pr}). The current world limit is
\(|d_n| < 1.8\times 10^{-26}\, e\cdot\mathrm{cm}\) (90\% C.L.) \cite{Abel:2020pzs}, corresponding to the familiar benchmark
\begin{equation}
|\bar\theta|\lesssim 10^{-10},
\label{eq:thetaBarBound_intro}
\end{equation}
up to hadronic input uncertainties.

QCD also exhibits spontaneous chiral symmetry breaking at low energies. Below \(\Lambda_{\rm QCD}\) the theory
develops a quark condensate
\begin{equation}
\langle \bar q q \rangle = \langle \bar q_L q_R + \bar q_R q_L\rangle \neq 0,
\label{eq:condensate_intro}
\end{equation}
triggering
\begin{equation}
\SU(N_f)_L\times \SU(N_f)_R\times \U(1)_V
\;\longrightarrow\;
\SU(N_f)_V\times \U(1)_V,
\label{eq:SSBpattern_intro}
\end{equation}
and producing \(N_f^2-1\) (pseudo-)Nambu--Goldstone bosons. The  \(\U(1)_A\) symmetry is explicitly broken 
because of the anomaly \eqref{eq:ABJfull_intro}. Accordingly, the would-be singlet Goldstone mode is lifted and is
identified with the heavy \(\eta'\).

In the t' Hooft large number of colors limint (large-\(N\)) the anomaly is suppressed, and one obtains the Witten--Veneziano relation
\cite{Witten:1979vv,Veneziano:1979ec} linking the \(\eta'\) mass to the pure Yang--Mills topological susceptibility,
\begin{equation}
m_{\eta'}^2
\simeq
\frac{2N_f}{f_\pi^2}\,\chi_{\rm YM},
\qquad
\chi_{\rm YM}\equiv \int d^4x\, \langle 0|\, T\{\mathcal{Q}(x)\mathcal{Q}(0)\}\,|0\rangle_{\rm YM}.
\label{eq:WVrelationIntro_intro}
\end{equation}
Here the label $YM$ signifies that these quantities are computed in the Yang-Mills theory (i.e. in the pure glue theory).

\bigskip

Explaining why the invariant combination \eqref{eq:thetaBar_intro} satisfies the bound
\eqref{eq:thetaBarBound_intro} is the \emph{strong \(CP\) problem}. 
It is instructive to compare this puzzle to other open questions in particle physics.
Unlike the naturalness problem of the Higgs mass~\cite{tHooft:1979rat,Susskind:1978ms}, 
where quadratically divergent radiative corrections destabilize the electroweak scale 
and demand new physics near the TeV scale, the strong \(CP\) problem is radiatively stable: 
Ellis and Gaillard showed that \(\bar\theta\) runs only very mildly under the renormalization 
group within the Standard Model~\cite{Ellis:1978hq}, so a small value at some high scale 
remains small at low energies. The puzzle is therefore one of initial conditions rather than 
quantum instability, why should the UV theory select \(\bar\theta \lesssim 10^{-10}\) 
in the first place? In this sense the strong \(CP\) problem is arguably more severe: 
while the Higgs mass could in principle be fine-tuned at tree level with corrections 
absorbed order by order in perturbation theory, \(\bar\theta\) is a physical, 
basis-independent observable whose smallness lacks even this technical refuge.

The comparison with neutrino masses is also illuminating. The observation of neutrino 
oscillations demands physics beyond the Standard Model, but the most popular extensions are 
minimal: for example, right-handed neutrinos with tiny Yukawa couplings or higher-dimensional 
operators suppressed by a large scale, the celebrated seesaw 
mechanism~\cite{Minkowski:1977sc,Yanagida:1979as,Mohapatra:1979ia}. 
In contrast, the strong \(CP\) problem  is already present in the Standard Model 
and requires a dynamical or symmetry-based explanation for its suppression. Similarly, while 
baryogenesis and dark matter provide compelling evidence for new physics, they specify 
\emph{what} is missing without pinpointing \emph{where} in parameter space the Standard Model needs to be extended.  Sakharov's conditions for baryogenesis~\cite{Sakharov:1967dj}, baryon number violation, 
\(C\) and \(CP\) violation, and departure from thermal equilibrium, can in principle be 
satisfied within interesting extensions of the Standard Model, as exemplified by 
leptogenesis~\cite{Fukugita:1986hr}. Likewise, the dark matter problem, first identified 
by Zwicky's observations of the Coma cluster~\cite{Zwicky:1933gu,Zwicky:1937zza} and now 
supported by overwhelming cosmological evidence~\cite{Bertone:2004pz,Jungman:1995df}, 
tells us that roughly \(85\% \) of the matter in the universe is non-baryonic, but leaves 
the mass scale and interactions of the dark sector largely unconstrained.

An interesting parallel arises in Grand Unified Theories, where the \emph{doublet-triplet 
splitting problem}~\cite{Dimopoulos:1981zb,Sakai:1981pk} demands an explanation for why 
the electroweak Higgs doublets remain light while their color-triplet partners in the same 
GUT multiplet must be superheavy to avoid rapid proton decay.  Like the strong \(CP\) problem, 
this requires a fine-tuning of one part in \(10^{14}\) in minimal \(\mathrm{SU}(5)\), and 
various mechanisms, the missing partner mechanism~\cite{Masiero:1982fe,Grinstein:1982um}, 
the Dimopoulos--Wilczek (missing VEV) mechanism~\cite{Dimopoulos:1981zu}, or orbifold 
projections in extra dimensions~\cite{Kawamura:2000ev,Hall:2001xr}, have been proposed 
to address it. However, while doublet-triplet splitting concerns ratios of mass scales, 
the strong \(CP\) problem identifies a specific dimensionless parameter that is 
experimentally constrained to be at least ten orders of magnitude smaller than its 
natural \(\mathcal{O}(1)\) expectation.

Of course, all of these problems are interesting and worth exploring in their own right, 
and one should remain humble about what Nature ultimately chooses. It is entirely possible 
that some, or even all, of these apparent fine-tunings are simply accidents, with no 
deeper dynamical explanation. The electroweak scale may be tuned, \(\bar\theta\) may be 
small by chance, and the cosmological constant may just happen to take its observed value. 
Such a scenario would be disappointing from the perspective of theoretical elegance, but 
Nature is under no obligation to satisfy our aesthetic preferences. With this caveat in 
mind, historically, three broad classes of ideas have been pursued the address the strong \(CP\) problem:
(i) a massless quark, (ii) \(CP\) as an exact UV symmetry broken spontaneously, and
(iii) a Peccei--Quinn symmetry and the axion, dynamically relaxing the vacuum to \(\bar\theta=0\)
\cite{Peccei:1977hh,Peccei:1977ur,Weinberg:1977ma,Wilczek:1977pj,Kim:1979if,Shifman:1979if,Zhitnitsky:1980tq,Dine:1981rt}.
For modern axion EFT and phenomenology discussions that we will use later, see also
\cite{GrillidiCortona:2015jxo,DiLuzio:2020wdo}.

\bigskip

\section{Chiral effective theory without the \texorpdfstring{$U(1)_A$}{U(1)A} anomaly}

In this section we construct the low-energy effective theory describing the spontaneous breaking of chiral symmetry in QCD, under the assumption that the axial $U(1)_A$ symmetry is exact. This fictitious limit is theoretically instructive and provides a useful baseline for understanding how the axial anomaly modifies the spectrum in real QCD.
\subsection{Pattern of spontaneous symmetry breaking}

The QCD Lagrangian with $N_f$ massless quarks possesses a global 
$G = U(N_f)_L \times U(N_f)_R$ chiral symmetry. The formation of a quark 
condensate $\langle \bar{q}_R q_L \rangle \neq 0$ spontaneously breaks this 
symmetry to the diagonal (vector) subgroup $H = U(N_f)_V$, yielding the 
symmetry breaking pattern
\begin{equation}
G = U(N_f)_L \times U(N_f)_R 
\;\xrightarrow{\;\mathrm{SSB}\;} 
H = U(N_f)_V.
\label{eq:SSBpattern}
\end{equation}
Since $\dim G = 2N_f^2$ and $\dim H = N_f^2$, the coset space $G/H$ has 
dimension $N_f^2$, corresponding to $N_f^2$ Nambu--Goldstone bosons. In the 
absence of the axial anomaly, this spectrum includes not only the usual 
pseudoscalar octet (for $N_f = 3$), but also a singlet state associated 
with $U(1)_A$.

\subsection{The coset space and nonlinear realization}

The Goldstone bosons parametrize the coset space $G/H$, which can be 
identified with the space of left cosets $gH$ for $g \in G$. Following the 
formalism of Callan, Coleman, Wess, and Zumino 
(CCWZ)~\cite{Coleman:1969sm,Callan:1969sn}, we introduce a representative 
$\xi(\Phi)$ of each coset, where $\Phi$ collectively denotes the Goldstone 
fields. A convenient choice is
\begin{equation}
\xi(\Phi) = \exp\!\left[\frac{i\Phi}{f_\pi}\right],
\label{eq:xi_def}
\end{equation}
where $\Phi(x)$ is a Hermitian matrix collecting the pseudoscalar fields,
\begin{equation}
\Phi = \pi^a T^a + \frac{\eta_0}{\sqrt{2N_f}}\,\mathbf{1}.
\end{equation}
Here $T^a$ are the generators of $SU(N_f)$, normalized as 
$\Tr[T^a T^b] = \frac{1}{2}\delta^{ab}$, and $\eta_0$ is the 
$U(1)_A$ singlet.

The full chiral field $U(x) \in U(N_f)$, which transforms linearly under 
$G$ as
\begin{equation}
U \;\longrightarrow\; g_L\,U\,g_R^\dagger,
\qquad g_L, g_R \in U(N_f)_{L,R},
\label{eq:Utransform}
\end{equation}
is related to $\xi$ by
\begin{equation}
U = \xi^2.
\label{eq:U_xi_relation}
\end{equation}

\subsection{Nonlinear realization and the compensator field}

While $U$ transforms linearly under $G$, the field $\xi$ transforms 
nonlinearly. Under a chiral transformation $(g_L, g_R) \in G$, the coset 
representative $\xi(\Phi)$ is mapped to a new element that generically lies 
outside the chosen representative section. To restore the standard form, we 
must apply a compensating $H$-transformation. This defines the 
\emph{compensator field} $h(\Phi, g_L, g_R) \in H$ through the constraint
\begin{equation}
g_L \, \xi(\Phi) \, h^\dagger(\Phi, g_L, g_R) 
= \xi(\Phi') 
= h(\Phi, g_L, g_R) \, \xi(\Phi) \, g_R^\dagger,
\label{eq:compensator}
\end{equation}
where $\Phi'$ denotes the transformed Goldstone fields. This equation 
uniquely determines $h$ as a function of $\Phi$ and $(g_L, g_R)$, with 
$h \in U(N_f)_V$.

The transformation law \eqref{eq:compensator} encapsulates the essence of 
the nonlinear realization: the full chiral group $G$ acts on the Goldstone 
fields $\Phi$, but only the unbroken subgroup $H$ acts linearly on matter 
fields (such as baryons) that transform in representations of $H$. This 
provides a systematic framework for coupling Goldstone bosons to other 
fields while maintaining chiral invariance.

For transformations in the unbroken subgroup, $g_L = g_R = g_V \in H$, the 
compensator reduces to $h = g_V$, and the Goldstone fields transform 
linearly:
\begin{equation}
\xi \;\longrightarrow\; g_V \, \xi \, g_V^\dagger.
\end{equation}
For axial transformations the compensator 
becomes field-dependent, inducing a nonlinear shift of the Goldstone bosons.

\subsection{Chiral covariant objects}

To construct chirally invariant Lagrangians coupling Goldstones to matter 
fields, we introduce the Maurer--Cartan form
\begin{equation}
\xi^\dagger \partial_\mu \xi,
\end{equation}
which decomposes into vector and axial-vector components:
\begin{align}
V_\mu &= \frac{1}{2}\bigl(\xi^\dagger \partial_\mu \xi 
+ \xi \, \partial_\mu \xi^\dagger\bigr), 
\label{eq:V_mu}\\[4pt]
A_\mu &= \frac{i}{2}\bigl(\xi^\dagger \partial_\mu \xi 
- \xi \, \partial_\mu \xi^\dagger\bigr).
\label{eq:A_mu}
\end{align}

Under a chiral transformation, these objects have well-defined 
transformation properties. The vector combination $V_\mu$ transforms as a 
gauge connection under $H$:
\begin{equation}
V_\mu \;\longrightarrow\; h \, V_\mu \, h^\dagger 
+ h \, \partial_\mu h^\dagger,
\label{eq:V_transform}
\end{equation}
while the axial-vector combination transforms homogeneously:
\begin{equation}
A_\mu \;\longrightarrow\; h \, A_\mu \, h^\dagger.
\label{eq:A_transform}
\end{equation}

The inhomogeneous transformation \eqref{eq:V_transform} allows us to 
define a chiral covariant derivative. For a matter field $\psi$ 
transforming in a representation $R$ of $H$, i.e., 
$\psi \to D_R(h)\,\psi$, the covariant derivative is
\begin{equation}
D_\mu \psi = \partial_\mu \psi + V_\mu^{(R)} \psi,
\end{equation}
where $V_\mu^{(R)}$ denotes $V_\mu$ in the representation $R$. For baryon 
fields $B$ in the adjoint representation of $SU(N_f)_V$, this becomes
\begin{equation}
D_\mu B = \partial_\mu B + [V_\mu, B],
\label{eq:baryon_cov_deriv}
\end{equation}
which transforms covariantly: $D_\mu B \to h \, (D_\mu B) \, h^\dagger$.

We derive the equivalence between the kinetic term written in terms of 
$U$ and the form involving the axial-vector building block $A_\mu$:
\begin{equation}
\frac{1}{4}\Tr[\partial_\mu U \, \partial^\mu U^\dagger] 
= \Tr[A_\mu A^\mu].
\label{eq:kinetic_identity}
\end{equation}
 From $U=\xi^2$ we have
\begin{equation}
\partial_\mu U = (\partial_\mu \xi)\,\xi + \xi\,(\partial_\mu \xi).
\end{equation}
Multiplying on the left and right by $\xi^\dagger$ gives
\begin{equation}
\xi^\dagger (\partial_\mu U)\,\xi^\dagger
= \xi^\dagger(\partial_\mu \xi) + (\partial_\mu \xi)\,\xi^\dagger.
\label{eq:xi_dU_xi}
\end{equation}
Using unitarity, $(\partial_\mu \xi)\,\xi^\dagger + \xi\,(\partial_\mu \xi^\dagger)=0$, we may rewrite
\begin{equation}
(\partial_\mu \xi)\,\xi^\dagger = -\,\xi\,(\partial_\mu \xi^\dagger),
\end{equation}
so that \eqref{eq:xi_dU_xi} becomes
\begin{equation}
\xi^\dagger (\partial_\mu U)\,\xi^\dagger
= \xi^\dagger(\partial_\mu \xi) - \xi\,(\partial_\mu \xi^\dagger).
\end{equation}
Comparing with the definition \eqref{eq:A_mu}, we obtain the useful identity
\begin{equation}
A_\mu = \frac{i}{2}\,\xi^\dagger (\partial_\mu U)\,\xi^\dagger.
\label{eq:A_mu_dU}
\end{equation}

Taking the trace of $A_\mu A^\mu$ and using $\xi^{\dagger 2}=U^\dagger$ together with cyclicity yields
\begin{equation}
\Tr[A_\mu A^\mu]
= -\frac{1}{4}\Tr\!\left[\partial_\mu U\,U^\dagger\,\partial^\mu U\,U^\dagger\right].
\label{eq:TrAA_intermediate}
\end{equation}
On the other hand, unitarity of $U$ implies $U^\dagger U=\mathbf{1}$ and hence
\begin{equation}
\partial^\mu U^\dagger = -\,U^\dagger (\partial^\mu U)\,U^\dagger.
\label{eq:dUdagger_identity}
\end{equation}
Therefore,
\begin{equation}
\Tr[\partial_\mu U \,\partial^\mu U^\dagger]
= -\Tr\!\left[\partial_\mu U\,U^\dagger\,\partial^\mu U\,U^\dagger\right].
\label{eq:TrdUdUdagger_intermediate}
\end{equation}
Combining \eqref{eq:TrAA_intermediate} and \eqref{eq:TrdUdUdagger_intermediate} gives the desired result
\eqref{eq:kinetic_identity}.

\subsection{Leading-order meson Lagrangian}

The leading-order effective Lagrangian for the Goldstone bosons, at 
$\mathcal{O}(p^2)$ in the chiral power counting, is most compactly written 
in terms of the linear field $U = \xi^2$:
\begin{equation}
\mathcal{L}_{\mathrm{LO}}
= \frac{f_\pi^2}{4}\,\Tr\!\left[\partial_\mu U\,\partial^\mu U^\dagger\right]
+ \frac{f_\pi^2 B_0}{2}\,
\Tr\!\left[\mathcal{M}\,U^\dagger + U\,\mathcal{M}^\dagger\right],
\label{eq:chiralLO}
\end{equation}
where $\mathcal{M}$ is the quark mass matrix and $f_\pi \simeq 92$--$93$ MeV 
is the pion decay constant in the chiral limit. The parameter $B_0$ is 
related to the quark condensate via
\begin{equation}
\langle \bar{q} q \rangle = -f_\pi^2 B_0.
\end{equation}

Equivalently, using the building blocks \eqref{eq:V_mu}--\eqref{eq:A_mu}, 
the kinetic term can be written as
\begin{equation}
\frac{f_\pi^2}{4}\,\Tr\!\left[\partial_\mu U\,\partial^\mu U^\dagger\right]
= f_\pi^2\,\Tr\!\left[A_\mu A^\mu\right],
\end{equation}
making the role of the non-abelian axial current manifest.

The potential is minimized for
\begin{equation}
\langle U \rangle = \mathbf{1},
\qquad \text{i.e.,} \qquad 
\langle \xi \rangle = \mathbf{1},
\label{eq:Uvev}
\end{equation}
which spontaneously breaks chiral symmetry according to 
\eqref{eq:SSBpattern}.

\subsection{Vacuum alignment and the GMOR relation}

Including explicit chiral symmetry breaking through the diagonal quark 
mass matrix $\mathcal{M}_{ij} = m_i\,\delta_{ij}$, and expanding the 
effective Lagrangian to quadratic order in the Goldstone fields, one 
obtains the pseudoscalar mass spectrum. The masses satisfy the 
Gell-Mann--Oakes--Renner (GMOR) relation~\cite{GellMann:1968rz}
\begin{equation}
M_\pi^2 f_\pi^2 = -(m_u + m_d)\,\langle \bar{q} q \rangle 
= (m_u + m_d)\,B_0 f_\pi^2,
\label{eq:GMOR}
\end{equation}
which directly links the pseudoscalar masses to the quark condensate and 
quark masses.

\subsection{Explicit form for \texorpdfstring{$N_f = 3$}{Nf = 3}}

For three light flavors, the Goldstone matrix takes the explicit form
\begin{equation}
\Phi = \frac{1}{\sqrt{2}}
\begin{pmatrix}
\dfrac{\pi^0}{\sqrt{2}} + \dfrac{\eta_8}{\sqrt{6}} + \dfrac{\eta_0}{\sqrt{3}} 
& \pi^+ & K^+ \\[8pt]
\pi^- 
& -\dfrac{\pi^0}{\sqrt{2}} + \dfrac{\eta_8}{\sqrt{6}} + \dfrac{\eta_0}{\sqrt{3}} 
& K^0 \\[8pt]
K^- & \bar{K}^0 
& -\dfrac{2\eta_8}{\sqrt{6}} + \dfrac{\eta_0}{\sqrt{3}}
\end{pmatrix}.
\label{eq:Phi_Nf3}
\end{equation}
In this anomaly-free theory, the singlet field $\eta_0$ remains massless 
(up to quark mass effects) and combines with the octet to form a full 
nonet of pseudoscalar Goldstone bosons. The resolution of this {\it $U(1)_A$ 
problem}, the observed large mass of the $\eta'$, requires 
incorporating the axial anomaly, which we address in subsequent sections.

\bigskip
\section{Implementing the axial anomaly with a background field \texorpdfstring{$\mathcal{Q}(x)$}{Q(x)}}

In this section we show how the axial $U(1)_A$ anomaly can be incorporated consistently into the chiral effective theory. The construction follows the large $N$ approach developed by Witten and Veneziano and provides a controlled low-energy realization of the anomalous Ward identity.

\subsection{Effective implementation of the anomaly}
In QCD  the divergence of the singlet axial current satisfies
\begin{equation}
\partial_\mu J^{\mu}_5 = 2N_f\, \mathcal{Q}(x) \ ,
\end{equation}
which stems from \eqref{eq:ABJfull_intro} when setting to zero the quark mass matrix. It cannot be reproduced by a purely local chiral Lagrangian built only from $U(x)$. The anomaly must therefore be encoded through an additional pseudoscalar degree of freedom.

To this end, we introduce a background field $\mathcal{Q}(x)$ with the quantum numbers of the topological charge density and add to the effective theory the term
\begin{equation}
\mathcal{L}_{\mathrm{anom}}
=
-\,\frac{i}{2}\,\mathcal{Q}(x)\,
\mathrm{Tr}\!\left[\ln U - \ln U^\dagger\right].
\label{eq:anomterm}
\end{equation}
Under a global axial rotation
\begin{equation}
U \;\longrightarrow\; e^{2i\beta}U,
\qquad
U^\dagger \;\longrightarrow\; e^{-2i\beta}U^\dagger,
\end{equation}
we have
\begin{equation}
\mathrm{Tr}\!\left[\ln U - \ln U^\dagger\right]
\longrightarrow
\mathrm{Tr}\!\left[\ln U - \ln U^\dagger\right]
+4i\beta\,N_f,
\end{equation}
and therefore
\begin{equation}
\delta \mathcal{L}_{\mathrm{anom}}
=
-\,\frac{i}{2}\,\mathcal{Q}(x)\,(4i\beta N_f)
=
+\,2N_f\,\beta\,\mathcal{Q}(x).
\end{equation}
Identifying $\delta\mathcal{L}=\beta\,\partial_\mu J^{\mu}_5$ reproduces
\begin{equation}
\partial_\mu J^{\mu}_5=2N_f\,\mathcal{Q}(x),
\end{equation}
i.e.\ the anomalous Ward identity with the same normalization as in \eqref{eq:ABJfull_intro}.

\subsection{Integrating out the pseudoscalar background field}
The field $\mathcal{Q}(x)$ represents short-distance gluonic dynamics and is heavy compared to the pseudoscalar mesons. At large $N$, higher powers of $\mathcal{Q}(x)$ are suppressed, and the leading gluonic contribution is quadratic. We therefore consider the effective Lagrangian
\begin{align}
\mathcal{L}
={}&
\frac{f_\pi^2}{4}\,\mathrm{Tr}\!\left[\partial_\mu U\,\partial^\mu U^\dagger\right]
+\frac{f_\pi^2 B_0}{2}\,\mathrm{Tr}\!\left[\mathcal{M}\,U^\dagger + U\,\mathcal{M}^\dagger\right]
\nonumber\\
&- \frac{i}{2}\, \mathcal{Q}(x)\,\mathrm{Tr}\!\left[\ln U - \ln U^\dagger\right]
+\frac{\mathcal{Q}^2(x)}{2\chi_{\mathrm{top}}}
-\theta\,\mathcal{Q}(x),
\label{eq:LwithQ}
\end{align}
where $\chi_{\mathrm{top}}$ has mass dimension four. In the large $N$ limit, $\chi_{\mathrm{top}}=\mathcal{O}(N^0)$ and coincides with the pure Yang--Mills topological susceptibility $\chi_{\mathrm{YM}}$.

Since $\mathcal{Q}(x)$ has no kinetic term, it can be eliminated algebraically by its equation of motion:
\begin{equation}
\frac{\delta\mathcal{L}}{\delta \mathcal{Q}(x)} = 0
\qquad\Rightarrow\qquad
\frac{\mathcal{Q}(x)}{\chi_{\mathrm{top}}}
- \frac{i}{2}\,\mathrm{Tr}\!\left(\ln U - \ln U^\dagger\right)
-\theta = 0,
\end{equation}
hence
\begin{equation}
\mathcal{Q}(x)=\chi_{\mathrm{top}}
\left[
\theta + \frac{i}{2}\,\mathrm{Tr}\!\left(\ln U - \ln U^\dagger\right)
\right].
\label{eq:Qsolution}
\end{equation}
Substituting this back yields the effective $\theta$-dependent chiral theory
\begin{align}
\mathcal{L}
={}&\frac{f_\pi^2}{4}\,\mathrm{Tr}\!\left[\partial_\mu U\,\partial^\mu U^\dagger\right]
+\frac{f_\pi^2 B_0}{2}\,\mathrm{Tr}\!\left[\mathcal{M}\,U^\dagger + U\,\mathcal{M}^\dagger\right]
\nonumber\\
&-\frac{\chi_{\mathrm{top}}}{2}
\left[
\theta + \frac{i}{2}\,\mathrm{Tr}\!\left(\ln U - \ln U^\dagger\right)
\right]^2.
\label{eq:LeffAfterQ}
\end{align}
This last term couples the topological angle $\theta$ to the pseudoscalar singlet field and generates a mass for the would-be Goldstone boson associated with spontaneous breaking of $U(1)_A$ before taking into account of the axial anomaly.
\section{Vacuum alignment, the physical angle \texorpdfstring{$\bar{\theta}$}{theta-bar}, and special cases}
In this section we investigate the vacuum properties of the theory and how they depend on the topological sector of QCD.  

\subsection{General vacuum configuration and minimization}
In presence of $\theta$ the ground state solution $\langle U \rangle = \mathbf{1}$ must be revisited.  For diagonal quark masses, the most general constant vacuum configuration is
\begin{equation}
\langle U_{ij}\rangle
=
e^{i\phi_i}\,\delta_{ij}
=
\begin{pmatrix}
e^{i\phi_1} & 0 & \cdots & 0\\
0 & e^{i\phi_2} & \cdots & 0\\
\vdots & \vdots & \ddots & \vdots\\
0 & 0 & \cdots & e^{i\phi_{N_f}}
\end{pmatrix}.
\end{equation}
At the vacuum, $\mathrm{Tr}[\ln U - \ln U^\dagger] = 2i\sum_i \phi_i$, so the bracket in \eqref{eq:LeffAfterQ} becomes
\begin{equation}
\theta + \frac{i}{2}(2i\sum_i \phi_i) = \theta - \sum_i \phi_i.
\end{equation}
The vacuum energy density extracted from \eqref{eq:LeffAfterQ} is therefore
\begin{equation}
E
=
\frac{\chi_{\mathrm{top}}}{2}\Big(\theta - \sum_{i=1}^{N_f}\phi_i\Big)^2
- f_\pi^2 B_0 \sum_{i=1}^{N_f} m_i\cos\phi_i.
\label{eq:Ebar}
\end{equation}

Minimizing $E$ with respect to each $\phi_i$ yields the Dashen equations
\begin{equation}
f_\pi^2 B_0\, m_i\sin\phi_i
=
\chi_{\mathrm{top}}\Big(\theta - \sum_{j=1}^{N_f}\phi_j\Big),
\qquad i=1,\dots,N_f.
\label{eq:Dashen}
\end{equation}
Using the GMOR relation per single quark $i$ we have taht for each pseudo Goldstone $M_{\pi,i}^2 = 2B_0 m_i$ and the above can be written as
\begin{equation}
\frac{f_\pi^2 M_{\pi,i}^2}{2}\sin\phi_i
=
\chi_{\mathrm{top}}\Big(\theta - \sum_{j=1}^{N_f}\phi_j\Big).
\label{eq:Dashen_M}
\end{equation}

The combination
\begin{equation}
\bar{\theta} \equiv \theta - \sum_i\phi_i
\end{equation}
is the physical $CP$-violating parameter of the low-energy theory.

\subsection{Parametrization around the vacuum} 
It is useful to factor out the vacuum phases by writing
\begin{equation}
U_{ij}=e^{i\phi_i}\,V_{ij},
\qquad\qquad
\langle V_{ij}\rangle=\delta_{ij}\,,
\end{equation}
with
\begin{equation}
V\,V^\dagger=\mathbf{1},
\qquad\qquad
V(x)=\exp\!\left(\frac{2i\,\Phi}{f_\pi}\right).
\end{equation}
\begin{equation}
\mathrm{Tr}[\ln U - \ln U^\dagger] = \mathrm{Tr}[\ln V - \ln V^\dagger] + 2i\sum_{i=1}^{N_f}\phi_i.
\end{equation}

Substituting into the anomaly potential:
\begin{align}
\mathcal{L}_{\mathrm{anom}} &= -\frac{\chi_{\mathrm{top}}}{2}\left[\theta + \frac{i}{2}\left(\mathrm{Tr}[\ln V - \ln V^\dagger] + 2i\sum_i\phi_i\right)\right]^2 \nonumber\\[6pt]
&= -\frac{\chi_{\mathrm{top}}}{2}\left[\underbrace{\theta - \sum_i\phi_i}_{\bar{\theta}} + \frac{i}{2}\mathrm{Tr}[\ln V - \ln V^\dagger]\right]^2.
\end{align}

By expanding the square 
\begin{align}
\mathcal{L}_{\mathrm{anom}} &= -\frac{\chi_{\mathrm{top}}}{2}\bar{\theta}^2 
- \frac{i\chi_{\mathrm{top}}}{2}\bar{\theta}\,\mathrm{Tr}[\ln V - \ln V^\dagger] 
+\frac{\chi_{\mathrm{top}}}{8}\left(\mathrm{Tr}[\ln V - \ln V^\dagger]\right)^2.
\label{eq:Lanom_expanded}
\end{align}

Including the kinetic and mass terms:
\begin{align}
\mathcal{L} ={}&
\frac{f_\pi^2}{4}\,\mathrm{Tr}\!\left[\partial_\mu V\,\partial^\mu V^\dagger\right]
+ \frac{f_\pi^2 B_0}{2}\,\mathrm{Tr}\!\left[\mathcal{M}\left(V^\dagger + V - 2\cdot\mathbf{1}\right)\right]
\nonumber\\[6pt]
&- \frac{i f_\pi^2 B_0}{2}\,\mathrm{Tr}\!\left[(\mathcal{M}\sin\phi)\,(V - V^\dagger)\right]
+ f_\pi^2 B_0 \sum_{i=1}^{N_f} m_i\cos\phi_i
\nonumber\\[6pt]
&+\frac{\chi_{\mathrm{top}}}{8}\left(\mathrm{Tr}\!\left[\ln V - \ln V^\dagger\right]\right)^2
- \frac{i\chi_{\mathrm{top}}}{2}\bar{\theta}\,\mathrm{Tr}\!\left[\ln V - \ln V^\dagger\right]
\nonumber\\[6pt]
&- \frac{\chi_{\mathrm{top}}}{2}\bar{\theta}^2,
\label{eq:L_intermediate}
\end{align}
where $(\mathcal{M}\sin\phi)_{ij} = m_i\sin\phi_i\,\delta_{ij}$ and $\bar{\theta} = \theta - \sum_i\phi_i$.
\vskip .2cm

Recalling that $V = \exp(2i\Phi/f_\pi)$:
\begin{equation}
\ln V - \ln V^\dagger = \frac{4i\Phi}{f_\pi} .
\end{equation}

Therefore:
\begin{equation}
\mathrm{Tr}[\ln V - \ln V^\dagger] = \frac{4i}{f_\pi}\mathrm{Tr}\Phi = \frac{4i}{f_\pi}\cdot\frac{\eta_0}{\sqrt{2N_f}}\cdot N_f = \frac{2i\sqrt{2N_f}}{f_\pi}\eta_0.
\end{equation}

The singlet mass term comes from:
\begin{align}
-\frac{\chi_{\mathrm{top}}}{8}\left(\mathrm{Tr}[\ln V - \ln V^\dagger]\right)^2 
&= -\frac{\chi_{\mathrm{top}}}{8}\left(\frac{2i\sqrt{2N_f}}{f_\pi}\eta_0\right)^2 \nonumber\\[6pt]
&= -\frac{\chi_{\mathrm{top}}}{8}\cdot\frac{-8N_f}{f_\pi^2}\eta_0^2 \nonumber\\[6pt]
&= +\frac{\chi_{\mathrm{top}} N_f}{f_\pi^2}\eta_0^2.
\end{align}

  Comparing with the standard form $\mathcal{L} \supset -\frac{1}{2}m^2\phi^2$:
\begin{equation}
 {m_{\eta_0}^2 = \frac{2\chi_{\mathrm{top}} N_f}{f_\pi^2}  } \ .
\label{eq:m0_chitop_relation} 
\end{equation}

\subsection{Final effective Lagrangian}

Putting everything together with the constraint $VV^\dagger = \mathbf{1}$:
\begin{align}
\mathcal{L} ={}&
\frac{f_\pi^2}{4}\,\mathrm{Tr}\!\left[\partial_\mu V\,\partial^\mu V^\dagger\right]
- \frac{\chi_{\mathrm{top}} N_f}{f_\pi^2}\,\eta_0^2
\nonumber\\[6pt]
&+ \frac{f_\pi^2 B_0}{2}\,\mathrm{Tr}\!\left[
\mathcal{M}(\theta)\left(\cos\!\left(\frac{2\Phi}{f_\pi}\right)-\mathbf{1}\right)
\right]
\nonumber\\[6pt]
&+ \chi_{\mathrm{top}}\,\bar{\theta}\,\mathrm{Tr}\!\left[
\sin\!\left(\frac{2\Phi}{f_\pi}\right) - \frac{2\Phi}{f_\pi}
\right],
\label{eq:vexp_final}
\end{align}
where
\begin{equation}
\mathcal{M}_{ij}(\theta) \equiv m_i\cos\phi_i(\theta)\,\delta_{ij}.
\end{equation}

\subsection{Solving for the vacuum angles $\boldsymbol{\phi_i(\theta)}$}
We now analyze the solutions of the Dashen equations
\begin{equation}\label{eq:Dashen_recap}
M_{\pi,i}^2 \,\sin\phi_i
\;=\;
\frac{\chi_{\mathrm{top}}}{f_\pi^2}\left(\theta - \sum_{j=1}^{N_f}\phi_j\right),
\qquad i=1,\dots,N_f,
\end{equation}
where $M_{\pi,i}^2 = 2B_0 m_i$ are the leading-order meson masses, which determine the vacuum alignment in the presence of the $\theta$ angle.

\subsubsection{$\theta$-periodicity of the theory}
The theory is invariant under shifts
\begin{equation}
\theta \;\longrightarrow\; \theta + 2\pi.
\end{equation}
Let $\{\hat{\phi}_i(\theta)\}$ be any solution of \eqref{eq:Dashen_recap} at angle $\theta$. We construct a solution at $\theta+2\pi$ by defining
\begin{equation}
\bar{\phi}_1(\theta+2\pi)
=
\hat{\phi}_1(\theta)+2\pi,
\qquad
\bar{\phi}_{i>1}(\theta+2\pi)
=
\hat{\phi}_i(\theta).
\label{eq:branch_shift}
\end{equation}
Using $\sin(x+2\pi)=\sin x$ and the defining equation for $\hat{\phi}_i(\theta)$, one verifies immediately that $\{\bar{\phi}_i\}$ satisfies \eqref{eq:Dashen_recap} at $\theta+2\pi$.
Although individual solutions $\phi_i(\theta)$ are multivalued, physical quantities depend only on $e^{i\phi_i}$ and are therefore $2\pi$-periodic in $\theta$.

\subsubsection{Vanishing strong CP violation at $\boldsymbol{\theta=0}$}
Strong $CP$ violation vanishes whenever
\begin{equation}
\bar{\theta} \;\equiv\; \theta - \sum_{i=1}^{N_f}\phi_i = 0.
\end{equation}

\paragraph{Case: $\theta=0$ (illustrated for $N_f=2$).}
For $\theta=0$, the Dashen equations become
\begin{align}
M_{\pi,1}^2 \sin\phi_1 &= -\frac{\chi_{\mathrm{top}}}{f_\pi^2}(\phi_1+\phi_2), \\
M_{\pi,2}^2 \sin\phi_2 &= -\frac{\chi_{\mathrm{top}}}{f_\pi^2}(\phi_1+\phi_2).
\end{align}
Subtracting yields
\begin{equation}
M_{\pi,1}^2\sin\phi_1 = M_{\pi,2}^2\sin\phi_2,
\end{equation}
and a manifest solution is
\begin{equation}
\phi_1=\phi_2=0,
\end{equation}
which implies $\bar{\theta}=0$ and hence a $CP$-conserving vacuum at $\theta=0$.
 The generalization to arbitrary $N_f$ is straightforward. 

\subsubsection{One massless quark removes $\boldsymbol{\theta}$}
If at least one quark is massless, the $\theta$ angle can be rotated away and strong $CP$ violation vanishes identically. Consider $N_f=2$ with
\begin{equation}
M_{\pi,1}^2=0,
\qquad
M_{\pi,2}^2 \neq 0.
\end{equation}
The Dashen equations read
\begin{align}
0
&=
\frac{\chi_{\mathrm{top}}}{f_\pi^2}\left(\theta - \phi_1 - \phi_2\right),
\label{eq:massless1}\\
M_{\pi,2}^2 \sin\phi_2
&=
\frac{\chi_{\mathrm{top}}}{f_\pi^2}\left(\theta - \phi_1 - \phi_2\right).
\label{eq:massless2}
\end{align}
From \eqref{eq:massless1} one obtains $\theta-(\phi_1+\phi_2)=0$, which inserted into \eqref{eq:massless2} gives $\phi_2=0$, hence $\phi_1=\theta$ and $\bar{\theta}=0$.

\subsection*{The special point $\boldsymbol{\theta=\pi}$ and the Dashen phenomenon}
\label{Dashen-phenomenon}
At $\theta=\pi$ the vacuum structure can become qualitatively different from generic values of $\theta$.
The reason is that $CP$ acts as $\theta\to-\theta$, so the point $\theta=\pi$ is mapped to itself modulo $2\pi$:
\begin{equation}
-\pi \equiv \pi \pmod{2\pi}.
\end{equation}
Hence $\theta=\pi$ is a $CP$-symmetric point of parameter space. Nevertheless, the ground state \emph{need not} respect $CP$ there: the theory can instead develop two degenerate vacua exchanged by $CP$. This is the Dashen phenomenon \cite{Dashen:1970et}.

A convenient way to diagnose spontaneous $CP$ breaking at $\theta=\pi$ is to examine whether there exist \emph{distinct degenerate} solutions of the Dashen equations with opposite values of a $CP$-odd order parameter (e.g.\ $\bar{\theta}$ or equivalently $\theta-\sum_i\phi_i$ once vacuum alignment is enforced). When such degenerate vacua exist, the vacuum selects one branch and $CP$ is spontaneously broken.

\paragraph{Two-flavor analysis in the large $\chi_{\mathrm{top}}$ limit.}
To display the mechanism explicitly, consider $N_f=2$ in the limit
\begin{equation}
\chi_{\mathrm{top}} \gg f_\pi^2 M_{\pi,1}^2, f_\pi^2 M_{\pi,2}^2.
\end{equation}
In this regime the energy \eqref{eq:Ebar} strongly penalizes deviations of $\theta-\phi_1-\phi_2$ from zero, so one must have, to leading order,
\begin{equation}
\theta \approx \phi_1+\phi_2.
\end{equation}
Imposing $\phi_2=\theta-\phi_1$, the minimization equations reduce to a single condition,
\begin{equation}
 M_{\pi,1}^2 \sin \phi_1 =  M_{\pi,2}^2 \sin( \theta - \phi_1 ) \ .
\label{min94}
\end{equation}
Solving \eqref{min94} yields
\begin{eqnarray}
\sin \phi_1 = \frac{ M_{\pi,2}^2 \sin \theta}{\sqrt{ M_{\pi,1}^4 +
     M_{\pi,2}^4 + 2 M_{\pi,1}^2  M_{\pi,2}^2 \cos \theta }} \ , \qquad 
\sin \phi_2 = \frac{ M_{\pi,1}^2 \sin \theta}{\sqrt{ M_{\pi,1}^4 +
     M_{\pi,2}^4 + 2 M_{\pi,1}^2 M_{\pi,2}^2 \cos \theta}}  \ ,
\label{sin}
\end{eqnarray}
and
\begin{eqnarray}
\cos \phi_1 = \frac{ M_{\pi,1}^2 +  M_{\pi,2}^2 \cos \theta}{\sqrt{ M_{\pi,1}^4 +
     M_{\pi,2}^4 + 2 M_{\pi,1}^2 M_{\pi,2}^2 \cos \theta }}\ , \qquad
\cos \phi_2 = \frac{ M_{\pi,2}^2 +  M_{\pi,1}^2 \cos \theta}{\sqrt{ M_{\pi,1}^4 +
     M_{\pi,2}^4 + 2 M_{\pi,1}^2  M_{\pi,2}^2 \cos \theta }} \ .
\label{cos}
\end{eqnarray}
Computing the associated energy gives
\begin{eqnarray}
E (\theta) = - \frac{f_{\pi}^{2}}{2}\sqrt{ M_{\pi,1}^4 +
     M_{\pi,2}^4 + 2 M_{\pi,1}^2  M_{\pi,2}^2 \cos \theta } \ .
\label{ene56}
\end{eqnarray}
For equal masses $( M_{\pi,1} =  M_{\pi,2} =  M_{\pi})$ this reduces to
\begin{eqnarray}
E (\theta) = - f_{\pi}^{2} M_{\pi}^2 \left| \cos \frac{\theta}{2} \right| \ .
\label{equama}
\end{eqnarray}
Both \eqref{ene56} and \eqref{equama} are $2\pi$-periodic in $\theta$, as required. However, in the equal-mass case \eqref{equama} exhibits a cusp at $\theta=\pi$ because $\cos(\theta/2)$ changes sign there. This non-analyticity signals the presence of two competing branches that exchange dominance at $\theta=\pi$.

\paragraph{First correction at large but finite $\chi_{\mathrm{top}}$.}
Having solved the minimization equation in the strict $\chi_{\mathrm{top}}\to\infty$ limit, we now compute the leading correction at large but finite $\chi_{\mathrm{top}}$ by returning to the full Dashen system
\begin{eqnarray}
 M_{\pi,1}^2 \sin \phi_1 =  M_{\pi,2}^2 \sin \phi_2 = \frac{\chi_{\mathrm{top}}}{f_\pi^2} ( \theta -
\phi_1 - \phi_2 ) \ .
\label{eq93}
\end{eqnarray}
We expand around the large $\chi_{\mathrm{top}}$ solution,
\begin{eqnarray}
\phi_{1,2} = {\bar{\phi}}_{1,2} + \epsilon\, \delta \phi_{1,2}\ , \qquad 
\epsilon = \frac{f_\pi^2  M_{\pi,1}  M_{\pi,2}}{\chi_{\mathrm{top}}} \ .
\label{ansa}
\end{eqnarray}
One deduces
\begin{eqnarray}
\phi_1 = \bar{\phi}_1 - \epsilon \frac{\sin \theta}{R^3}
\frac{ M_{\pi,2}^2 +   M_{\pi,1}^2 \cos \theta}{ M_{\pi,1}^2} \ , \qquad 
\phi_2 = \bar{\phi}_2 - \epsilon \frac{\sin \theta}{R^3}
\frac{ M_{\pi,1}^2 +   M_{\pi,2}^2 \cos \theta}{ M_{\pi,2}^2} \ ,
\label{sol76}
\end{eqnarray}
where ${\bar{\phi}}_{1,2}$ is the large $\chi_{\mathrm{top}}$ solution satisfying
\begin{eqnarray}
{\bar{\phi}}_{1} + {\bar{\phi}}_{2} = \theta \ , \qquad  R =
\sqrt{\frac{ M_{\pi,1}^4 +
     M_{\pi,2}^4 + 2 M_{\pi,1}^2  M_{\pi,2}^2 \cos \theta}{
 M_{\pi,1}^2 M_{\pi,2}^2 }} \ .
\label{sol82}
\end{eqnarray}
Using these expressions one can compute the $CP$-odd quantity
\begin{eqnarray}
\theta - \phi_1 - \phi_2 = \epsilon \frac{\sin \theta}{R} =
\frac{f_\pi^2  M_{\pi,1}^2 M_{\pi,2}^2 \sin \theta}{ \chi_{\mathrm{top}} \sqrt{ M_{\pi,1}^4 +
     M_{\pi,2}^4 + 2 M_{\pi,1}^2  M_{\pi,2}^2 \cos \theta }} \ .
\label{cpvio}
\end{eqnarray}
This contribution vanishes at $\theta =0$ and also if $ M_{\pi,1}^2$ and/or $ M_{\pi,2}^2$ are zero, as expected.

\paragraph{$CP$ at $\theta=\pi$.}
At $\theta=\pi$ one has $\sin\theta=0$, so \eqref{cpvio} vanishes \emph{unless} the equal-mass limit produces an enhanced sensitivity. Indeed, for $ M_{\pi,1} \neq  M_{\pi,2}$ the $CP$-odd quantity vanishes at $\theta=\pi$. But if $ M_{\pi,1}= M_{\pi,2}\equiv M_{\pi}$, one finds instead
\begin{eqnarray}
\theta - \phi_1 - \phi_2 = \frac{f_\pi^2  M_{\pi}^2}{\chi_{\mathrm{top}}} \sin \frac{\theta}{2} \ ,
\qquad\Rightarrow\qquad
\theta=\pi:\quad
\theta - \phi_1 - \phi_2 = \frac{f_\pi^2  M_{\pi}^2}{\chi_{\mathrm{top}}} \ .
\label{pi}
\end{eqnarray}
Thus for degenerate quark masses the vacuum at $\theta=\pi$ carries a nonzero $CP$-odd order parameter, and $CP$ is spontaneously violated. The two degenerate vacua correspond to the two branches related by $CP$ (equivalently by $\phi_i\to-\phi_i$ up to $2\pi$ shifts), and the cusp in \eqref{equama} reflects the branch exchange.

\paragraph{Physical interpretation and generalizations.}
The Dashen phenomenon at $\theta=\pi$ is best viewed as a competition between distinct vacuum branches whose energies cross at $\theta=\pi$. When a branch crossing occurs, the vacuum energy develops a cusp and the theory admits domain walls interpolating between the degenerate vacua. In multi-flavor QCD-like theories, whether $CP$ breaks at $\theta=\pi$ depends on the pattern of quark masses (or more generally on the effective mass parameters $ M_{\pi,i}^2$), and the vacuum structure can involve several branches.

In the following phenomenology appendix, the same $CP$-odd quantity $\theta-\sum_i\phi_i$ controls $CP$-violating mesonic and baryonic couplings, making the Dashen analysis directly relevant for observable amplitudes.

\vskip .3cm 
A refined analysis of spontaneous $CP$ breaking at $\theta=\pi$ and its implications for the axion potential has been presented in~\cite{DiVecchia:2017xpu}. Using the low-energy EFT, the authors  mapped out the regions of parameter space where $CP$ is spontaneously broken or unbroken at $\theta=\pi$. They showed that in the $CP$-broken region, first-order phase transitions occur as $\theta$ crosses $\pi$, while on the hypersurface separating the two regions, second-order phase transitions emerge, characterized by a vanishing pseudo-Nambu-Goldstone boson mass and a divergent topological susceptibility.

\section{Witten--Veneziano relation and topological susceptibility}
The Witten--Veneziano relation represents one of the most profound connections between the topological structure of QCD and the mass spectrum of hadrons. This relation, independently derived by Witten~\cite{Witten:1979vv} and Veneziano~\cite{Veneziano:1979ec}, provides a theoretical explanation for the large mass of the $\eta'$ meson and its connection to the $\U(1)_A$ anomaly.

\subsection{Topological susceptibility}
Recalling that the topological charge density is
\begin{equation}
\mathcal{Q}(x)\equiv \frac{g^2}{32\pi^2}G^a_{\mu\nu}\tilde{G}^{a\,\mu\nu},
\qquad
Q \equiv \int d^4x\, \mathcal{Q}(x)\,,
\end{equation}
and the vacuum energy density $E(\theta)$ is defined from the generating functional,
\begin{equation}
Z(\theta)=\int \mathcal{D}A\; e^{\,i\int d^4x\,\left(\mathcal{L}_{\rm YM}-\theta\,\mathcal{Q}(x)\right)}
\equiv e^{\,i V_4\,(-E(\theta))}\,,
\end{equation}
where $V_4$ is the four-volume. The (pure-glue) topological susceptibility is
\begin{equation}
\chi_{\rm YM}\equiv \left.\frac{d^2E(\theta)}{d\theta^2}\right|_{\theta=0}
=
\lim_{k\to 0} i\!\int d^4x\,e^{ik\cdot x}\,\langle 0|T\, \mathcal{Q}(x)\mathcal{Q}(0)|0\rangle_{\theta=0}\,.
\label{eq:chi_def}
\end{equation}
As we shall see, in the large $N$ framework, $\chi_{\rm YM}$ of \emph{pure} Yang--Mills controls the anomalous contribution to the singlet pseudoscalar mass and leads to the Witten--Veneziano relation \cite{Witten:1979vv,Veneziano:1979ec,DiVecchia:1980yfw}.

\subsection{A Gaussian ``pure glue'' effective model for $\mathcal{Q}(x)$ with a source}
To isolate the notion of susceptibility in a way that matches the effective-Lagrangian derivation, consider a minimal no-fermion (pure-glue) effective theory in which $\mathcal{Q}$ is treated as a heavy pseudoscalar variable and the leading gluonic term is quadratic:
\begin{equation}
\mathcal{L}[\mathcal{Q};\theta,J]=\frac{\mathcal{Q}^2}{2\chi_{\mathrm{top}}}-\theta\,\mathcal{Q}-i\,\mathcal{Q}\,J\,.
\label{eq:L_q_source}
\end{equation}
Here $J$ is an external source coupled linearly to $\mathcal{Q}$. In the underlying theory one may view $J$ as a bookkeeping device to generate correlators of $\mathcal{Q}$.  

\vskip .2cm
Since $\mathcal{Q}$ has no kinetic term, its equation of motion is local:
\begin{align}
 \frac{\delta\mathcal{L}}{\delta \mathcal{Q}} = 0
\quad &\Longrightarrow\quad
\frac{\mathcal{Q}}{\chi_{\mathrm{top}}}-\theta-iJ=0
\nonumber\\
&\Longrightarrow\quad
\mathcal{Q}= {\chi_{\mathrm{top}}} \,(\theta+iJ)\,.
\label{eq:q_eom}
\end{align}

\vskip .2cm
Insert \eqref{eq:q_eom} into \eqref{eq:L_q_source}:
\begin{align}
\mathcal{L}_{\rm EOM}(\theta,J)
&=
\frac{\chi_{\mathrm{top}}}{2}(\theta+iJ)^2-{\chi_{\mathrm{top}}}(\theta+iJ)^2
\nonumber\\
&=-\frac{\chi_{\mathrm{top}}}{2}\,(\theta+iJ)^2\,.
\label{eq:L_on_shell_q}
\end{align}
At $J=0$ the vacuum energy density is therefore
\begin{equation}
E(\theta)=\frac{\chi_{\mathrm{top}}}{2}\,\theta^2+\mathcal{O}(\theta^4)\,,
\end{equation}
and the pure-glue topological susceptibility is, at leading order in the t' Hooft large number of colors: 
\begin{equation}
\chi_{\rm YM}
\equiv
\left.\frac{d^2E}{d\theta^2}\right|_{\theta=0}
=
 {\chi_{\mathrm{top}}}\,.
\label{eq:chiYM_chitop_relation}
\end{equation}
This is the effective-theory avatar of the statement that pure Yang--Mills has a nonzero susceptibility at large $N$.

Using \eqref{eq:chiYM_chitop_relation} in \eqref{eq:m0_chitop_relation} immediately yields
\begin{equation}
m_0^2
=
\frac{2 N_f}{f_\pi^2}\chi_{\mathrm{top}}
=
\frac{2N_f}{f_\pi^2}\,\chi_{\rm YM}\,.
\label{eq:WV_chiral_limit}
\end{equation}
This is the Witten--Veneziano relation in the chiral limit and in the large $N$ ordering where $\chi_{\rm YM}$ is the appropriate susceptibility to use  \cite{Witten:1979vv,Veneziano:1979ec,DiVecchia:1980yfw}.

For $N_f=3$ with nonzero quark masses, the phenomenologically standard form is often written as
\begin{equation}
m_{\eta'}^2+m_\eta^2-2m_K^2
\simeq
\frac{2N_f}{f_\pi^2}\,\chi_{\rm YM}
\qquad (N_f=3),
\label{eq:WV_pheno}
\end{equation}
up to corrections suppressed by $1/N$ and by higher-order chiral effects. Equation \eqref{eq:WV_pheno} is the statement that the \emph{anomalous} piece of the singlet mass is controlled by pure-glue topology.

At large $N$  we have the scalings:
\begin{equation}
f_\pi^2\propto N,
\qquad
\chi_{\rm YM}\propto \mathcal{O}(1),
\end{equation}
so that
\begin{equation}
m_0^2\propto\frac{1}{N}.
\end{equation}
Thus, as $N\to\infty$, the $U(1)_A$ anomaly becomes parametrically small, $\eta'$ becomes a pseudo-Nambu--Goldstone boson, and the WV relation precisely encodes how the leading anomaly effect is tied to the pure-glue topological susceptibility  \cite{Witten:1979vv,Veneziano:1979ec}.

Recent lattice determinations of pure-gauge topological observables that connect to $\chi_{\rm YM}$ and related quantities include, e.g., \cite{Bonanno:2023hhp}

\bigskip

\section{Strong $CP$ violating mesonic amplitudes}

From the $CP$ violating term in Eq.~(\ref{eq:vexp_final}) we can extract a cubic term in the fields of the pseudoscalar mesons:
\begin{eqnarray}
- \frac{\chi_{\mathrm{top}} \bar{\theta}}{3 f_\pi^3} \mathrm{Tr} \left[ \Phi^3 \right] \quad \longrightarrow \quad -
\frac{\chi_{\mathrm{top}} \bar{\theta}}{\sqrt{3} f_\pi^3} \pi^{+} \pi^{-} \eta_8,
\label{cpviob}
\end{eqnarray}
from which we extract the decay amplitude $\eta_8 \rightarrow \pi^{+} \pi^{-}$:
\begin{eqnarray}
T ( \eta \rightarrow \pi^{+} \pi^{-}) =
\frac{\chi_{\mathrm{top}} \bar{\theta}}{\sqrt{3} f_\pi^3} = \frac{2 m_{\pi}^{2} (\theta )}{\sqrt{3} f_\pi} \cdot
\frac{ M_{\pi,1}^2 M_{\pi,2}^2 \sin \theta }{ M_{\pi,1}^4 +
     M_{\pi,2}^4 + 2 M_{\pi,1}^2  M_{\pi,2}^2 \cos \theta},
\label{deca}
\end{eqnarray}
where we have used \eqref{cpvio} and
\begin{eqnarray}
 m_{\pi}^{2} (\theta ) = \frac{ M_{\pi,1}^2 \cos \phi_1 +  M_{\pi,2}^2
\cos \phi_2}{2} = \frac{1}{2} \sqrt{ M_{\pi,1}^4 +
     M_{\pi,2}^4 + 2 M_{\pi,1}^2  M_{\pi,2}^2 \cos \theta }.
\label{mpi}
\end{eqnarray}
For small values of $\theta$ we get
\begin{eqnarray}
T ( \eta \rightarrow \pi^{+} \pi^{-})  \sim  \frac{2
  m_{\pi}^{2}}{\sqrt{3} f_\pi} \frac{\theta}{\left(
  \sqrt{\frac{m_1}{m_2}} + \sqrt{\frac{m_2}{m_1}} \right)^2},
\label{smathe}
\end{eqnarray}
where $m_i$ is the quark mass related to the meson mass through the GMOR relation $ M_{\pi,i}^2 = 2B_0 m_i$. Notice that we have identified $\eta_{8}$ with the particle state $\eta$.\footnote{The physical $\eta$ is the linear combination $\eta = \cos \varphi \, \eta_8 + \sin \varphi \, \eta_0$ with mixing angle   $\varphi \sim 11^\circ$ \cite{Feldmann:1998vh}.}

From the previous equation we get:
\begin{eqnarray}
\Gamma (  \eta \rightarrow \pi^{+} \pi^{-})  = \frac{\theta^2}{\left(
  \sqrt{\frac{m_1}{m_2}} + \sqrt{\frac{m_2}{m_1}} \right)^4 } \frac{m_{\pi}^4  \sqrt{m^2_{\eta} - 4m_{\pi}^2} }{12
 \pi f_\pi^2 m^2_{\eta}}.
\label{Gam}
\end{eqnarray}
Using $f_\pi = 93$~MeV, $m_{\pi} = 140$~MeV and $m_{\eta}= 548$~MeV we get
\begin{eqnarray}
\Gamma (  \eta \rightarrow \pi^{+} \pi^{-})  = \theta^2 \cdot 98.2 \,\,{\rm keV},
\label{amp82}
\end{eqnarray}
and
\begin{eqnarray}
\frac{\Gamma (\eta \rightarrow \pi^{+} \pi^{-} ) }{\Gamma_{tot}} =
68 \,\,\theta^2.
\label{ampli}
\end{eqnarray}
From experiments \cite{KLOE:2004vek} we have
\begin{eqnarray}
\frac{\Gamma (\eta \rightarrow \pi^{+} \pi^{-} ) }{\Gamma_{tot}} < 1.3
\cdot 10^{-5},
\label{exp}
\end{eqnarray}
yielding an upper limit $\theta < 4.4 \times 10^{-4}$. A much better limit comes from the neutron EDM.

The decay amplitude vanishes for $\theta=0$ and $\pi$ when $ M_{\pi,1}^2 \neq  M_{\pi,2}^2$. For equal masses, the process does not vanish at $\theta = \pi$. The three flavor case can be found in \cite{DiVecchia:2013swa}. 


\subsection{Meson masses as functions of $\theta$}

From the meson mass matrix in \eqref{eq:vexp_final} we derive \cite{DiVecchia:1980yfw,DiVecchia:2013swa}:
\begin{eqnarray}
m_{\pi^{0,\pm}}^{2} = \frac{ M_{\pi,1}^2 \cos \phi_1 +  M_{\pi,2}^2
\cos \phi_2}{2}, \qquad m_{K^{\pm}}^{2} =
\frac{ M_{\pi,1}^2 \cos \phi_1 + M_{3}^{2}
\cos \phi_3}{2},
\label{mass}
\end{eqnarray}
and
\begin{eqnarray}
m_{K^{0}, \bar{K}^{0}}^{2} =
\frac{ M_{\pi,2}^2 \cos \phi_2 + M_{3}^{2}
\cos \phi_3}{2}.
\label{mass2}
\end{eqnarray}
These relations imply
\begin{eqnarray}
R (\theta)  \equiv \frac{m^{2}_{K^0} - m^{2}_{K^+} - m^{2}_{\pi^0} +
  m^{2}_{\pi^+} }{m_{\pi}^{2}} = \frac{(  M_{\pi,2}^2 -  M_{\pi,1}^2)
(  M_{\pi,2}^2 +  M_{\pi,1}^2) }{ M_{\pi,1}^4 +
     M_{\pi,2}^4 + 2 M_{\pi,1}^2  M_{\pi,2}^2 \cos \theta },
\label{mas67}
\end{eqnarray}
with
\begin{eqnarray}
R (\theta =0 ) = \frac{ M_{\pi,2}^2 -  M_{\pi,1}^2 }{ M_{\pi,2}^2 +
   M_{\pi,1}^2 }, \qquad R (\theta =\pi ) = \frac{ M_{\pi,2}^2 +  M_{\pi,1}^2}{ M_{\pi,2}^2 -
   M_{\pi,1}^2 }.
\label{rat9}
\end{eqnarray}
Experimentally $R \simeq 0.26$, is consistent with small $\theta \approx 0$. The quark mass ratios are
\begin{eqnarray}
\frac{m_1}{m_2} = \frac{ M_{\pi,1}^2}{  M_{\pi,2}^2} \simeq 0.56, \qquad
\frac{m_3}{m_2} = \frac{M_{\pi,3}^{2}}{  M_{\pi,2}^2} \simeq 20.18.
\label{rat78}
\end{eqnarray}

\section{Strong CP violating amplitudes with baryons}
\label{sec:baryons}

In order to compute the $CP$ violating terms involving baryons we add to the effective Lagrangian terms involving baryons. This section provides a comprehensive treatment of the baryon chiral Lagrangian, the transformation properties of baryons under chiral symmetry, and the extraction of $CP$-violating observables.

\subsection{The baryon octet representation}

The baryons belong to an octet of $SU_V(3)$ and are described
by the following matrix:
\begin{eqnarray}
B = \left(\begin{array}{ccc}
\frac{\Sigma^{0}}{\sqrt{2}}+ \frac{\Lambda}{\sqrt{6}} & \Sigma^{+} & p \\
\Sigma^{-} & -\frac{\Sigma^{0}}{\sqrt{2}}+ \frac{\Lambda}{\sqrt{6}} & n \\
\Xi^{-} & {{\Xi}}^{0} &  - 2 \frac{\Lambda}{\sqrt{6}}
       \end{array}\right) \ .
\label{eq:ba}
\end{eqnarray}
Here $B$ is a Dirac spinor and, being a matter field, transforms naturally under the $SU(3)_V$ diagonal vector subgroup,
\begin{equation}
B \rightarrow h B h^{\dagger} \ , \quad {\rm with} \quad h \in SU(3)_V \ .
\label{eq:Btransform}
\end{equation} 

\subsection{The baryon chiral Lagrangian}

Recalling that $
U \rightarrow g_L \, U \, g_R^\dagger$ under a chiral transformation we have that the relevant Lagrangian involving baryons can be written as follows:
\begin{align}
\mathcal{L}_{\rm bar} &= \mathrm{Tr} \left[{\bar{B}} i \gamma^{\mu} D_{\mu} B \right] -
\alpha\, f_\pi \, \mathrm{Tr} \left[ \bar{\mathcal{L}} U \mathcal{R} U^{\dagger} + \bar{\mathcal{R}}
  U^{\dagger} \mathcal{L} U \right]
+ \delta \, f_\pi B_0 \, \mathrm{Tr} \left[ \bar{\mathcal{L}} U \mathcal{R} \mathcal{M}
+ \bar{\mathcal{R}} U^{\dagger} \mathcal{L} \mathcal{M}^{\dagger} \right] \nonumber\\
&\quad
+
\gamma \, f_\pi B_0 \, \mathrm{Tr} \left[ \bar{\mathcal{L}} \mathcal{M}^{\dagger} \mathcal{R} U^{\dagger} + \bar{\mathcal{R}}
  \mathcal{M} \mathcal{L} U \right].
\label{eq:bar}
\end{align}
with $ D_\mu B = \partial_\mu B + [V_\mu, B]$ which transforms covariantly: $D_\mu B \to h \, (D_\mu B) \, h^\dagger$ and $V_\mu$ defined in \eqref{eq:V_transform}. 
We have further defined the proper right and left globally transforming baryon fields by dressing the original baryon field $B$ by the Goldstone dependent quantity $\xi$ as follows:  
\begin{equation}
\mathcal{R} \equiv \frac{1+\gamma_5}{2}\xi^\dagger B \xi, \qquad 
\mathcal{L} \equiv \frac{1-\gamma_5}{2}\xi B \xi^\dagger, \qquad 
\xi\xi = U.
\end{equation}

\subsection{Rewriting in terms of $\xi$ and $B$}

Using the definitions of $\mathcal{L}$ and $\mathcal{R}$, we can rewrite each term. For the mass-independent term:
\begin{align}
\mathrm{Tr}[\bar{\mathcal{L}} U \mathcal{R} U^\dagger + \bar{\mathcal{R}} U^\dagger \mathcal{L} U]
&= \mathrm{Tr}\left[\bar{B}\frac{1-\gamma_5}{2}\xi^\dagger \cdot \xi^2 \cdot \xi^\dagger B \xi \cdot \xi^{-2}\right] + \text{h.c.} \nonumber\\
&= \mathrm{Tr}\left[\bar{B}\frac{1-\gamma_5}{2}B\right] + \mathrm{Tr}\left[\bar{B}\frac{1+\gamma_5}{2}B\right] \nonumber\\
&= \mathrm{Tr}[\bar{B}B].
\end{align}

For the mass-dependent terms, we use $U = \xi^2$ and the cyclic property of the trace:
\begin{align}
\mathrm{Tr}[\bar{\mathcal{L}} U \mathcal{R} \mathcal{M}] &= \mathrm{Tr}\left[\bar{B}\frac{1-\gamma_5}{2}\xi^\dagger \cdot \xi^2 \cdot \xi^\dagger B \xi \cdot \mathcal{M}\right] 
= \mathrm{Tr}\left[\bar{B}\frac{1-\gamma_5}{2} B \xi \mathcal{M} \xi\right], \\
\mathrm{Tr}[\bar{\mathcal{R}} U^\dagger \mathcal{L} \mathcal{M}^\dagger] &= \mathrm{Tr}\left[\bar{B}\frac{1+\gamma_5}{2} B \xi^\dagger \mathcal{M}^\dagger \xi^\dagger\right].
\end{align}

Combining and using $\frac{1\pm\gamma_5}{2} = \frac{1}{2}(1 \pm \gamma_5)$:
\begin{align}
\mathcal{L}_{\rm bar} &= \mathrm{Tr} \left[{\bar{B}} i \gamma^{\mu} D_{\mu} B \right] -
\alpha \, f_\pi \, \mathrm{Tr} \left[ {\bar{B}} B \right] \nonumber\\
&\quad +  \delta \, \frac{f_\pi B_0}{2} \, \mathrm{Tr} \left[ {\bar{B}} B  (\xi \mathcal{M} \xi + \xi^{\dagger}\mathcal{M}^{\dagger} \xi^{\dagger}) \right]
+  \delta \, \frac{f_\pi B_0}{2} \, \mathrm{Tr} \left[ {\bar{B}} \gamma_5 B  (\xi \mathcal{M} \xi - \xi^{\dagger}\mathcal{M}^{\dagger} \xi^{\dagger}) \right]  \nonumber \\
&\quad +   \gamma \, \frac{f_\pi B_0}{2} \, \mathrm{Tr} \left[ {\bar{B}}   (\xi \mathcal{M} \xi + \xi^{\dagger}\mathcal{M}^{\dagger} \xi^{\dagger})B \right] 
- \gamma \, \frac{f_\pi B_0}{2} \, \mathrm{Tr} \left[ {\bar{B}} \gamma_5  (\xi \mathcal{M} \xi - \xi^{\dagger}\mathcal{M}^{\dagger} \xi^{\dagger}) B \right].
\label{eq:bar_explicit}
\end{align}

\subsection{Incorporating the $\theta$-dependence}

Following our convention $U_{ij} = e^{i\phi_i}V_{ij}$ with $\langle V_{ij}\rangle = \delta_{ij}$, we write:
\begin{equation}
(\xi^2)_{ij} = U_{ij} = e^{i\frac{\phi_i}{2}} V_{ij} e^{i\frac{\phi_j}{2}},
\label{eq:var67}
\end{equation}
which implies for $\xi$:
\begin{equation}
\xi_{ij} = e^{i\frac{\phi_i}{2}} \nu_{im} h_{mj}^\dagger = h_{im} \nu_{mj} e^{i\frac{\phi_j}{2}}, \qquad \text{with} \qquad \nu = \exp\left(\frac{i\Phi}{f_\pi}\right).
\end{equation}
Here $h \in SU(N_f)_V$ is determined by requiring $\nu\nu = V$.

\subsubsection{Evaluating the mass combinations}

For diagonal mass matrix $\mathcal{M}_{ij} = m_i \delta_{ij}$, we compute:
\begin{align}
(\xi \mathcal{M} \xi)_{ij} &= \xi_{ik} m_k \xi_{kj} = e^{i\frac{\phi_i}{2}} (\nu h^\dagger)_{ik} m_k (h\nu)_{kj} e^{i\frac{\phi_j}{2}} \nonumber\\
&= e^{i\frac{\phi_i}{2}} (\nu m_k h^\dagger h \nu)_{ij} e^{i\frac{\phi_j}{2}} = e^{i\frac{\phi_i}{2}} (\nu \mathcal{M} \nu)_{ij} e^{i\frac{\phi_j}{2}}.
\end{align}

Similarly:
\begin{equation}
(\xi^\dagger \mathcal{M}^\dagger \xi^\dagger)_{ij} = e^{-i\frac{\phi_i}{2}} (\nu^\dagger \mathcal{M} \nu^\dagger)_{ij} e^{-i\frac{\phi_j}{2}}.
\end{equation}

Define the $\theta$-dependent mass matrix:
\begin{equation}
\mathcal{M}(\theta)_{ij} \equiv m_i \cos\phi_i(\theta)\, \delta_{ij}.
\end{equation}

Then:
\begin{align}
\xi \mathcal{M} \xi + \xi^\dagger \mathcal{M}^\dagger \xi^\dagger &= 2\, \nu \mathcal{M}(\theta) \nu \cdot \frac{e^{i\phi} + e^{-i\phi}}{2} + \ldots \nonumber\\
&= \nu \mathcal{M}(\theta) \nu + \nu^\dagger \mathcal{M}(\theta) \nu^\dagger \equiv 2\mathcal{M}_p(\theta),
\end{align}
and
\begin{align}
\xi \mathcal{M} \xi - \xi^\dagger \mathcal{M}^\dagger \xi^\dagger &= \nu \mathcal{M}(\theta) \nu - \nu^\dagger \mathcal{M}(\theta) \nu^\dagger \equiv 2\mathcal{M}_m(\theta),
\end{align}
where we defined:
\begin{equation}
\mathcal{M}_{p/m}(\theta) \equiv \frac{\nu \mathcal{M}(\theta) \nu \pm \nu^\dagger \mathcal{M}(\theta) \nu^\dagger}{2}.
\end{equation}

 The $CP$-violating terms deriving from the anomaly are obtained when making explicit use of the Dashen equations. This 
 generates the following $CP$-odd terms:
\begin{equation}
\mathcal{L}_{\text{CP-odd}} = \frac{\chi_{\mathrm{top}}}{f_\pi}\bar{\theta}\left[\delta\, \mathrm{Tr}\left(\bar{B}B \sin\frac{2\Phi}{f_\pi} - i\bar{B}\gamma_5 B \cos\frac{2\Phi}{f_\pi}\right) + \gamma\,\mathrm{Tr}\left(\bar{B}\sin\frac{2\Phi}{f_\pi} B + i\bar{B}\gamma_5 \cos\frac{2\Phi}{f_\pi} B\right)\right] \ .
\end{equation}

\subsection{Full $\theta$-dependent baryon Lagrangian}

Combining all contributions:
\begin{align}
\mathcal{L}_{\rm bar} &= \mathrm{Tr}\left[\bar{B} i\gamma^\mu D_\mu B\right] - \alpha f_\pi \mathrm{Tr}\left[\bar{B}B\right] \nonumber\\[6pt]
&\quad + f_\pi B_0\, \delta\, \mathrm{Tr}\left[\bar{B}B\, \mathcal{M}_p(\theta) + \bar{B}\gamma_5 B\, \mathcal{M}_m(\theta)\right] \nonumber\\[6pt]
&\quad + f_\pi B_0\, \gamma\, \mathrm{Tr}\left[\bar{B}\, \mathcal{M}_p(\theta) B - \bar{B}\gamma_5\, \mathcal{M}_m(\theta) B\right] \nonumber\\[6pt]
&\quad + \frac{\chi_{\mathrm{top}}}{f_\pi}\bar{\theta}\, \mathrm{Tr}\left[\delta\left(\bar{B}B \sin\frac{2\Phi}{f_\pi} - i\bar{B}\gamma_5 B \cos\frac{2\Phi}{f_\pi}\right)\right. \nonumber\\[6pt]
&\qquad\qquad\qquad\quad \left. + \gamma\left(\bar{B}\sin\frac{2\Phi}{f_\pi} B + i\bar{B}\gamma_5 \cos\frac{2\Phi}{f_\pi} B\right)\right],
\label{eq:efflag}
\end{align}
where $\bar{\theta} = \theta - \sum_i \phi_i$ and $\mathcal{M}_{p/m}(\theta) = \frac{1}{2}(\nu\mathcal{M}(\theta)\nu \pm \nu^\dagger\mathcal{M}(\theta)\nu^\dagger)$.

\subsection{Determination of the parameters $\alpha$, $\gamma$, and $\delta$}

The parameters $\alpha$, $\gamma$, and $\delta$ are determined by matching to baryon masses at $\theta = 0$. Setting $\phi_i = 0$ (so $\mathcal{M}(\theta=0) = \mathcal{M}$) and expanding to leading order in $\Phi$ (see Appendix~\ref{Appendix-baryon} for extra details):
\begin{equation}
\mathcal{M}_p(0) \approx \mathcal{M} + \mathcal{O}(\Phi^2), \qquad \mathcal{M}_m(0) \approx \frac{i}{f_\pi}[\Phi, \mathcal{M}] + \mathcal{O}(\Phi^3).
\end{equation}

From the baryon mass spectrum one obtains:
\begin{align}
\alpha &= \frac{1}{f_\pi}\left[m_\Sigma + \frac{M^2}{ M_{\pi,3}^2 - m_\pi^2}(2m_\Sigma - m_\Xi - m_N)\right], \label{eq:alpha}\\[6pt]
\gamma &= \frac{1}{f_\pi( M_{\pi,3}^2 - m_\pi^2)}(m_\Sigma - m_\Xi), \label{eq:gamma}\\[6pt]
\delta &= \frac{1}{f_\pi( M_{\pi,3}^2 - m_\pi^2)}(m_\Sigma - m_N), \label{eq:delta}
\end{align}
where $m_\pi^2 = ( M_{\pi,1}^2 +  M_{\pi,2}^2)/2$ is the average light quark mass parameter.

The baryon masses satisfy the Gell-Mann--Okubo mass formula:
\begin{equation}
3m_\Lambda + m_\Sigma = 2(m_\Xi + m_N).
\label{eq:gmo}
\end{equation}

\subsection{The CP-violating pion-nucleon coupling}

From the Lagrangian \eqref{eq:efflag}, expanding to linear order in $\Phi$ and extracting the $\pi N$ interaction:
\begin{equation}
\mathcal{L}_{\pi NN} = \bar{N}\left[ig_{\pi NN}\gamma_5 + \bar{g}_{\pi NN}\right]\vec{\tau}\cdot\vec{\pi}\, N,
\label{eq:pinc}
\end{equation}
where $g_{\pi NN}$ is the $CP$-conserving coupling and $\bar{g}_{\pi NN}$ is $CP$-violating.

The $CP$-violating coupling arises from the $\bar{\theta}$-dependent terms. Using current algebra methods~\cite{Crewther:1977ce}, the $CP$-odd pion-nucleon coupling can be expressed as:
\begin{equation}
\bar{g}_{\pi NN} \approx \frac{m_u m_d}{m_u + m_d} \frac{m_\Xi - m_\Sigma}{f_\pi m_N} \bar{\theta} \approx 0.02\, \bar{\theta}.
\label{eq:gbar_current_algebra}
\end{equation}
This result has been refined using QCD sum rules~\cite{Pospelov:2005pr}, yielding consistent estimates for the $CP$-odd coupling.

At leading order:
\begin{equation}
\bar{g}_{\pi NN} = -\frac{\chi_{\mathrm{top}}\bar{\theta}}{f_\pi^2( M_{\pi,3}^2 - m_{\pi}^2)}\frac{m_\Xi - m_\Sigma}{f_\pi}.
\label{eq:gbar_full}
\end{equation}

Using the solution for $\bar{\theta}$ from the Dashen equations:
\begin{equation}
\bar{\theta} = \theta - \sum_i\phi_i = \frac{f_\pi^2  M_{\pi,1}^2  M_{\pi,2}^2 \sin\theta}{\chi_{\mathrm{top}}\sqrt{ M_{\pi,1}^4 +  M_{\pi,2}^4 + 2 M_{\pi,1}^2  M_{\pi,2}^2\cos\theta}},
\end{equation}
we obtain:
\begin{equation}
\bar{g}_{\pi NN} = -\frac{ M_{\pi,1}^2  M_{\pi,2}^2 \sin\theta}{( M_{\pi,3}^2 - m_{\pi}^2)\sqrt{ M_{\pi,1}^4 +  M_{\pi,2}^4 + 2 M_{\pi,1}^2  M_{\pi,2}^2\cos\theta}}\frac{m_\Xi - m_\Sigma}{f_\pi}.
\end{equation}

For small $\theta$:
\begin{equation}
\bar{g}_{\pi NN} = -2\theta \frac{m_1 m_2}{(2m_3 - m_1 - m_2)(m_1 + m_2)}\frac{m_\Xi - m_\Sigma}{f_\pi},
\label{eq:gbar_quark}
\end{equation}
where we used $M_{\pi,i}^2 = 2B_0 m_i$ and again $m_\pi^2 = (M_{\pi,1}^2 + M_{\pi,2}^2)/2 $.

\subsection{The Goldberger--Treiman relation}

For the $CP$ preserving coupling one must add new operators dictated by current algebra involving derivative couplings with the mesons. This leads to
 \begin{eqnarray}
 f_\pi g_{\pi NN} \simeq g_A m_N  \ ,
\label{eq:gt}
\end{eqnarray}
where $g_A \approx 1.27$ is the axial coupling constant of the nucleon. This is the Goldberger--Treiman relation. Setting $g_A = 1$ gives:
\begin{equation}
g_{\pi NN} \approx \frac{m_N}{f_\pi} \approx \frac{939~\text{MeV}}{93~\text{MeV}} \approx 10.1,
\end{equation}
which is close to the experimental value $g_{\pi NN} \approx 13.1$. The Goldberger--Treiman relation holds up to corrections that vanish in the chiral limit.

One vertex in the nEDM calculation is the $CP$-conserving $\pi NN$ coupling:
\begin{equation}
\mathcal{V}_{\rm CP-even} = i g_{\pi NN} \bar{N} \gamma_5 \vec{\tau} \cdot \vec{\pi} N,
\end{equation}
while the other vertex is the $CP$-violating coupling:
\begin{equation}
\mathcal{V}_{\rm CP-odd} = \bar{g}_{\pi NN} \bar{N} \vec{\tau} \cdot \vec{\pi} N.
\end{equation}

The key observation is that the EDM requires both a $CP$-violating coupling (to provide the necessary $CP$-odd phase) and an electromagnetic coupling (to couple to the external photon). At one loop, this is achieved by having one $CP$-even and one $CP$-odd vertex, with the photon coupling either to the intermediate proton or to the charged pion.

\subsection{Calculation of the nEDM}

Following the classic analysis of Crewther, Di Vecchia, Veneziano, and Witten \cite{Crewther:1977ce}, the dominant contribution comes from the pion loop with one $CP$-conserving and one $CP$-violating vertex. The result is:
\begin{equation}
d_n = \frac{e \, g_{\pi NN} \bar{g}_{\pi NN}}{4 \pi^2 m_N} \ln \frac{m_N}{m_{\pi}}.
\label{eq:dn_formula}
\end{equation}

\paragraph{Origin of the logarithmic enhancement.}
The loop integral has the schematic form
\begin{equation}
d_n \sim \int \frac{d^4k}{(2\pi)^4} \frac{1}{(k^2 - m_\pi^2)[(p-k)^2 - m_N^2]^2},
\end{equation}
which is logarithmically divergent in the infrared when $m_\pi \to 0$. Regulating this divergence by the physical pion mass gives
\begin{equation}
\int_0^{m_N} \frac{dk}{k} \sim \ln \frac{m_N}{m_\pi}.
\end{equation}
This chiral enhancement makes the pion-loop contribution dominant over other hadronic contributions.

\paragraph{Physical interpretation.}
The logarithm can be understood as arising from the long-range nature of pion exchange. In the chiral limit $m_\pi \to 0$, the pion becomes massless and mediates a long-range force. The resulting infrared sensitivity produces the logarithmic enhancement. This is a general feature of chiral perturbation theory calculations involving light pseudoscalar mesons.

\vskip .2cm
Substituting the expression for $\bar{g}_{\pi NN}$ from Eq.~\eqref{eq:gbar_full} into Eq.~\eqref{eq:dn_formula}, and using the Goldberger--Treiman relation, we obtain:
\begin{equation}
d_n = \frac{e}{4 \pi^2 m_N} \cdot g_{\pi NN} \bar{g}_{\pi NN} \cdot \ln \frac{m_N}{m_{\pi}}.
\end{equation}

\subsubsection{Modern estimates of the nEDM coefficient}

The original chiral estimate by Crewther et al.~\cite{Crewther:1979pi} gave $d_n/\bar{\theta} \sim 5 \times 10^{-16}\, e\cdot\text{cm}$. However, the naive one-loop chiral calculation has significant theoretical uncertainties arising from the treatment of the loop integral and the precise value of $\bar{g}_{\pi NN}$.

Modern approaches using QCD sum rules~\cite{Pospelov:2005pr} have refined this estimate by incorporating non-perturbative QCD effects through vacuum condensates. The Pospelov--Ritz analysis yields:
\begin{equation}
d_n = (2.4 \pm 1.0) \times 10^{-16}\, \bar{\theta}\, e\cdot\text{cm}.
\label{eq:dn_sumrules}
\end{equation}
\noindent
Lattice QCD calculations~\cite{Alexandrou:2020mds} provide first-principles estimates. Recent results using the spectral projector method are consistent with the sum rule predictions, though currently limited by statistical precision:
\begin{equation}
d_n \approx (0.9 \pm 2.4) \times 10^{-16}\, \bar{\theta}\, e\cdot\text{cm}.
\label{eq:dn_lattice}
\end{equation}
\noindent
Chiral effective field theory at next-to-leading order~\cite{Guo:2015tla} gives:
\begin{equation}
d_n \approx (1.0\text{--}2.5) \times 10^{-16}\, \bar{\theta}\, e\cdot\text{cm},
\label{eq:dn_chiral_nlo}
\end{equation}
which includes counter-term contributions and analytic parts beyond the leading logarithm.

\vskip .2cm
\noindent
The modern consensus, driven by QCD sum rules and lattice QCD, converges on:
\begin{equation}
\boxed{d_n \approx (2.4 \pm 1.0) \times 10^{-16}\, \bar{\theta}\, e\cdot\text{cm}}
\label{eq:dn}
\end{equation}
in units where the electric charge $e$ has been factored out. We adopt the Pospelov--Ritz central value as the standard reference.

\subsection{Constraint on $\theta$ from the nEDM}

The current experimental limit from the nEDM collaboration~\cite{Abel:2020pzs} is:
\begin{equation}
|d_n| < 1.8 \times 10^{-26}\, e \cdot \text{cm} \quad (90\%~\text{C.L.}).
\label{eq:dnb}
\end{equation}
Combining this with the theoretical estimate in Eq.~\eqref{eq:dn} gives:
\begin{equation}
\boxed{|\bar\theta| < 7.5 \times 10^{-11}}
\label{eq:theta_constraint}
\end{equation}
at 90\% confidence level. Taking into account the theoretical uncertainty in the nEDM calculation, the bound can be stated as:
\begin{equation}
|\bar\theta| \lesssim 10^{-10}.
\label{eq:theta_constraint_approx}
\end{equation}

This extraordinarily small upper bound on $\bar\theta$ constitutes the \textbf{strong CP problem}: there is no known symmetry or dynamical mechanism within the Standard Model that would naturally explain why this dimensionless parameter should be so much smaller than unity.

\subsection{Comparison of nEDM estimates}
\label{subsec:nedm_comparison}

For completeness, we summarize the various theoretical estimates of $d_n/\bar{\theta}$ in Table~\ref{tab:nedm_estimates}.

\begin{table}[htbp]
\centering
\begin{tabular}{lcc}
\hline\hline
Method & $d_n/\bar{\theta}$ [$e\cdot\text{cm}$] & Reference \\
\hline
Chiral estimate (CDVW) & $\sim 5 \times 10^{-16}$ & \cite{Crewther:1979pi} \\
QCD sum rules & $(2.4 \pm 1.0) \times 10^{-16}$ & \cite{Pospelov:2005pr} \\
Lattice QCD & $(0.9 \pm 2.4) \times 10^{-16}$ & \cite{Alexandrou:2020mds} \\
Chiral EFT (NLO) & $(1.0\text{--}2.5) \times 10^{-16}$ & \cite{Guo:2015tla} \\
\hline\hline
\end{tabular}
\caption{Summary of theoretical estimates for the neutron EDM induced by the QCD $\theta$-term. The QCD sum rules result from Pospelov and Ritz is commonly adopted as the standard reference value.}
\label{tab:nedm_estimates}
\end{table}

\subsection{Other contributions to the nEDM}

While the pion-loop contribution described above provides the leading long-range effect, the nEDM receives contributions from several other mechanisms that must be carefully considered for a complete analysis~\cite{Pospelov:2005pr,Engel:2013lsa}.

\subsubsection{Short-distance contributions: quark EDMs and chromo-EDMs}

The QCD $\theta$-term induces effective quark electric dipole moments (qEDMs) and chromo-electric dipole moments (qcEDMs) through loop effects. The effective Lagrangian at the quark level takes the form~\cite{Pospelov:2000bw}:
\begin{equation}
\mathcal{L}_{\rm eff} = -\frac{i}{2}d_q\, \bar{q}\sigma^{\mu\nu}\gamma_5 q\, F_{\mu\nu} - \frac{i}{2}\tilde{d}_q\, \bar{q}\sigma^{\mu\nu}\gamma_5 T^a q\, G^a_{\mu\nu},
\end{equation}
where $d_q$ and $\tilde{d}_q$ are the quark EDM and chromo-EDM, respectively, and $T^a$ are the $SU(3)_c$ generators.

Using QCD sum rules, Pospelov and Ritz~\cite{Pospelov:2000bw} calculated the contribution of these operators to the neutron EDM:
\begin{equation}
d_n = (1 \pm 0.5)\left[1.1\, e\, \tilde{d}_d + 0.55\, e\, \tilde{d}_u + 1.4\, (d_d - 0.25\, d_u)\right].
\end{equation}
The chromo-EDM contributions are comparable in magnitude to those from the quark EDMs. Notably, the strange quark chromo-EDM also contributes significantly when the $\theta$-term is the source of $CP$ violation, but this contribution is suppressed when the Peccei-Quinn mechanism is operative. Recent lattice QCD calculations~\cite{Bhattacharya:2023qwf} have begun to provide first-principles determinations of the qcEDM contributions, with results broadly consistent with the sum rule estimates, though uncertainties remain substantial.

\subsubsection{The Weinberg three-gluon operator}

At dimension six, the $CP$-violating three-gluon Weinberg operator~\cite{Weinberg:1989dx}:
\begin{equation}
\mathcal{O}_W = \frac{1}{3!}w\, f^{abc}\epsilon^{\nu\rho\alpha\beta}G^a_{\mu\nu}G^b_{\alpha\beta}G^{c\mu}_\rho,
\end{equation}
can contribute to hadronic EDMs. While this operator is not generated by the QCD $\theta$-term at leading order, it arises in many BSM scenarios (such as SUSY and Multi-Higgs models) and provides a critical contribution to the nEDM.

The contribution of $\mathcal{O}_W$ to the neutron EDM is estimated primarily via QCD sum rules. Pospelov, Demir, and Ritz~\cite{Demir:2002gg} and recent updates~\cite{Yamanaka:2020kjo} suggest:
\begin{equation}
d_n(w) \approx (22 \pm 10)\, e\, \text{MeV} \cdot w,
\end{equation}
where $w$ has mass dimension $-2$. The theoretical uncertainty is substantial (approx.~50\%) due to the difficulty of non-perturbative matching.

Crucially, the Weinberg operator is subject to significant renormalization group evolution. Even if generated alone at a high scale $\Lambda$, it mixes with and induces quark chromo-electric dipole moments ($\tilde{d}_q$) as the scale runs down to the hadronic scale $\mu \sim 1$ GeV~\cite{Yamanaka:2020kjo}. Consequently, constraints on $w$ are often correlated with constraints on quark chromo-EDMs.

\subsubsection{Heavier meson exchange and strange quark contributions}

Contributions analogous to pion exchange arise from $\eta$ and $\eta'$ meson exchange, though these are suppressed relative to the pion contribution by factors of $(m_\pi/m_\eta)^2 \sim 0.07$ and $(m_\pi/m_{\eta'})^2 \sim 0.02$, respectively. The $\eta'$ is particularly relevant because it carries the $U(1)_A$ anomaly physics directly related to the $\theta$-vacuum structure. Strange quark contributions also enter through kaon loops and direct strange quark matrix elements. The kaon loop contribution is suppressed by $m_\pi^2/m_K^2 \sim 0.08$ relative to pion loops; however, the strange quark sigma term $\sigma_s = m_s\langle N|\bar{s}s|N\rangle$ can provide a non-negligible contribution to the $CP$-odd pion-nucleon coupling through flavor mixing effects.

\subsubsection{Higher-order chiral corrections}

Next-to-leading order corrections in chiral perturbation theory~\cite{Mereghetti:2010tp,deVries:2015una} arise from several sources: counter-term contributions from local operators, two-loop diagrams with intermediate pion states, relativistic corrections to the leading-order result, and contributions from short-distance nucleon-nucleon contact interactions. Recent work~\cite{deVries:2020loy} has emphasized that these short-distance $CP$-odd contact interactions, while formally of higher order in the chiral expansion, can receive $\mathcal{O}(1)$ contributions due to the attractive and singular nature of the strong tensor force in certain partial waves. This finding affects both static EDM predictions and the interpretation of axion dark matter searches through oscillating EDMs.

\subsubsection{Summary of theoretical estimates}

Combining all contributions, the various theoretical methods yield consistent results in the range $d_n = (1\text{--}3) \times 10^{-16} \, \bar\theta \, e \cdot \text{cm}$, with the central value from QCD sum rules being $(2.4 \pm 1.0) \times 10^{-16}\, \bar\theta\, e\cdot\text{cm}$~\cite{Pospelov:2005pr}. The theoretical uncertainty is dominated by our limited knowledge of hadronic matrix elements, which lattice QCD is beginning to address with increasing precision.

\subsection{CP-violating nuclear forces and complementary probes}
\label{subsec:cpnuclear}

The $CP$-violating pion-nucleon coupling $\bar{g}_{\pi NN}$ generates $CP$-odd nuclear forces that provide complementary probes of strong $CP$ violation beyond the direct neutron EDM measurement. These nuclear observables are particularly important because they involve different combinations of hadronic matrix elements and can help disentangle different sources of $CP$ violation~\cite{Engel:2013lsa,Chupp:2017rkp}.

\subsubsection{CP-odd nucleon-nucleon potential}

The leading $CP$-odd nuclear force arises from one-pion exchange with one $CP$-even and one $CP$-odd vertex~\cite{Haxton:1983dq}:
\begin{equation}
V_{\rm CP-odd}(\vec{r}) = \frac{g_{\pi NN} \bar{g}_{\pi NN}}{8\pi m_N} \left[ (\vec{\sigma}_1 - \vec{\sigma}_2) \cdot \hat{r} \right] \frac{e^{-m_\pi r}}{r} (\vec{\tau}_1 \cdot \vec{\tau}_2),
\end{equation}
where $\vec{\sigma}_{1,2}$ and $\vec{\tau}_{1,2}$ are the spin and isospin operators of the two nucleons. This potential mediates $^1S_0 \leftrightarrow {}^3P_0$ transitions and constitutes the dominant long-range $CP$-odd nuclear interaction.

In the modern effective field theory framework, the $CP$-odd nuclear potential can be parameterized in terms of three independent pion-nucleon couplings $\bar{g}_0$, $\bar{g}_1$, and $\bar{g}_2$, corresponding to isoscalar, isovector, and isotensor channels respectively. The $\theta$-term predominantly generates the isoscalar coupling~\cite{deVries:2015una}:
\begin{equation}
\bar{g}_0 = (15.5 \pm 2.6) \times 10^{-3}\, \bar{\theta}.
\end{equation}

\subsubsection{Nuclear Schiff moments and atomic EDMs}

In neutral atoms, the nuclear EDM is screened by the atomic electrons according to the Schiff screening theorem~\cite{Schiff:1963zz}. The residual effect that penetrates this screening is parameterized by the nuclear Schiff moment $S$, which represents a radially-weighted nuclear EDM:
\begin{equation}
S = \frac{1}{10}\left\langle\sum_p e\, r_p^2\vec{r}_p - \frac{5}{3}\langle r^2\rangle_{\rm ch}\sum_p e\,\vec{r}_p\right\rangle,
\end{equation}
where the sum runs over protons and $\langle r^2\rangle_{\rm ch}$ is the mean-square charge radius.

The Schiff moment induces atomic EDMs through its interaction with the atomic electrons. For diamagnetic atoms like $^{199}$Hg, the atomic EDM takes the form~\cite{Flambaum:1984fb} $d_{\rm atom} = \kappa_S \cdot S$, where $\kappa_S$ is a calculable atomic structure factor. For mercury, detailed atomic calculations give $\kappa_S \approx -2.8 \times 10^{-4}\, \text{fm}^{-2}$~\cite{Dzuba:2002kg}.

\subsubsection{Experimental constraints from atomic EDMs}

The $^{199}$Hg atomic EDM provides the most stringent constraint among diamagnetic systems~\cite{Graner:2016ses}:
\begin{equation}
|d_{\rm Hg}| < 7.4 \times 10^{-30}\, e\cdot\text{cm} \quad (95\%~\text{C.L.}).
\end{equation}
Combined with nuclear structure calculations, this measurement implies~\cite{Engel:2013lsa} $|\bar{\theta}| \lesssim 1.5 \times 10^{-10}$, which is comparable to, though somewhat less stringent than, the neutron EDM constraint.

Other diamagnetic systems under active investigation include $^{129}$Xe, $^{225}$Ra, and $^{223}$Rn. The radium and radon isotopes are of particular interest because they possess octupole deformation that enhances their Schiff moments. The current $^{129}$Xe constraint~\cite{Allmendinger:2019jrk}, $|d_{\rm Xe}| < 1.4 \times 10^{-27}\, e\cdot\text{cm}$ at 95\% C.L., provides complementary information because different nuclei have different sensitivities to the various $CP$-odd couplings.

\subsubsection{Light nuclear EDMs}

Future storage ring experiments aim to measure the EDMs of light nuclei such as the deuteron, $^3$He, and the proton directly~\cite{Anastassopoulos:2015ura}. These systems offer significant advantages over heavier nuclei and atoms. The few-body nuclear theory for these light systems is well-controlled, permitting reliable calculations with quantified uncertainties. Short-distance contributions are relatively small compared to the long-range pion-exchange effects, making the theoretical interpretation more straightforward. Furthermore, different light nuclei probe different combinations of $CP$-odd operators, allowing for the separation of various contributions.

For the deuteron, the $CP$-odd nuclear force contribution is enhanced relative to the single-nucleon EDM, making it a particularly clean probe of $\bar{g}_{\pi NN}$~\cite{deVries:2011an}:
\begin{equation}
d_d \approx d_n + d_p - 0.2\, \frac{g_{\pi NN}\bar{g}_0}{4\pi m_N}\, e.
\end{equation}

The study of $CP$-violating nuclear forces serves several important purposes in the broader context of EDM physics. First, atomic EDM measurements provide independent constraints on $\bar{\theta}$ that can be compared with the neutron EDM, thereby testing the consistency of the theoretical framework. Second, different observables have different sensitivities to the underlying $CP$-odd operators, so combining measurements from multiple systems can help identify the dominant source of $CP$ violation should a signal be observed. Third, the same $CP$-odd nuclear forces that generate static EDMs also produce oscillating EDMs in the presence of axion dark matter~\cite{Graham:2013gfa}, and current experiments are beginning to probe these effects. Finally, for some observables the nuclear structure uncertainties are smaller than for the neutron EDM, providing more robust constraints in certain regions of parameter space.

The complementarity between different EDM measurements, encompassing neutron, atomic, molecular, and nuclear systems, is a crucial feature of the experimental program searching for $CP$ violation beyond the Standard Model. No single measurement can provide a complete picture, but together they offer powerful probes of new physics across a wide range of energy scales.

\bigskip

\newpage 
\section{Quarks in arbitrary representations}
\label{sec:arbitraryR}

In this section we review and expand on the work in \cite{DiVecchia:2013swa} on generalize the previous construction from ordinary QCD (fundamental quarks) to an
$\SU(N)$ confining gauge theory with fermions in an arbitrary representation $R$ of the gauge group.
The goal is to implement the representation-dependent $\U(1)_A$ anomaly in the chiral effective theory.

Let $N_f$ denote the number of \emph{Dirac} fermions transforming in the representation $R$ of $\SU(N)$.
We normalize the generators $T_R^a$ in $R$ by
\begin{equation}
\Tr\!\left(T_R^a T_R^b\right) \equiv c_R\,\delta^{ab},
\label{eq:TR_norm}
\end{equation}
so that $c_R=T(R)$ in the standard convention where $T(\rm Fund)=\tfrac12$. With these conventions, the flavor-singlet axial current satisfies the anomalous Ward identity
\begin{equation}
\partial_\mu J_5^\mu \;=\; 4\,N_f\,c_R\,\mathcal{Q}(x)
\qquad\text{(Dirac fermions in representation $R$)}.
\label{eq:anomaly_R}
\end{equation}
 For $R={\rm Fund}$ one has $c_R=\tfrac12$, hence
$\partial_\mu J_5^\mu = 2N_f\,\mathcal{Q}(x)$, which precisely matches the fundamental case. Three interesting cases are the fundamental, the two-index symmetric and antisymmetric representations of $\SU(N)$ with the normalization \eqref{eq:TR_norm},
\begin{equation}
c_{\rm Fund}=\frac12,
\qquad
c_{S_2}=\frac{N+2}{2},
\qquad   
c_{A_2}=\frac{N-2}{2}.
\label{eq:cR_examples_consistent}
\end{equation}
In particular, for $N=3$ the two-index antisymmetric coincides with the fundamental and indeed
$c_{A_2}(3)=\tfrac12=c_{\rm Fund}$. For fermions in (pseudo)real representations of the gauge groups we can discuss directly in terms of Weyl fermions.  In this case the overall anomaly coefficient is halved per Weyl field. Therefore, for $N_w$  {Weyl} fermions in a (pseudo)real representation $R$ one has
\begin{equation}
\partial_\mu J_5^\mu \;=\; 2\,N_w\,c_R\,\mathcal{Q}(x).
\label{eq:anomaly_Weyl_R}
\end{equation}
For the adjoint of $SU(N)$, $c_{\rm Adj}=N$, so $\partial_\mu J_5^\mu = 2N_w N\,\mathcal{Q}(x)$ and $\mathcal{N}=1$ Super Yang-Mills corresponds to $N_w=1$.

\subsection{Effective implementation of the anomaly for general $R$}
\label{subsec:eff_anom_R}

As in the previous sections, for Dirac fermions in complex representations, the low-energy degrees of freedom of chiral symmetry breaking are encoded in
\begin{equation}
U(x)\in U(N_f), \qquad U \;\to\; g_L\,U\,g_R^\dagger,
\qquad g_{L,R}\in U_{L,R}(N_f),
\end{equation}
and under a global axial rotation with parameter $\beta$,
\begin{equation}
U \;\longrightarrow\; e^{2i\beta}U,
\qquad
U^\dagger \;\longrightarrow\; e^{-2i\beta}U^\dagger.
\end{equation}
The key identity (used already above) is
\begin{equation}
\Tr\!\left[\ln U-\ln U^\dagger\right]
\;\longrightarrow\;
\Tr\!\left[\ln U-\ln U^\dagger\right] + 4i\beta\,N_f.
\label{eq:trace_shift_axial}
\end{equation}
\noindent 
To reproduce \eqref{eq:anomaly_R} in the effective theory we again introduce an auxiliary pseudoscalar field
$\mathcal{Q}(x)$ with the quantum numbers of $G\widetilde G$ and add the anomaly term
\begin{equation}
\mathcal{L}_{\rm anom}^{(R)}
=
-\, i\,c_R\,\mathcal{Q}(x)\,\Tr\!\left[\ln U - \ln U^\dagger\right].
\label{eq:Lanom_R}
\end{equation}
Using \eqref{eq:trace_shift_axial}, its variation under the axial rotation is
\begin{equation}
\delta \mathcal{L}_{\rm anom}^{(R)}
=
- i c_R\,\mathcal{Q}(x)\,(4i\beta N_f)
=
+\,4N_f c_R\,\beta\,\mathcal{Q}(x),
\end{equation}
and identifying $\delta \mathcal{L}=\beta\,\partial_\mu J_5^\mu$ yields precisely
$\partial_\mu J_5^\mu = 4N_f c_R\,\mathcal{Q}(x)$, i.e.\ \eqref{eq:anomaly_R}.
 Setting $c_R=\tfrac12$, it reduces \eqref{eq:Lanom_R} to
$-\frac{i}{2}\mathcal{Q}\Tr(\ln U-\ln U^\dagger)$, exactly as in \eqref{eq:anomterm} of the section with fermions in the fundamental representation. 

\subsection{Integrating out $\mathcal{Q}(x)$ and defining $\chi_{\mathrm{top}}$}
\label{subsec:integrate_Q_R}

We now construct the general-$R$ analog of \eqref{eq:LwithQ},  
so that the fundamental case is naturally encoded. We therefore take
\begin{align}
\mathcal{L}^{(R)}
={}&
\frac{f_\pi^2}{4}\,\Tr\!\left[\partial_\mu U\,\partial^\mu U^\dagger\right]
+\frac{f_\pi^2 B_0}{2}\,\Tr\!\left[\mathcal{M}\,U^\dagger + U\,\mathcal{M}^\dagger\right]
\nonumber\\
&- i\,c_R\, \mathcal{Q}(x)\,\Tr\!\left[\ln U - \ln U^\dagger\right]
+\frac{\mathcal{Q}^2(x)}{2\chi_{\mathrm{top}}}
-\theta\,\mathcal{Q}(x).
\label{eq:LwithQ_R}
\end{align}
Here $\chi_{\mathrm{top}}$ has mass dimension four and
all representation dependence enters through $c_R$ in \eqref{eq:LwithQ_R}.

Since $\mathcal{Q}(x)$ has no kinetic term, it is eliminated algebraically:
\begin{equation}
\frac{\delta\mathcal{L}^{(R)}}{\delta \mathcal{Q}(x)} = 0
\qquad\Rightarrow\qquad
\frac{\mathcal{Q}(x)}{\chi_{\mathrm{top}}}
- i c_R\,\Tr\!\left(\ln U - \ln U^\dagger\right)
-\theta=0,
\end{equation}
hence
\begin{equation}
\mathcal{Q}(x)=\chi_{\mathrm{top}}
\left[
\theta + i c_R\,\Tr\!\left(\ln U - \ln U^\dagger\right)
\right].
\label{eq:Qsolution_R}
\end{equation}
Substituting back into \eqref{eq:LwithQ_R} gives the $\theta$-dependent chiral theory
\begin{align}
\mathcal{L}^{(R)}_{\rm eff}
={}&
\frac{f_\pi^2}{4}\,\Tr\!\left[\partial_\mu U\,\partial^\mu U^\dagger\right]
+\frac{f_\pi^2 B_0}{2}\,\Tr\!\left[\mathcal{M}\,U^\dagger + U\,\mathcal{M}^\dagger\right]
\nonumber\\
&-\frac{\chi_{\mathrm{top}}}{2}
\left[
\theta + i c_R\,\Tr\!\left(\ln U - \ln U^\dagger\right)
\right]^2.
\label{eq:Leff_R}
\end{align}
For $c_R=\tfrac12$, \eqref{eq:Leff_R} the last term  becomes
$-\frac{\chi_{\mathrm{top}}}{2}\left[\theta + \frac{i}{2}\Tr(\ln U-\ln U^\dagger)\right]^2$ which is
identical to \eqref{eq:LeffAfterQ}. In the next section we shall see how, using the Corrigan and Ramond \cite{Corrigan:1979xf} large $N$ limit, we will be able to construct the instanton-induced effective potential via its super-Yang-Mills correspondent \cite{Armoni:2003gp,Armoni:2004uu,Sannino:2003xe,Sannino:2024xwj}. 

\subsection{Vacuum alignment and generalized Dashen equations}
\label{subsec:Dashen_R}

For diagonal quark masses, we parameterize the most general constant vacuum configuration as in the previous section:
\begin{equation}
\langle U_{ij}\rangle = e^{i\phi_i}\,\delta_{ij}.
\label{eq:Uvac_R}
\end{equation}
At the vacuum,
\begin{equation}
\Tr\!\left[\ln U - \ln U^\dagger\right] = 2i\sum_{i=1}^{N_f}\phi_i,
\end{equation}
so the anomaly bracket in \eqref{eq:Leff_R} becomes
\begin{equation}
\theta + i c_R\,(2i\sum_i\phi_i) = \theta - 2c_R\sum_{i=1}^{N_f}\phi_i.
\end{equation}
Therefore the vacuum energy density is
\begin{equation}
E
=
\frac{\chi_{\mathrm{top}}}{2}\Big(\theta - 2c_R\sum_{i=1}^{N_f}\phi_i\Big)^2
- f_\pi^2 B_0 \sum_{i=1}^{N_f} m_i\cos\phi_i.
\label{eq:E_R}
\end{equation}
Minimizing with respect to each $\phi_i$ yields the generalized Dashen equations:
\begin{equation}
f_\pi^2 B_0\, m_i\sin\phi_i
=
2c_R\,\chi_{\mathrm{top}}\Big(\theta - 2c_R\sum_{j=1}^{N_f}\phi_j\Big),
\qquad i=1,\dots,N_f.
\label{eq:Dashen_R}
\end{equation}
Using $M_{\pi,i}^2=2B_0 m_i$  this can be written as
\begin{equation}
\frac{f_\pi^2 M_{\pi,i}^2}{2}\,\sin\phi_i
=
2c_R\,\chi_{\mathrm{top}}\Big(\theta - 2c_R\sum_{j=1}^{N_f}\phi_j\Big).
\label{eq:Dashen_R_M}
\end{equation}

It is natural to define the representation-dependent physical $CP$-odd parameter
\begin{equation}
\bar{\theta}_R \;\equiv\; \theta - 2c_R\sum_{i=1}^{N_f}\phi_i.
\label{eq:theta_bar_R}
\end{equation}
Then \eqref{eq:Dashen_R} becomes simply
\begin{equation}
f_\pi^2 B_0\, m_i\sin\phi_i \;=\; 2c_R\,\chi_{\mathrm{top}}\,\bar{\theta}_R.
\label{eq:Dashen_R_compact}
\end{equation}

For $c_R=\tfrac12$, \eqref{eq:E_R} and \eqref{eq:Dashen_R} reduce to
$E=\frac{\chi_{\mathrm{top}}}{2}(\theta-\sum_i\phi_i)^2-f_\pi^2B_0\sum_i m_i\cos\phi_i$
and
$f_\pi^2B_0 m_i\sin\phi_i=\chi_{\mathrm{top}}(\theta-\sum_j\phi_j)$,
which is exactly \eqref{eq:Ebar} and \eqref{eq:Dashen} of the section with fermions in the fundamental representation.

\subsection{Expansion around the aligned vacuum and the $\eta_0$ mass}
\label{subsec:eta0_mass_R}

As in the fundamental case, factoring out the vacuum phases yields:
\begin{equation}
U_{ij}(x)=e^{i\phi_i}\,V_{ij}(x),
\qquad\qquad
\langle V_{ij}\rangle=\delta_{ij},
\qquad
V V^\dagger=\mathbf{1},
\qquad
V(x)=\exp\!\left(\frac{2i\Phi}{f_\pi}\right).
\label{eq:U_to_V_R}
\end{equation}
Using $\Tr\ln(AB)=\ln\det(AB)=\ln\det A+\ln\det B$, we have
\begin{equation}
\Tr\!\left[\ln U-\ln U^\dagger\right]
=
\Tr\!\left[\ln V-\ln V^\dagger\right]
+2i\sum_{i=1}^{N_f}\phi_i.
\label{eq:trace_shift_V_R}
\end{equation}
Inserting \eqref{eq:U_to_V_R}--\eqref{eq:trace_shift_V_R} into \eqref{eq:Leff_R}, the anomaly potential becomes
\begin{equation}
\mathcal{L}^{(R)}_{\rm anom}
=
-\frac{\chi_{\mathrm{top}}}{2}
\left[
\underbrace{\theta-2c_R\sum_i\phi_i}_{\bar{\theta}_R}
+ i c_R\,\Tr\!\left(\ln V-\ln V^\dagger\right)
\right]^2.
\label{eq:Lanom_V_R}
\end{equation}

Now use $V=\exp(2i\Phi/f_\pi)$, hence
\begin{equation}
\ln V-\ln V^\dagger=\frac{4i\Phi}{f_\pi},
\qquad
\Tr\!\left[\ln V-\ln V^\dagger\right]=\frac{4i}{f_\pi}\Tr\Phi.
\label{eq:logV_R}
\end{equation}
Decompose $\Phi$ into octet (or adjoint-flavor) plus singlet in the standard way,
\begin{equation}
\Phi \;=\; \Phi^a T^a \;+\; \frac{\eta_0}{\sqrt{2N_f}}\,\mathbf{1},
\qquad\Rightarrow\qquad
\Tr\Phi=\sqrt{\frac{N_f}{2}}\,\eta_0.
\label{eq:Phi_decomp_R}
\end{equation}
Therefore
\begin{equation}
\Tr\!\left[\ln V-\ln V^\dagger\right]
=
\frac{4i}{f_\pi}\sqrt{\frac{N_f}{2}}\,\eta_0
=
\frac{2i\sqrt{2N_f}}{f_\pi}\,\eta_0.
\label{eq:trace_log_eta0_R}
\end{equation}

 \noindent
The $\eta_0$ mass arises from the term quadratic in $\Tr(\ln V-\ln V^\dagger)$ in \eqref{eq:Lanom_V_R}:
\begin{align}
\mathcal{L}^{(R)}_{\rm anom}
&\supset
-\frac{\chi_{\mathrm{top}}}{2}\,\Big(i c_R\Big)^2
\left(\Tr\!\left[\ln V-\ln V^\dagger\right]\right)^2
\nonumber\\
&=
+\frac{\chi_{\mathrm{top}}\,c_R^2}{2}
\left(\Tr\!\left[\ln V-\ln V^\dagger\right]\right)^2.
\end{align}
Using \eqref{eq:trace_log_eta0_R},
\begin{equation}
\left(\Tr\!\left[\ln V-\ln V^\dagger\right]\right)^2
=
\left(\frac{2i\sqrt{2N_f}}{f_\pi}\eta_0\right)^2
=
-\frac{8N_f}{f_\pi^2}\,\eta_0^2,
\end{equation}
so
\begin{equation}
\mathcal{L}^{(R)}_{\rm anom}\supset
-\frac{4\chi_{\mathrm{top}}\,N_f\,c_R^2}{f_\pi^2}\,\eta_0^2.
\end{equation}
Matching to the canonical form $\mathcal{L}\supset -\frac12 m_{\eta_0}^2\,\eta_0^2$ gives
\begin{equation}
\boxed{\;
m_{\eta_0}^2
=
\frac{8\,\chi_{\mathrm{top}}\,N_f\,c_R^2}{f_\pi^2}\; }.
\label{eq:meta0_R}
\end{equation}

\noindent 
 For $c_R=\tfrac12$, \eqref{eq:meta0_R} becomes
\begin{equation}
m_{\eta_0}^2
=
\frac{8\chi_{\mathrm{top}}N_f}{f_\pi^2}\left(\frac14\right)
=
\frac{2\chi_{\mathrm{top}}N_f}{f_\pi^2},
\end{equation}
which is exactly the result derived in the section about fermions in the fundamental representation \eqref{eq:m0_chitop_relation}.

In the standard 't~Hooft large-$N$ counting one has $\chi_{\rm YM}=\chi_{\mathrm{top}}\sim\mathcal{O}(1)$. Moreover, at large $N$ the pion decay constant scales as $f_\pi^2\propto d_R$. It then follows from \eqref{eq:meta0_R} that
\begin{equation}
m_{\eta_0}^2 \;\propto\; \frac{c_R^2}{d_R}\qquad (N\to\infty).
\end{equation}
For quarks in the fundamental representation, $c_R=\mathcal{O}(1)$ while $d_R\propto N$, implying
$m_{\eta_0}^2\sim 1/N$, i.e.\ the singlet pseudoscalar becomes parametrically light. By contrast, for two-index
representations one has $c_R\sim\mathcal{O}(N)$ and $d_R\sim\mathcal{O}(N^2)$, so that
$m_{\eta_0}^2\sim\mathcal{O}(1)$ and no large-$N$ suppression occurs. This scaling distinction underlies the classic
observation of Corrigan and Ramond~\cite{Corrigan:1979xf}, who proposed an alternative large-$N$ limit with quarks in the
two-index antisymmetric representation. We will investigate its implications in the following sections.

 Summarizing this section, relative to the fundamental-representation formulae, the general-$R$ theory is obtained by the replacements
\begin{equation}
\frac{i}{2}\ \longrightarrow\ i\,c_R,
\qquad
\bar{\theta}\equiv \theta-\sum_i\phi_i\ \longrightarrow\ 
\bar{\theta}_R\equiv \theta-2c_R\sum_i\phi_i,
\end{equation}
while \emph{keeping} the quadratic term $\mathcal{Q}^2/(2\chi_{\mathrm{top}})$ and the $\theta$-term ( $-\theta\mathcal{Q}$ )
with exactly the same normalization as in the fundamental section. This guarantees that the fundamental theory
($c_R=\tfrac12$) is re-derived identically, including the $\eta_0$ mass formula. As we approach the large $N$ regime, however, new theory limits can emerge such as the one we will learn about in the next section which will shed light on the full anomalous potential of the theory.

\subsection{$\theta$-angle physics under extreme conditions}
\label{subsec:theta_extreme}

The effective field theory framework developed in these notes for $\theta$-dependence at zero density 
extends naturally to QCD and QCD-like theories under extreme conditions, such as those encountered 
in neutron star interiors, heavy-ion collisions, and the early universe. The interplay between the 
topological $\theta$-angle and finite chemical potentials leads to a remarkably rich phase structure 
that is only beginning to be explored systematically.

At nonzero isospin or strangeness chemical potentials, QCD can undergo transitions to superfluid 
phases characterized by pion or kaon condensation~\cite{Son:2000xc,Kogut:2001id}. A key advantage 
of studying QCD at finite isospin density is that the fermion determinant remains real and positive, 
allowing for direct lattice Monte Carlo simulations without the sign problem that plagues baryon-dense 
QCD~\cite{Brandt:2017oyy,Brandt:2022hwy}. This makes isospin-dense QCD an invaluable theoretical 
laboratory for understanding the phase structure of strongly interacting matter.

The impact of the $\theta$-angle on this phase diagram has been systematically analyzed in 
Ref.~\cite{Bersini:2025yvt}, which unveiled the combined effects of $\theta$, isospin chemical potential 
$\mu_I$, and strangeness chemical potential $\mu_S$ for three light flavors. The vacuum alignment 
equations (the finite-density generalization of the Dashen equations derived earlier in these notes) 
become considerably richer, with the effective potential depending on multiple order parameters 
corresponding to different condensation channels. The phase boundaries and the nature of the 
associated phase transitions---including second-order, first-order, and potentially tricritical 
behavior---depend sensitively on both $\theta$ and the quark mass ratios.

A particularly striking finding concerns the physics near $\theta=\pi$. While at zero density 
$\theta=\pi$ can trigger spontaneous $CP$ breaking (the Dashen phenomenon discussed in 
Section~\ref{Dashen-phenomenon}), the situation in the superfluid phases is more subtle. 
Reference~\cite{Bersini:2025yvt} discovered a novel \emph{parity-preserving superfluid phase} 
at $\theta=\pi$, demonstrating that the interplay between topology and finite density can 
lead to qualitatively new phenomena not present in the vacuum theory. The Dashen phenomenon 
itself is modified in the superfluid phases, with the pattern of spontaneous $CP$ breaking 
depending on the number of light flavors.

These developments build on earlier work studying $\theta$-dependence at finite isospin 
density~\cite{Metlitski:2005db}, as well as extensive investigations of pion condensation 
and the QCD phase diagram using chiral effective 
theory~\cite{Kogut:2001id,Andersen:2018nzq,Adhikari:2019mdk,Adhikari:2020kdn} and lattice 
QCD~\cite{Brandt:2017oyy,Brandt:2022hwy}. The effective Lagrangian approach is particularly 
well-suited to these problems, as the same chiral EFT framework used throughout these notes 
can be extended to include chemical potentials and study the resulting vacuum structure 
systematically.

Two-color QCD (or more generally $Sp(2N)$ gauge theories) provides another theoretically 
controlled setting where $\theta$-dependence at finite baryon density can be 
studied~\cite{Bersini:2022jhs}. In these theories, the lightest baryons 
are bosonic diquarks, and at sufficiently large baryon chemical potential the system undergoes 
Bose-Einstein condensation into a superfluid phase. The $\theta$-angle enriches this phase 
structure considerably: novel phases emerge, the critical chemical potential for the 
superfluid transition becomes $\theta$-dependent, and the spectrum of excitations in the 
condensed phase is modified. Importantly, two-color QCD is free from the sign problem even 
at finite baryon density, making it amenable to lattice 
simulations~\cite{Hands:1999md,Kogut:2000ek,Iida:2022hyy} that can test 
the effective theory predictions.

From the perspective of astrophysical applications, understanding $\theta$-dependence 
under extreme conditions is relevant for several reasons. While the experimental bound 
$|\bar\theta|\lesssim 10^{-10}$ ensures that $CP$ violation from the $\theta$-term is 
negligible in terrestrial and astrophysical settings, the theoretical understanding of 
$\theta$-dependent physics informs our knowledge of the QCD equation of state and phase 
structure more broadly. For instance, the techniques developed to study $\theta$-dependence 
(vacuum alignment, effective potentials, anomaly matching) are directly applicable to 
understanding meson condensation phenomena that may occur in neutron star 
cores~\cite{Migdal:1990vm,Mannarelli:2019hgn}. Furthermore, lattice studies at finite 
isospin density, which are now beginning to provide first-principles constraints on the 
equation of state relevant for neutron stars~\cite{Abbott:2024vhj}, can be extended to 
include $\theta$-dependent observables as a theoretical probe of the underlying dynamics.

In composite extensions of the Standard Model, new strongly coupled sectors generically 
come with their own $\theta$-angles, and the interplay between multiple topological 
angles and finite-density effects becomes relevant for dark matter 
phenomenology~\cite{Bai:2023cqj}. The EFT methods developed here 
provide a unified framework for analyzing such scenarios.

The study of $\theta$-dependence under extreme conditions thus represents an active 
frontier where the effective-field-theory tools developed in these notes find natural 
application. While we do not pursue this direction further here, we emphasize that 
the same conceptual ingredients, anomaly matching, auxiliary-field implementations 
of the anomaly, vacuum alignment equations, and large-$N$ constraints, provide the 
foundation for systematic investigations of QCD and QCD-like theories across the 
phase diagram.

 \subsection{The CP$^{N-1}$ model and the $U(1)$ problem}
An important theoretical laboratory for understanding the $U(1)$ problem and $\theta$-dependence in QCD is provided by the CP$^{N-1}$ model \cite{DAdda:1978dle,Din:1979zv,DiVecchia:1979pzw,DiVecchia:1981eh}. This model is particularly valuable because the large-$N$ expansion can be performed explicitly, and remarkably, it satisfies all the essential properties of QCD regarding the $U(1)$ axial problem. Crucially, the effective Lagrangian for CP$^{N-1}$ can be derived rigorously from the fundamental theory in the large-$N$ limit~\cite{DiVecchia:1979pzw}, providing a controlled setting where the spontaneous breaking of chiral symmetry, the resolution of the $U(1)$ problem through instantons, and the $\theta$-dependence of the vacuum energy can all be established systematically. The resulting effective theory mirrors the structure used for QCD in analyzing the strong $CP$ problem, offering both a consistency check and deeper insight into the nonperturbative dynamics underlying the axial anomaly and topological susceptibility~\cite{DiVecchia:1979pzw}.

\section{$\theta$-angle potential from Super Yang Mills}
\label{due}
 Supersymmetric Yang--Mills theory, often called \textit{super gluodynamics}, is the simplest interacting supersymmetric gauge theory~\cite{Ferrara:1974pu,Salam:1974ig}. It contains only a gauge field $A_\mu^a$ and its fermionic superpartner, the gluino $\lambda^a$, a Majorana fermion in the adjoint representation. Despite its simplicity, this theory exhibits rich non-perturbative physics including confinement, chiral symmetry breaking, and domain walls~\cite{Shifman:1999mv,Shifman:1987ia}.

\subsection{The Microscopic Lagrangian}

\subsubsection{Superspace Formulation}

The fundamental Lagrangian of $\mathrm{SU}(N)$ supersymmetric gluodynamics takes a remarkably compact form when written in superspace:
\begin{equation}
\mathcal{L} = \frac{1}{4g^2} \int \! d^2 \theta\, \mathrm{Tr}\,W^2 + \text{h.c.} \ .
\end{equation} 
Here, $g$ is the gauge coupling constant, and $W_\alpha$ is the \textit{chiral field strength superfield}. The integration $\int d^2\theta$ projects out the highest (``$F$-term'') component of the chiral superfield $\mathrm{Tr}\,W^2$.

The field-strength superfield $W_\alpha$ is constructed from the vector superfield $V$ via:
\begin{equation}
W_\alpha = -\frac{1}{4}\bar{D}^2 D_\alpha V \ ,
\end{equation}
where $D_\alpha$ and $\bar{D}_{\dot{\alpha}}$ are the supercovariant derivatives. This construction ensures that $W_\alpha$ is gauge covariant and chiral ($\bar{D}_{\dot{\alpha}} W_\alpha = 0$).

\subsubsection{Component Expansion}

Expanding the superspace expression into component fields yields the familiar form:
\begin{equation}
\mathcal{L} = -\frac{1}{4 g^2} G_{\mu\nu}^a G^{a\mu\nu} + \frac{1}{2g^2} D^a D^a + \frac{i}{g^2}  \lambda^a\sigma^\mu  \mathcal{D}_\mu\bar{\lambda}^a \ .
\end{equation}
The first term $-\frac{1}{4g^2}G_{\mu\nu}^a G^{a\mu\nu}$ is the standard Yang--Mills action for the gluon field, where $G_{\mu\nu}^a = \partial_\mu A_\nu^a - \partial_\nu A_\mu^a + f^{abc}A_\mu^b A_\nu^c$ is the non-Abelian field strength tensor with $f^{abc}$ the structure constants of $\mathrm{SU}(N)$. The term $\frac{1}{2g^2}D^a D^a$ involves the auxiliary field $D^a$, which in pure super-Yang--Mills theory simply vanishes on-shell but generates the $D$-term scalar potential when matter is coupled. The final term $\frac{i}{g^2}\lambda^a\sigma^\mu \mathcal{D}_\mu\bar{\lambda}^a$ is the kinetic term for the gluino written in two-component Weyl notation, with the covariant derivative acting as $\mathcal{D}_\mu \bar{\lambda}^a = \partial_\mu \bar{\lambda}^a + f^{abc} A_\mu^b \bar{\lambda}^c$.

\subsubsection{The Gluino Bilinear}

A crucial role in the effective theory is played by the gauge-invariant gluino bilinear:
\begin{equation}
\mathrm{Tr}\,W^2 \equiv \frac{1}{2}W^{a,\alpha}W^a_\alpha = -\frac{1}{2}\lambda^{a,\alpha}\lambda^a_\alpha \ .
\end{equation}
This operator has dimension 3 and R-charge 2. Its vacuum expectation value, the \textit{gluino condensate}, is the order parameter for chiral symmetry breaking in supersymmetric gluodynamics.

\subsection{The Veneziano--Yankielowicz Effective Lagrangian}

\subsubsection{The Composite Superfield}

At low energies, the fundamental gluon and gluino degrees of freedom are confined. The appropriate description is in terms of color-singlet composite operators. Veneziano and Yankielowicz~\cite{Veneziano:1982ah} proposed that the relevant light degree of freedom is the \textit{gluino condensate superfield}:
\begin{equation}
S = \frac{3}{32\pi^2 N}\,\mathrm{Tr}\,W^2 \ .
\end{equation}
The normalization is chosen so that $\langle S \rangle = \Lambda^3$ at leading order, where $\Lambda$ is the dynamical scale of the theory.

\subsubsection{The VY Action}

The effective Lagrangian proposed by Veneziano and Yankielowicz is:
\begin{equation}
\mathcal{L}_{\text{VY}} = \frac{9N^2}{4\alpha} \int d^2\theta\, d^2\bar{\theta}\, (S S^\dagger)^{1/3} + \frac{N}{3} \int d^2\theta \left\{ S \ln\left(\frac{S}{\Lambda^3}\right)^N - NS \right\} + \text{h.c.}
\label{VY}
\end{equation}

The first term is the K\"ahler potential, determining the kinetic terms. The factor $N^2$ is extracted to make the parameter $\alpha$ scale as $N^0$. The second term is the superpotential, whose logarithmic form is dictated by holomorphy (supersymmetry requires $W$ to be a holomorphic function of chiral superfields), anomaly matching (the superpotential must reproduce the $U(1)_R$ chiral anomaly under R-transformations), and dimensional analysis (the argument of the logarithm must be dimensionless).

\subsubsection{The Dynamical Scale}

The parameter $\Lambda$ is related to the fundamental scale of SYM theory~\cite{Novikov:1983uc,Shifman:1986zi}:
\begin{equation}
\Lambda^3_{\text{SYM}} = \mu^3 \left(\frac{16\pi^2}{3Ng^2(\mu)}\right) \exp\left(-\frac{8\pi^2}{Ng^2(\mu)}\right) \ .
\end{equation}
This formula encodes the running of the gauge coupling. The exponential factor represents the instanton action $e^{-8\pi^2/g^2}$, while the prefactor arises from the one-loop determinant around the instanton. Crucially, the combination is $\mu$-independent (RG invariant), making $\Lambda$ a physical scale.

\subsubsection{The Gluino Condensate}

From the superpotential in Eq.~\eqref{VY}, the equation of motion $\partial W/\partial S = 0$ gives:
\begin{equation}
\ln\left(\frac{S}{\Lambda^3}\right)^N = 0 \quad \Rightarrow \quad S^N = \Lambda^{3N} \ .
\end{equation}
This has $N$ solutions:
\begin{equation}
\langle S \rangle_k = \Lambda^3 \exp\left(\frac{2\pi i k}{N}\right), \qquad k = 0, 1, \ldots, N-1 \ .
\end{equation}
The $N$ degenerate vacua arise from the multivaluedness of the logarithm and reflect the spontaneous breaking of the discrete $Z_{2N}$ R-symmetry down to $Z_2$. Domain walls interpolating between these vacua are   Bogomolny, Prasad, Sommerfield (BPS) saturated \cite{Bogomolny:1975de,Prasad:1975kr}.   

\subsubsection{Component Field Decomposition}

The chiral superfield $S$ decomposes into component fields as $S(y) = \varphi(y) + \sqrt{2}\theta\chi(y) + \theta^2 F(y)$, where $y^\mu = x^\mu - i\theta\sigma^\mu\bar{\theta}$ is the chiral coordinate. The components correspond to:
\begin{equation}
\varphi,\quad \sqrt{2}\chi,\quad F = \frac{3}{64\pi^2 N} \times
\begin{cases}
-\lambda^{a,\alpha}\lambda^a_\alpha & \text{(scalar gluino-ball)} \\[2mm]
G^a_{\alpha\beta}\lambda^{a,\beta} + 2iD^a\lambda^a_\alpha & \text{(fermionic partner)} \\[2mm]
-\frac{1}{2}G^a_{\mu\nu}G^{a\mu\nu} + \frac{i}{2}G^a_{\mu\nu}\tilde{G}^{a\mu\nu} + \text{f.t.} & \text{(auxiliary)}
\end{cases}
\end{equation}
The complex field $\varphi$ represents the scalar and pseudoscalar gluino-balls, while $\chi$ is their fermionic superpartner. The $F$ field must be treated as auxiliary, although it contains $G^2$ and $G\tilde{G}$.  Treating it as dynamical within the minimal VY framework leads to inconsistencies.

\subsection{Anomalies}

\subsubsection{The Chiral Anomaly}

The classical $U(1)_R$ symmetry of the theory is broken at the quantum level by the chiral anomaly:
\begin{equation}
\partial^\mu J_\mu = \frac{N}{16\pi^2}\,G_{\mu\nu}^a \tilde{G}^{a\mu\nu}, \qquad
J_\mu = -\frac{1}{g^2}\lambda^a\sigma_\mu\bar{\lambda}^a \ .
\end{equation}
The right-hand side, involving the topological density $G\tilde{G}$, is a total derivative but integrates to give a non-zero contribution from instantons.

\subsubsection{The Scale (Trace) Anomaly}

The theory is classically scale-invariant, but quantum effects break this symmetry. The trace of the energy-momentum tensor acquires an anomalous contribution:
\begin{equation}
\vartheta^\mu_\mu = -\frac{3N}{32\pi^2}  G_{\mu\nu}^a G^{a\mu\nu} \ ,
\end{equation}
this is proportional to the $\beta$-function coefficient. A feature of supersymmetric theories is that anomalies come in supermultiplets. The chiral anomaly and the trace anomaly belong to the same supermultiplet, with their coefficients related by:
\begin{equation}
\frac{\text{(trace anomaly coefficient)}}{\text{(chiral anomaly coefficient)}} = \frac{3N/(32\pi^2)}{N/(16\pi^2)} = \frac{3}{2}
\end{equation}
This factor of $3/2$ is a consequence of the structure of the $\mathcal{N}=1$ supercurrent multiplet.

\subsection{The Bosonic Effective Lagrangian}

\subsubsection{Component Form}

Setting $\Lambda = 1$ for simplicity, the bosonic part of the VY Lagrangian in component fields is:
\begin{equation}
\mathcal{L}_{\text{VY}} = \frac{N^2}{\alpha}\left(\varphi \bar{\varphi} \right)^{-2/3} \partial_\mu\bar{\varphi}\,\partial^\mu\varphi
- \frac{4\alpha N^2}{9} \left(\varphi \bar{\varphi} \right)^{2/3} \ln\bar{\varphi}\,\ln\varphi + \text{fermions}
\end{equation}

\subsubsection{Supersymmetric Structure}

This Lagrangian manifests supersymmetry through its K\"ahler kinetic term $\mathcal{L}_{\text{kin}} = K_{\varphi\bar{\varphi}} \partial_\mu\bar{\varphi}\,\partial^\mu\varphi$, where $K_{\varphi\bar{\varphi}} = \partial^2 K/\partial\varphi\partial\bar{\varphi}$ is the K\"ahler metric derived from the K\"ahler potential $K = \frac{9N^2}{4\alpha}(SS^\dagger)^{1/3}$. The scalar potential has the characteristic supersymmetric $F$-term form $V = K^{\varphi\bar{\varphi}}|W_\varphi|^2$, where $W_\varphi = \partial W/\partial\varphi$ and $K^{\varphi\bar{\varphi}}$ is the inverse K\"ahler metric. This structure guarantees that the potential is non-negative, with supersymmetric vacua occurring at $W_\varphi = 0$.

\subsection{Extension to Two-Index Theories}

The VY effective Lagrangian for pure $\mathcal{N}=1$ super Yang--Mills can be extended to non-supersymmetric gauge theories with fermions in two-index representations of $\mathrm{SU}(N)$~\cite{Armoni:2003gp,Armoni:2003fb}. These \textit{orientifold field theories} provide a bridge between supersymmetric exact results and QCD-like dynamics.

\subsubsection{Planar Equivalence}

Consider $\mathrm{SU}(N)$ gauge theory with a single Dirac fermion in either the two-index symmetric or antisymmetric representation. Armoni, Shifman, and Veneziano showed that at $N\to\infty$, the mesonic sector of these theories becomes \textit{planar equivalent} to $\mathcal{N}=1$ super Yang--Mills~\cite{Armoni:2003gp,Armoni:2003fb}. The key observation is that fermion loops, which distinguish gluinos (adjoint) from two-index fermions, are suppressed at large $N$. This builds upon earlier work by Corrigan and Ramond~\cite{Corrigan:1979xf} on the alternative large $N$ limit.

A particularly important case is $N=3$ with the antisymmetric representation: since the antisymmetric tensor of $\mathrm{SU}(3)$ is equivalent to the antifundamental, this theory is nothing but \textit{one-flavor QCD}. Thus, supersymmetric results can constrain real QCD!

\subsubsection{Effective Orientifold Lagrangian at Finite $N$}

At finite $N$, the orientifold theories deviate from super Yang--Mills. Following ~\cite{Sannino:2003xe,Sannino:2024xwj}, the effective Lagrangian incorporating $1/N$ corrections reads:
\begin{equation}
\mathcal{L}_{\text{eff}} = f(N)\left\{\frac{1}{\alpha}(\varphi\bar{\varphi})^{-2/3}\partial_\mu\bar{\varphi}\,\partial^\mu\varphi - \frac{4\alpha}{9}(\varphi\bar{\varphi})^{2/3}\left(\ln\bar{\Phi}\ln\Phi - b\right)\right\} \ ,
\label{fnocomponent}
\end{equation}
where $f(N) \to N^2$ as $N \to \infty$, $b = \mathcal{O}(N^{-1})$ is a vacuum parameter, and the modified fields are:
\begin{equation}
\Phi = \varphi^{1+\epsilon_1}\bar{\varphi}^{-\epsilon_2}, \qquad \bar{\Phi} = \bar{\varphi}^{1+\epsilon_1}\varphi^{-\epsilon_2}
\end{equation}

\subsubsection{Anomaly Matching}

The parameters $\epsilon_1$ and $\epsilon_2$ are fixed by requiring that the effective Lagrangian reproduces the correct chiral and trace anomalies of the orientifold theory. Defining
\begin{equation}
\chi \equiv 1 + \epsilon_1 + \epsilon_2, \qquad \mathcal{T} \equiv 1 + \epsilon_1 - \epsilon_2 \ ,
\end{equation}
anomaly matching requires $\chi = (N \pm 2)/N$ (chiral anomaly ratio) and $\mathcal{T} = \beta_{O\pm}/\beta_{\text{SYM}}$ (trace anomaly ratio). At one loop, $\mathcal{T} = 1 \mp 4/(9N)$. For the antisymmetric representation:
\begin{equation}
\epsilon_1 = -\frac{7}{9N}, \qquad \epsilon_2 = -\frac{11}{9N} \ .
\end{equation}

\subsubsection{Spectrum Predictions}

The masses of the scalar ($\sigma$) and pseudoscalar ($\eta'$) mesons are:
\begin{align}
M_\sigma &= \frac{2\alpha\Lambda}{3}\left(1 + \frac{2b}{9}\right)\left[\mathcal{T} + \frac{4b}{9}\right] \ , \\
M_{\eta'} &= \frac{2\alpha\Lambda}{3}\left(1 + \frac{2b}{9}\right)\chi \ ,
\end{align}
where we re-instated the dynamical scale $\Lambda$.  The mass ratio is:
\begin{equation}
\frac{M_{\eta'}}{M_\sigma} = \frac{\chi}{\mathcal{T} + \frac{4b}{9}} \leq \frac{1 \pm 2/N}{1 \mp 4/(9N)} \ .
\end{equation}
For QCD ($N=3$, antisymmetric), this gives $M_{\eta'}/M_\sigma \lesssim 0.29$, remarkably close to lattice results $0.356(54)$~\cite{DellaMorte:2023ylq,Martins:2023kcj}.

\subsection{Effects of Fermion Mass and $\theta$-Angle}

\subsubsection{Soft SUSY Breaking via Fermion Mass}

A fermion mass $m$ can be incorporated as a soft supersymmetry-breaking term~\cite{Girardello:1981wz}. Following Masiero and Veneziano~\cite{Masiero:1984ss}, the mass term adds:
\begin{equation}
\Delta\mathcal{L} = \frac{4m}{3\lambda}N^2(\varphi + \bar{\varphi}) \ .
\end{equation}
This preserves the holomorphic structure at leading order, making the mass corrections \textit{protected} by supersymmetry.

\subsubsection{Modified Vacuum and Spectrum}

The full effective Lagrangian at finite $N$ with a fermion mass term is~\cite{Sannino:2003xe}:
\begin{equation}
\mathcal{L} = f(N)\left\{\frac{1}{\alpha}(\varphi\bar{\varphi})^{-2/3}\partial_\mu\bar{\varphi}\partial^\mu\varphi - \frac{4\alpha}{9}(\varphi\bar{\varphi})^{2/3}\left(\ln\bar{\Phi}\ln\Phi - b\right)\right\} + \frac{4m}{3\lambda}N^2(\varphi + \bar{\varphi}) \ , 
\label{eq:fullLagrangian}
\end{equation}
where the modified fields encoding finite-$N$ corrections are $\Phi = \varphi^{1+\epsilon_1}\bar{\varphi}^{-\epsilon_2}$ and $\bar{\Phi} = \bar{\varphi}^{1+\epsilon_1}\varphi^{-\epsilon_2}$, with $\epsilon_1 = -\frac{7}{9N}$ and $\epsilon_2 = -\frac{11}{9N}$ for the antisymmetric representation.

The vacuum expectation value is (see Appendix~\ref{app:vacuum} for the detailed derivation):
\begin{equation}
\langle\varphi\rangle = \Lambda^3\left(1 + \frac{2}{3}b + \frac{3m}{\alpha\lambda\Lambda}\right) + \mathcal{O}(m^2, N^{-2}, mN^{-1}) \ .
\end{equation}
The vacuum energy density is:
\begin{equation}
\mathcal{E}_{\text{vac}} = -\frac{4\alpha f}{9}b\Lambda^4 - \frac{8N^2}{3\lambda}m\Lambda^3 + \mathcal{O}(m^2, N^0, mN) \ .
\end{equation}
The gluon condensate acquires a mass-dependent piece:
\begin{equation}
\frac{\langle G^a_{\mu\nu}G^{a\mu\nu}\rangle}{64\pi^2} = \frac{4Nm}{3\lambda}\Lambda^3 + \frac{8\alpha Nb}{27}\Lambda^4 + \mathcal{O}(m^2, N^{-1}, mN^0) \ .
\end{equation}
The scalar and pseudoscalar masses at finite $N$ and $m$ are (see Appendix~\ref{app:spectrum} for the complete derivation):
\begin{align}
M_\sigma &= \frac{2\alpha}{3}\Lambda\left[1 + \frac{4}{9N} + \frac{2}{3}b + \frac{7}{2}\frac{m}{\alpha\lambda\Lambda} + \mathcal{O}(m^2, N^{-2}, mN^{-1})\right] \ ,\\[0.5em]
M_{\eta'} &= \frac{2\alpha}{3}\Lambda\left[1 - \frac{2}{N} + \frac{2}{9}b + \frac{5}{2}\frac{m}{\alpha\lambda\Lambda} + \mathcal{O}(m^2, N^{-2}, mN^{-1})\right] \ .
\end{align}
Both masses receive contributions from the vacuum parameter $b$, but with different coefficients ($\frac{2}{3}$ for the scalar, $\frac{2}{9}$ for the pseudoscalar). This reflects the fact that the scalar field shares quantum numbers with the vacuum and is thus more sensitive to vacuum structure.

The pseudoscalar-to-scalar mass ratio is:
\begin{equation}
\frac{M_{\eta'}}{M_\sigma} = 1 - \frac{22}{9N} - \frac{4}{9}b - \frac{m}{\alpha\lambda\Lambda} + \mathcal{O}(m^2, N^{-2}, mN^{-1}) \ .
\end{equation}
The fermion mass reinforces the finite-$N$ effect, further decreasing the ratio below unity.

\subsubsection{$\theta$-Angle Dependence}

The vacuum angle $\theta$ is introduced at the fundamental level through the topological term $\mathcal{L}_\theta = \frac{\theta}{32\pi^2}G_{\mu\nu}^a\tilde{G}^{a\mu\nu}$. In the orientifold theory with massless fermions, $\theta$ is unobservable---it can be rotated away by a chiral transformation of the fermion fields. However, the vacuum structure and the fermion condensate do depend on $\theta$.

To incorporate $\theta$ into the effective Lagrangian~\eqref{fnocomponent}, one performs the replacements~\cite{Sannino:2003xe} $\varphi \to \varphi\exp(-i\theta/(N-2))$ and $\Phi \to \Phi\exp(-i\theta/N)$ for the antisymmetric representation. The different factors $(N-2)$ and $N$ reflect the distinct transformation properties under chiral rotations of the fermion bilinear $\varphi$ and the modified field $\Phi$.

The spontaneously broken discrete chiral symmetry $Z_{2(N-2)}$ leads to $(N-2)$ degenerate vacua. To properly implement this at the effective Lagrangian level, one must introduce branch Lagrangians labeled by an integer $n$~\cite{Kovner:1997ca,Sannino:2003xe}:
\begin{equation}
\mathcal{L}^{(n)} = f(N)N^{-2}\left\{\frac{N^2}{\alpha}(\varphi\bar{\varphi})^{-2/3}\partial_\mu\bar{\varphi}\partial^\mu\varphi - \frac{4\alpha}{9}(\varphi\bar{\varphi})^{2/3}\left[\left|\ln\bar{\Phi}^N + 2\pi in + i\theta\right|^2 - N^2 b\right]\right\} \ .
\end{equation}
The parameter $n$ effectively shifts the vacuum angle $\theta \to \theta + 2\pi n$. The full $Z_{2(N-2)}$ symmetry is restored when the partition function sums over all branches: $\mathcal{Z} = \sum_{n=-\infty}^{\infty}\int\mathcal{D}\varphi\,e^{i\int d^4x\,\mathcal{L}^{(n)}}$.
When a fermion mass $m\neq 0$ is present, the $\theta$-angle becomes physical and the vacuum energy acquires $\theta$-dependence. For the antisymmetric representation:
\begin{equation}
\mathcal{E}_{\text{vac}}(\theta) = \frac{8N^2}{3\lambda}m\Lambda^3\min_k\left\{-\cos\left[\frac{\theta + 2\pi k}{N-2}\right]\right\} - \frac{4\alpha f}{9}b\Lambda^4 \ .
\end{equation}
The $(N-2)$-fold vacuum degeneracy present at $m=0$ is lifted by the mass term, selecting a unique vacuum. The formula exhibits the expected $2\pi$ periodicity in $\theta$. For the two-index symmetric representation, the corresponding formulas involve $(N+2)$ instead of $(N-2)$.

The $\theta$-angle also affects the mass spectrum (see Appendix~\ref{app:theta} for the complete derivation). The vacuum expectation value acquires a phase:
\begin{equation}
\langle\varphi\rangle_k = \Lambda^3\exp\left[i\frac{\theta + 2\pi k}{N-2}\right]\left(1 + \frac{2}{3}b + \frac{3m}{\alpha\lambda\Lambda}\cos\left[\frac{\theta+2\pi k}{N-2}\right]\right) + \mathcal{O}(m^2, N^{-2}) \ .
\end{equation}
The masses become $\theta$-dependent:
\begin{align}
M_\sigma(\theta) &= \frac{2\alpha}{3}\Lambda\left[1 + \frac{4}{9N} + \frac{2}{3}b + \frac{7}{2}\frac{m}{\alpha\lambda\Lambda}\cos\left(\frac{\theta}{N-2}\right) + \mathcal{O}(m^2, N^{-2}, mN^{-1})\right] \ ,\\[0.5em]
M_{\eta'}(\theta) &= \frac{2\alpha}{3}\Lambda\left[1 - \frac{2}{N} + \frac{2}{9}b + \frac{5}{2}\frac{m}{\alpha\lambda\Lambda}\cos\left(\frac{\theta}{N-2}\right) + \mathcal{O}(m^2, N^{-2}, mN^{-1})\right] \ ,
\end{align}
where we selected the $k=0$ branch. These formulas apply when $\theta = 0$ or $\theta = \pi(N-2)$, where $CP$ is preserved and the scalar and pseudoscalar are mass eigenstates. At generic values of $\theta$, $CP$ is violated and scalar-pseudoscalar mixing occurs, as we discuss in detail in Appendix~\ref{app:theta}.

The mass ratio at finite $\theta$ (in the $CP$-conserving cases) is:
\begin{equation}
\frac{M_{\eta'}(\theta)}{M_\sigma(\theta)} = 1 - \frac{22}{9N} - \frac{4}{9}b - \frac{m}{\alpha\lambda\Lambda}\cos\left(\frac{\theta}{N-2}\right) + \mathcal{O}(m^2, N^{-2}, mN^{-1}) \ .
\end{equation}
At $\theta = 0$ the ratio reduces to the result derived earlier. At $\theta = \pi(N-2)$ the cosine term changes sign, partially counteracting the mass contribution. At generic $\theta \neq 0, \pi$, $CP$ is explicitly broken and the scalar and pseudoscalar states mix. The physical mass eigenstates are linear combinations with $\theta$-dependent mixing angles and all physical quantities are periodic in $\theta$ with period $2\pi$, as required by the underlying gauge theory.

For the symmetric representation, the mass ratio formula is analogous with the replacements $b \to \hat{b}$ and $(N-2) \to (N+2)$:
\begin{equation}
\frac{M_{\eta'}(\theta)}{M_\sigma(\theta)}\bigg|_{\text{sym}} = 1 + \frac{22}{9N} - \frac{4}{9}\hat{b} - \frac{m}{\alpha\lambda\Lambda}\cos\left(\frac{\theta}{N+2}\right) + \mathcal{O}(m^2, N^{-2}, mN^{-1}) \ .
\end{equation}
Here we have neglected the mixing term assuming $\displaystyle{N \rightarrow \infty,m/\Lambda \rightarrow 0}$ and $\displaystyle{Nm/\Lambda \ll 1}$.   The details of the computation for the vacuum and spectrum of the theory  are reported in Appendix~\ref{app:vacuum}.

\vskip .3cm 
In this section we learnt that the Veneziano--Yankielowicz effective Lagrangian provides a powerful description of the low-energy physics of $\mathcal{N}=1$ supersymmetric Yang--Mills theory. It is written in terms of the composite gluino-ball superfield $S \propto \mathrm{Tr}\,W^2$, with the logarithmic superpotential fixed by holomorphy and anomaly matching. The $N$ degenerate vacua arise from the multivaluedness of the complex logarithm, and the anomaly structure is encoded in a supersymmetric multiplet with the characteristic factor of $3/2$ between trace and chiral anomalies. The component Lagrangian exhibits the K\"ahler structure required by supersymmetry. 

The extension to orientifold field theories connects these exact supersymmetric results to QCD-like theories at finite $N$. The effective Lagrangian, constrained by anomaly matching, yields predictions for the meson spectrum that agree remarkably well with lattice simulations, even for $N=3$. The fermion mass and $\theta$-angle dependence provide additional handles for testing the supersymmetric connection and probing the vacuum structure of strongly coupled gauge theories.

\section{A recent claim of ``no strong $CP$'' and what it means in EFT language}
\label{sec:extra_phase_xi}
 
The recent analysis of Ref.~\cite{Ai:2020ptm} revisits the logic according to which the $\theta$-term 
is a total derivative at the level of the local Lagrangian, yet it becomes physical through nontrivial 
gauge-field topology, the existence of topological sectors, and the structure of the QCD vacuum.
The authors investigate this from the standpoint of how the sum over sectors is implemented in the 
functional integral and, in particular, the order in which one takes the infinite-volume limit and 
performs the interference (sum) over winding sectors labeled by $\Delta_n$.

The authors' main technical object is the set of fermionic correlation functions in the presence of 
massive quarks with complex mass phases and a general vacuum angle. They evaluate these correlators 
 using a dilute instanton-gas framework, where instantons generate the familiar multi-fermion 't~Hooft 
vertex, and  via a more general argument based on factorization/cluster decomposition and the index 
theorem controlling the phase of the fermion determinant in a fixed topological class~\cite{Ai:2020ptm}. 
In their presentation, the usual rephasing-invariant combination $\bar\theta=\theta+\arg\det M$ arises, 
but they argue that, if one first evaluates each fixed-$\Delta_n$ sector in the infinite-volume limit 
and only then sums over $\Delta_n$, the phases induced by the quark masses and by topological effects 
become aligned. In the low-energy chiral description, this alignment is expressed as a particular choice 
for the phase $\xi$ multiplying the anomalous determinant interaction $\sim e^{-i\xi}\det U + \text{h.c.}$, 
such that $CP$-odd observables vanish even when $\bar\theta\neq 0$. In other words: strong interactions 
would exhibit no $CP$ violation without setting $\theta=0$ and without invoking a Peccei--Quinn 
mechanism~\cite{Ai:2020ptm}.

The purpose of this section is twofold. First, I will restate their low-energy effective description 
in the language and conventions of these lecture notes, making fully explicit which operator structures 
and spurion transformations are assumed and how $\theta$-dependence is represented in the EFT. Second, 
I will analyze their prescription purely from the EFT viewpoint, i.e.\ as a statement about degrees of 
freedom and constraints in the infrared theory. The conclusion will be that, once translated into 
standard EFT logic, the \emph{no strong $CP$} outcome corresponds to imposing an additional constraint 
that removes physical $\bar\theta$-dependence. From the effective point of view, this is equivalent to 
adding a \emph{non-propagating axion-like degree of freedom}  whose sole 
role is to enforce the alignment that eliminates $CP$-odd effects---an input that is not guaranteed by 
QCD itself.

\subsection{Low-energy realization: mass term plus an anomalous determinant term}
Similarly to our previous construction, the authors in \cite{Ai:2020ptm} employ a leading order chiral effective theory for the light pseudoscalars. These are packaged into
the unitary $N_f\times N_f$ matrix $U$. The Lagrangian for $N_f=3$ (light $u,d,s$) reads:
\begin{equation}
\mathcal{L}_{\chi}
=
\frac{f_\pi^2}{4}\Tr(\partial_\mu U\,\partial^\mu U^\dagger)
+\frac{f_\pi^2 B_0}{2}\Tr(\mathcal{M}^\dagger U+U^\dagger \mathcal{M})
+|\Delta| f_\pi^4\Big(e^{-i\xi}\det U + e^{+i\xi}\det U^\dagger\Big),
\label{eq:Lchi_det}
\end{equation}
where $\mathcal{M}=\mathrm{diag}(m_u e^{-i\alpha_u},\, m_d e^{-i\alpha_d},\, m_s e^{-i\alpha_s})$ is the (complex) quark mass matrix.
The last term is the anomalous $\mathrm{U}(1)_A$-breaking interaction (a low-energy avatar of the 't~Hooft interaction),
and it contains an \emph{a priori independent} phase $\xi$.
\vskip .2cm
{\it What is the ``extra phase'' $\xi$?} In this approach, $\xi$ is the phase multiplying the determinant interaction in \eqref{eq:Lchi_det}.
It is \emph{not} fixed by chiral symmetry alone: symmetry only constrains how $\xi$ must \emph{transform}
under anomalous $\mathrm{U}(1)_A$ rephasings so that \eqref{eq:Lchi_det} is consistent with the ultraviolet selection rule.
The standard phenomenological choice is to identify $\xi$ with the microscopic $\theta$ angle.
However, \cite{Ai:2020ptm} emphasizes that (given their treatment of topological sectors in the infinite-volume limit)
one can enforce invariance also with a different choice for $\xi$, namely $\xi=-\bar\alpha$ with
\begin{equation}
\bar\alpha \equiv \alpha_u+\alpha_d+\alpha_s,
\qquad
\text{so that}\quad
\xi+\bar\alpha = 0.
\label{eq:alphabar_def}
\end{equation}
Their key point is then: if $\xi=-\bar\alpha$, all $CP$-violating phases in \eqref{eq:Lchi_det} can be removed
by a field redefinition, so there is no strong-$CP$ violation in this framework.

\subsection{Vacuum alignment and the coupled stationarity equations}
The LO potential associated with \eqref{eq:Lchi_det} is (up to an additive constant)
\begin{equation}
V(U)
=
- f_\pi^2 B_0\,\Re \Tr(\mathcal{M^\dagger}U)
- 2|\Delta| f_\pi^4\, \Re\!\Big(e^{-i\xi}\det U\Big).
\label{eq:V_basic}
\end{equation}
For diagonal $\mathcal{M}$ the vacuum $U_0\equiv\langle U\rangle$ is diagonal as well, and we parametrize
\begin{equation}
U_0=\mathrm{diag} \,\!\big(e^{i\phi_u},\,e^{i\phi_d},\,e^{i\phi_s}\big),
\label{eq:U0_diag}
\end{equation}
so that $\det U_0=e^{i(\phi_u+\phi_d+\phi_s)}$. Computing the mass term contribution to the energy of the system we derive:
\[
\Tr(\mathcal{M}^\dagger U_0)=\sum_{i=u,d,s} m_i e^{i(\alpha_i+\phi_i)},
\qquad
\Tr(U_0^\dagger \mathcal{M})=\sum_{i=u,d,s} m_i e^{-i(\alpha_i+\phi_i)}. 
\]
Hence
\[
\Tr(\mathcal{M}^\dagger U_0+U_0^\dagger \mathcal{M} )=2\sum_i m_i\cos(\alpha_i+\phi_i),
\]
and
\begin{equation}
E_{\text{mass}}=
-\frac{f_\pi^2 B_0}{2}\Tr(\mathcal{M}^\dagger U_0+U_0^\dagger \mathcal{M})
=
-f_\pi^2 B_0\sum_i m_i\cos(\alpha_i+\phi_i).
\label{eq:Vmass_app}
\end{equation}
The anomalous term yields: 
\[
e^{-i\xi}\det U_0 + e^{+i\xi}\det U_0^\dagger
=
e^{i(\Delta\phi-\xi)}+e^{-i(\Delta\phi-\xi)}
=2\cos(\Delta\phi-\xi),
\]
therefore
\begin{equation}
E_{\det}=
-|\Delta|f_\pi^4\Big(e^{-i\xi}\det U_0 + e^{+i\xi}\det U_0^\dagger\Big)
=
-2|\Delta|f_\pi^4 \cos(\Delta\phi-\xi).
\label{eq:Vdet_app}
\end{equation}
The full energy of the ground state is: 
\begin{equation}
E(\phi)=
-f_\pi^2 B_0\sum_i m_i\cos(\alpha_i+\phi_i)
-2|\Delta|f_\pi^4 \cos(\Delta\phi-\xi).
\label{eq:Vtotal_app}
\end{equation}

\subsection{Derivation of the stationarity equations}

 We now differentiate $E(\phi)$ with respect to each vacuum angle. The determinant contribution depends only on the \emph{sum} of angles, which is why the three equations are coupled. Differentiate \eqref{eq:Vtotal_app} w.r.t.\ each $\phi_i$:
\[
\frac{\partial}{\partial\phi_i}\Big[-f_\pi^2B_0\,m_i\cos(\alpha_i+\phi_i)\Big]
=
+f_\pi^2B_0\,m_i\sin(\alpha_i+\phi_i),
\]
and using $\partial_{\phi_i}\Delta\phi=1$,
\[
\frac{\partial}{\partial\phi_i}\Big[-2|\Delta|f_\pi^4\cos(\Delta\phi-\xi)\Big]
=
+2|\Delta|f_\pi^4\sin(\Delta\phi-\xi).
\]
Thus $\partial V/\partial\phi_i=0$ gives the Dashen equations
\begin{equation}
f_\pi^2B_0\,m_i\sin(\alpha_i+\phi_i)
+2|\Delta|f_\pi^4\sin(\Delta\phi-\xi)=0,
\qquad i=u,d,s.
\label{eq:stat_raw_app}
\end{equation}
Define shifted angles $\varphi_i\equiv \alpha_i+\phi_i$. Then

\[
\Delta\phi-\xi
=
(\varphi_u+\varphi_d+\varphi_s)-(\xi+\bar\alpha)
=
(\varphi_u+\varphi_d+\varphi_s)-\delta,
\]
In other words the variables $\varphi_i$ absorb the quark mass phases into the vacuum alignment angles. In these variables, the only remaining external phase appears through the single combination $\delta=\xi+\bar\alpha$. Therefore,  \eqref{eq:stat_raw_app} becomes
\begin{equation}
B_0 f_\pi^2\, m_i \sin\varphi_i
-2|\Delta|f_\pi^4\,\sin\!\Big(\delta-\varphi_u-\varphi_d-\varphi_s\Big)=0,
\qquad i=u,d,s.
\label{eq:S112_app}
\end{equation}
 Since the anomalous term enters identically in each stationary condition by subtracting two equations, it cancels  and yields an immediate relation among the mass-weighted sines.

\noindent
In fact, letting $S\equiv \delta-\varphi_u-\varphi_d-\varphi_s$,  each equation reads
\[
B_0 f_\pi^2\, m_i \sin\varphi_i = 2|\Delta|f_\pi^4\sin S.
\]
Subtracting any two of the three equations eliminates the common $\sin S$ term, e.g.
\[
B_0 f_\pi^2\big(m_u\sin\varphi_u-m_d\sin\varphi_d\big)=0
\quad\Rightarrow\quad
m_u\sin\varphi_u=m_d\sin\varphi_d,
\]
and similarly $m_u\sin\varphi_u=m_s\sin\varphi_s$. Hence
\begin{equation}
m_u\sin\varphi_u = m_d\sin\varphi_d = m_s\sin\varphi_s.
\label{eq:msin_equal_app}
\end{equation}

\subsection{Phase-locking constraint in the determinant-dominated regime}

 When the determinant term dominates, it first pins the collective phase $\sum_i\varphi_i$ (``phase locking''). The mass term then selects the individual $\varphi_i$ consistent with this constraint.

Assume the hierarchy
\begin{equation}
|\Delta| f_\pi^4 \gg B_0 f_\pi^2\, m_i \qquad (i=u,d,s).
\label{eq:hierarchy_app}
\end{equation}
Then the dominant contribution to the vacuum energy is $V_{\det}$, and minimizing \eqref{eq:Vdet_app} gives
\[
\Delta\phi-\xi=0\quad (\mathrm{mod}\ 2\pi).
\]
In terms of shifted variables this is
\begin{equation}
\varphi_u+\varphi_d+\varphi_s = \delta \quad (\mathrm{mod}\ 2\pi).
\label{eq:phase_lock_app}
\end{equation}

\subsection{Two-flavour closed-form solution}

We start by dropping the strange quark and have
\begin{equation}
\delta_{ud}\equiv \xi+\alpha_u+\alpha_d,
\qquad
\varphi_d=\delta_{ud}-\varphi_u
\quad (\mathrm{mod}\ 2\pi).
\label{eq:deltaud_app}
\end{equation}
We further introduce the common quantity K
\begin{equation}
m_u\sin\varphi_u=m_d\sin\varphi_d \equiv K,
\label{eq:Kdef_app}
\end{equation}
such that
\begin{equation}
\sin\varphi_u=\frac{K}{m_u},\qquad \sin\varphi_d=\frac{K}{m_d}.
\label{eq:sins_app}
\end{equation}

Using $\cos(\varphi_u+\varphi_d)=\cos\delta_{ud}$:
\[
\cos\delta_{ud}
=
\cos\varphi_u\cos\varphi_d-\sin\varphi_u\sin\varphi_d
=
\sqrt{1-\frac{K^2}{m_u^2}}\sqrt{1-\frac{K^2}{m_d^2}}-\frac{K^2}{m_um_d} \ ,
\]
we arrive at 
\begin{equation}
\sqrt{1-\frac{K^2}{m_u^2}}\sqrt{1-\frac{K^2}{m_d^2}}
=
\cos\delta_{ud}+\frac{K^2}{m_um_d}.
\label{eq:root_prod_app}
\end{equation}
By squaring \eqref{eq:root_prod_app} and simplifying we obtain
\[
\sin^2\delta_{ud}
=
\frac{K^2}{m_u^2}+\frac{K^2}{m_d^2}+\frac{2K^2\cos\delta_{ud}}{m_um_d}.
\]
Thus
\begin{equation}
K=
\frac{\sin\delta_{ud}}{\sqrt{\frac{1}{m_u^2}+\frac{1}{m_d^2}+\frac{2\cos\delta_{ud}}{m_um_d}}}.
\label{eq:Ksol_app}
\end{equation}
Since $K=m_u\sin\varphi_u=m_d\sin\varphi_d$, we obtain
\begin{equation}
m_{u,d}\sin\varphi_{u,d}
=
\frac{\sin(\xi+\alpha_u+\alpha_d)}
{\sqrt{\dfrac{1}{m_u^2}+\dfrac{1}{m_d^2}+\dfrac{2\cos(\xi+\alpha_u+\alpha_d)}{m_um_d}}}.
\label{eq:S113_app}
\end{equation}

\subsection{Three flavours: small-angle solution}
For small $\delta$ one has $\varphi_i\ll 1$, so $\sin\varphi_i\simeq\varphi_i$. The coupled transcendental system becomes linear and is solved conveniently in terms of the reduced mass $\tilde m$. Assume $\delta=\xi+\bar\alpha\ll 1$ so $\varphi_i\ll 1$ and $\sin\varphi_i\simeq \varphi_i$.
Then \eqref{eq:msin_equal_app} implies
\begin{equation}
m_u\varphi_u=m_d\varphi_d=m_s\varphi_s\equiv C.
\label{eq:Cdef_app}
\end{equation}
Phase locking \eqref{eq:phase_lock_app} gives $\varphi_u+\varphi_d+\varphi_s=\delta$, hence
\[
\frac{C}{m_u}+\frac{C}{m_d}+\frac{C}{m_s}=\delta
\quad\Rightarrow\quad
C=
\frac{m_um_dm_s}{m_um_d+m_dm_s+m_sm_u}\,\delta.
\]
Define
\begin{equation}
\tilde m\equiv \frac{m_um_dm_s}{m_um_d+m_dm_s+m_sm_u},
\label{eq:mtilde_app}
\end{equation}
so $C=\tilde m\,\delta$, and therefore
\begin{equation}
m_i\varphi_i=\tilde m\,(\xi+\alpha_u+\alpha_d+\alpha_s),
\qquad i=u,d,s.
\label{eq:S114_app}
\end{equation}
 
\subsection{How \texorpdfstring{$\xi$}{xi} appears in the auxiliary-field formulation and the would-be no strong-$CP$ solution}
In our earlier approach, the topological charge density \(\mathcal{Q}(x)\) is introduced as an auxiliary field.
For convenience, we recall the effective Lagrangian of \eqref{eq:LwithQ} in the form
\begin{align}
\mathcal{L}
={}&
\frac{f_\pi^2}{4}\,\Tr\!\left(\partial_\mu U\,\partial^\mu U^\dagger\right)
+\frac{f_\pi^2 B_0}{2}\,\Tr\!\left(\mathcal{M}\,U^\dagger + U\,\mathcal{M}^\dagger\right)
\nonumber\\
&- \frac{i}{2}\, \mathcal{Q}(x)\,\Tr\!\left(\ln U - \ln U^\dagger\right)
+\frac{\mathcal{Q}^2(x)}{2\chi_{\mathrm{top}}}
-\theta\,\mathcal{Q}(x),
\label{eq:LwithQn}
\end{align}
where \(\chi_{\mathrm{top}}\) denotes the topological susceptibility in the pure-gauge (Yang--Mills) sector in the
sense specified earlier, and \(\mathcal{M}\) is the (generally complex) quark mass matrix.
The field \(\mathcal{Q}(x)\) has no kinetic term and therefore does not propagate. Eliminating \(\mathcal{Q}(x)\) by its algebraic equation of motion,
\begin{equation}
\frac{\partial \mathcal{L}}{\partial \mathcal{Q}}=0
\qquad\Rightarrow\qquad
\mathcal{Q}(x)=\chi_{\mathrm{top}}
\left[
\theta+\frac{i}{2}\Tr\!\left(\ln U-\ln U^\dagger\right)
\right],
\label{eq:Qeom_recap}
\end{equation}
and substituting back into \eqref{eq:LwithQn} yields the standard quadratic anomalous potential,
\begin{align}
\mathcal{L}
\supset
-\frac{\chi_{\mathrm{top}}}{2}
\left[
\theta+\frac{i}{2}\Tr\!\left(\ln U-\ln U^\dagger\right)
\right]^2.
\label{eq:LeffAfterQ2}
\end{align}

To incorporate the phase \(\xi\) that multiplies the determinant interaction in \eqref{eq:Lchi_det},
one may equivalently shift the \(\theta\)-source by a constant,
\begin{equation}
\theta \;\longrightarrow\; \theta-\xi,
\label{eq:theta_shift_for_xin}
\end{equation}
so that the quadratic potential depends on
\((\theta-\xi)+\frac{i}{2}\Tr(\ln U-\ln U^\dagger)\).
In the determinant language \eqref{eq:Lchi_det}, the same information is carried directly by the phase multiplying
\(\det U\) and \(\det U^\dagger\).
We stress that the quadratic form \eqref{eq:LeffAfterQ2} and the  determinant interaction in
\eqref{eq:Lchi_det} need not be globally identical as functions of \(\arg\det U\): beyond the small-field regime
they encode the anomalous physics differently. However, the bookkeeping of $CP$-violating phases is the same in
both descriptions: the effective potential depends on the invariant combination built from \(\theta\) and the
\(\mathrm{U}(1)_A\) phase of \(\det U\).
In what follows we work in the convention where the phases in the quark mass matrix have already been absorbed
into the definition of the vacuum angles \(\phi_i\).

Once an a priori arbitrary phase \(\xi\) is introduced in this manner, the effective vacuum equations are modified
accordingly.
Denoting the vacuum angles by \(\phi_i\) (as in the preceding subsection), the Dashen equations take the form
\begin{equation}
\frac{\partial E}{\partial\phi_i} = 0
\qquad\Rightarrow\qquad
M_{\pi,i}^{2}\sin\phi_i
=
\frac{\chi_{\mathrm{top}}}{f_\pi^2}
\left(\theta - \sum_{j=1}^{N_f}\phi_j -\xi \right),
\qquad i=1,\ldots,N_f.
\label{eq:phieom}
\end{equation}
If, in addition, one treats \(\xi\) as a quantity to be extremized at the level of the vacuum energy,
then one obtains an extra stationarity condition,
\begin{equation}
\frac{\partial E}{\partial\xi} = 0
\qquad\Rightarrow\qquad
\theta - \sum_{j=1}^{N_f}\phi_j - \xi= 0.
\label{eq:axioneom_phase}
\end{equation}
Inserting \eqref{eq:axioneom_phase} into \eqref{eq:phieom} forces
\(\sin\phi_i=0\) for all \(i\), so the $CP$-conserving minimum is obtained at \(\phi_i=0\) (mod \(2\pi\)) and
\begin{equation}
\boxed{\ 
\bar\theta \;\equiv\; \theta - \sum_{j=1}^{N_f}\phi_j -  \xi
\;=\; 0\quad (\mathrm{mod}\ 2\pi)\,.
\ }
\label{eq:thetazero}
\end{equation}

From the effective Lagrangian point of view, the logic above is formally identical to the axion mechanism:
\(\xi\) plays the role of a $CP$-odd phase that can relax the physical \(\bar\theta\) to zero by an equation of motion.
The crucial difference is that, in the Peccei--Quinn solution, \(\xi\) is not an arbitrary parameter but the phase
associated with a propagating dynamical field (the axion), with its own kinetic term and ultraviolet origin.
In the present auxiliary-field setup, \(\xi\) is not a degree of freedom of QCD, and there is no derivation within
QCD that compels one to extremize the vacuum energy with respect to \(\xi\) as it is for the axion case~\footnote{The authors, referring to their work in \cite{Ai:2025quf}, maintain that the topological charge $\cal Q$, in the EFT, is not subjected to topological constraints and is identified with the one in a subvolume with open boundary conditions that  fluctuate freely. As I have shown, this amounts to include a non-propagating axion. }.

\vskip .3cm 
Treating \(\xi\) as a variable is therefore an additional assumption that effectively adds a non-propagating
axion-like  sector to the infrared theory. Since this extra ingredient is not mandated by
QCD, it does not constitute a general resolution of the strong $CP$ problem within QCD itself, and the strong $CP$
problem remains.

\vskip .2cm 
\subsection{An overview of the ongoing debate}
\label{sec:literature_review}

The claim that the strong $CP$ problem may be \textit{illusory} has generated significant discussion in the recent literature. Below, we provide a brief overview of the main contributions and their central arguments. The framework of Ref.~\cite{Ai:2020ptm} has been extended in subsequent works~\cite{Ai:2022htq,Ai:2024cnp,Ai:2024vfa}. Reference~\cite{Ai:2024vfa} reformulates the argument using canonical quantization, claiming that when all gauge transformations are treated as redundancies, consistency with the gauge-group structure of QCD leaves no room for physical $CP$-odd phases. The key technical assertion throughout this body of work is that topological quantization requires finite Euclidean action (and hence saddle-point dominance), implying that the infinite-volume limit must be taken \emph{before} summing over topological sectors.

Several groups have criticized these conclusions. A detailed response is provided in Ref.~\cite{Albandea:2024fui}, which studies the quantum rotor, a simple mechanical analogue of the $\theta$-vacuum, and demonstrates that all physical quantities of interest can be extracted from \emph{local} correlators, whose $\theta$-dependence is insensitive to the order of limits. The authors argue that while global observables (integrals over all of spacetime) may be sensitive to the disputed limiting procedure, such observables are unphysical. Conversely, the topological susceptibility, the proton mass, and the neutron electric dipole moment are local quantities, unaffected by boundary conditions at spatial infinity.

In a recent note~\cite{Khoze:2025auv}, Khoze provides an instructive reexamination of instanton methods. Within the dilute instanton gas approximation (DIGA), the physically relevant connected part of any correlator consists solely of single-instanton and single-anti-instanton contributions. The author shows that the expressions derived in Refs.~\cite{Ai:2020ptm,Ai:2022htq} do not coincide with these standard single-instanton results, in particular, they exhibit a structure that lacks the expected factorization into disconnected contributions. The conclusion is that the $\theta$-dependence of instanton-dominated observables remains intrinsically non-trivial whenever instanton physics is operative.

A comprehensive critical assessment is given in Ref.~\cite{Benabou:2025viy}, which revisits both gauged-discrete-symmetry solutions and the ``no strong $CP$'' claim. The authors demonstrate that a nonzero neutron electric dipole moment at finite $\bar\theta$ follows directly from well-understood QCD dynamics, and that the order-of-limits argument does not eliminate $CP$ violation from local observables. They also address arguments based on the low-energy chiral effective theory, showing that choices of couplings compatible with spurion analysis generically lead to $CP$ violation unless additional assumptions are imposed---assumptions equivalent to adding a non-propagating axion-like degree of freedom, as discussed earlier in this section. The authors of the original papers have responded to these criticisms in Ref.~\cite{Ai:2025quf}, maintaining that local correlators and the topological susceptibility can be defined consistently within their framework, and that the low-energy effective theory alone cannot establish $CP$ violation without importing specific choices that are not uniquely fixed by symmetry. 

Additional perspectives on the interplay between topology, gauge structure, and $CP$ violation are offered in Refs.~\cite{Gamboa:2025hxa,Williams:2026cec}, which reformulate the problem in terms of infrared dressing and global holonomy. These works confirm that the vacuum angle encodes global (rather than local) information about the theory, but argue that this does not render $\theta$ unobservable; rather, it determines which infrared representation of the theory is realized. Most recently,  the authors in Ref~\cite{Aghaie:2026pkf} analyze quantum mechanics on a ring, arguing that consistent results require summing over all topological sectors  {before} taking the large-time limit, thereby supporting the conventional path integral formulation and suggesting that the strong $CP$ problem persists in QCD.

Lattice QCD studies provide independent information. Reference~\cite{Dragos:2019oxn} explicitly confirms the existence of the strong $CP$ problem by computing the neutron electric dipole moment using gradient-flow methods; the result is consistent with standard expectations for a nonzero $\bar\theta$. Alternatively, Ref.~\cite{Schierholz:2025tns} proposes that nonperturbative screening limits the vacuum angle to $\theta=0$ at macroscopic distances. However, this scenario would challenge the established Witten--Veneziano relation. 

Most recently, Bhattacharya~\cite{Bhattacharya:2025qsk} argues that even in finite volumes and with different boundary conditions, locally defined observables far from the boundary recover the standard $\theta$-dependence in the large-volume limit, leaving the standard understanding of the strong $CP$ problem intact. From the lattice viewpoint, the note emphasizes three points:
\begin{itemize}
    \item[(i)] Although topology is subtle at finite lattice spacing, near the continuum limit one recovers robust topological sectors under mild smoothness assumptions~\cite{Luscher:1981zq,Phillips:1986qd,Luscher:2010we,Alexandrou:2017hqw}.
    \item[(ii)] Lattice simulations are routinely performed with different boundary conditions, consistent with the exponential decoupling of boundary effects in a gapped theory~\cite{Kronfeld:1990qu,Luscher:2011kk,Gattringer:2021xrb,vanBaal:1982ag}.
    \item[(iii)] Working at fixed global topological charge does not reproduce generic finite-$\theta$ physics directly. Rather, fixed topological charge correlation functions approach $\theta=0$ physics with controlled $1/V$ corrections; one must allow the distribution of the topological charge to broaden with volume to recover finite-$\theta$ results~\cite{Brower:2003yx,Guo:2015tla}. See also explicit lattice evidence that flowed definitions drive the topological charge close to integers on typical ensembles~\cite{Bhattacharya:2021lol}.
\end{itemize}

\vskip .5cm
Overall, the strong $CP$ problem appears genuine: the combination $\bar\theta=\theta+\arg\det \mathcal{M}$ is a physical, rephasing-invariant quantity whose smallness requires explanation. While the works of Ai et~al.\ raise interesting questions about the correct treatment of topological sectors in the infinite-volume limit, the counterarguments summarized above indicate that local, measurable observables retain their expected $\theta$-dependence. From the effective field theory standpoint, eliminating $CP$ violation without invoking a dynamical relaxation mechanism (such as the axion) requires additional assumptions that go beyond QCD itself.
 \newpage 

\section{Peccei--Quinn symmetry and the QCD axion}
\label{sec:PQ_axion}

In the preceding discussion on chiral symmetry and the axial anomaly, we saw that the classical
$U(1)_A$ current is not conserved in QCD: nonperturbative gauge configurations generate $\partial_\mu J_5^\mu = 2N_f\, \mathcal{Q}(x)$.  The same anomaly that resolves the would-be ninth flavor singlet Goldstone boson problem also implies that QCD admits a $CP$-odd vacuum angle $\theta$ multiplying $\mathcal{Q}(x)$. Because $\theta$ is periodic yet physical, it provides a new source of $CP$ violation whose observed absence is the strong $CP$ problem.

The Peccei--Quinn idea to resolve the strong $CP$ problem can be viewed as the minimal dynamical extension of that anomalous structure. Rather than attempting to enforce $CP$ by hand, one enlarges the chiral symmetry content of the theory by a global $\U(1)_{\rm PQ}$ whose only explicit breaking is \emph{again} the same QCD anomaly. When $\U(1)_{\rm PQ}$ is spontaneously broken at a high scale $f_a$, the associated phase field (the axion) enters the low-energy theory exactly where the anomaly already appears, namely through the topological density $\mathcal{Q}(x)$.

Consequently, the physically relevant parameter becomes the single invariant combination
\begin{equation}
\bar{\theta}=\theta + \arg\det\mathcal{M} - c_{\rm PQ}\,\frac{a}{f_a},
\end{equation}
where $c_{\rm PQ}$ is the QCD anomaly coefficient of $\U(1)_{\rm PQ}$ (with $|c_{\rm PQ}|$ being the domain-wall number $N_{DW}$) and $f_a$ the axion decay constant. The nonperturbative QCD vacuum energy that depends on $\theta$ in the anomalous chiral effective theory now depends on the new $\bar{\theta}$. The axion mechanism is therefore best understood as an \emph{anomaly-driven vacuum alignment} statement: the same dynamics responsible for anomalous chiral symmetry breaking generates an axion potential whose minimum lies at the $CP$-even point $\bar{\theta}=0$ (mod $2\pi$), thereby cancelling $\bar\theta$ dynamically.



\subsection{Introducing a Peccei--Quinn field}

Introduce a complex scalar $\mathcal{N}$ that spontaneously breaks a global $\U(1)_{\rm PQ}$. At energies well below the PQ radial mode, we retain only its phase, the axion, parameterized as
\begin{equation}
\mathcal{N}(x) = \frac{f_a}{\sqrt{2}}\exp\!\left(\frac{i\sqrt{2}\,a(x)}{f_a}\right),
\qquad
\frac{1}{2}\,\partial_\mu\mathcal{N}\,\partial^\mu\mathcal{N}^\dagger
= \frac{1}{2}\,(\partial_\mu a)^2,
\label{eq:Nparam}
\end{equation}
so the axion field $a(x)$ is canonically normalized.

Let $c_{\rm PQ}$ denote the \emph{QCD anomaly coefficient} of $\U(1)_{\rm PQ}$, defined by the mixed anomaly between the PQ current and gluons. In a UV completion where $\U(1)_{\rm PQ}$ acts on heavy quarks $Q$ with PQ charges $X_Q$, one has
\begin{equation}
c_{\rm PQ} = \sum_Q X_Q,
\end{equation}
summed over all colored fermions charged under $\U(1)_{\rm PQ}$. This coefficient determines the strength of the axion coupling to QCD.

When $\U(1)_{\rm PQ}$ is spontaneously broken, the axion field $a$ is an angular variable with fundamental period $2\pi f_a$. However, the QCD-induced potential $V(a) \propto 1 - \cos(c_{\rm PQ} a/f_a + \bar\theta)$ has $|c_{\rm PQ}|$ degenerate minima within this period. If $|c_{\rm PQ}| > 1$, domain walls form during the cosmological PQ phase transition, interpolating between these degenerate vacua. The \emph{domain wall number} $N_{DW} \equiv |c_{\rm PQ}|$ counts these vacua and is a crucial parameter for axion cosmology: models with $N_{DW} > 1$ face the ``domain wall problem'' unless inflation occurs after PQ breaking or the degeneracy is lifted by additional physics. For simplicity of notation, we henceforth write $c_{\rm PQ}$ for the anomaly coefficient, keeping in mind that $N_{DW}= |c_{\rm PQ}|$.

Assuming the only explicit breaking of $\U(1)_{\rm PQ}$ is the QCD anomaly, the combined effective Lagrangian is
\begin{align}
\mathcal{L}
={}&
\frac{f_\pi^2}{4}\,\mathrm{Tr}\!\left[\partial_\mu U\,\partial^\mu U^\dagger\right]
+\frac{1}{2}\,\partial_\mu\mathcal{N}\,\partial^\mu\mathcal{N}^\dagger
+\frac{f_\pi^2 B_0}{2}\,\mathrm{Tr}\!\left[\mathcal{M}(U + U^\dagger)\right]
\nonumber\\
&+\frac{\mathcal{Q}^{2}}{2\chi_{\mathrm{top}}}
-\theta\,\mathcal{Q}
-\frac{i}{2}\,\mathcal{Q}\left[
\mathrm{Tr}\!\left(\ln U - \ln U^\dagger\right)
+ c_{\rm PQ}\,\ln\!\left(\frac{\mathcal{N}}{\mathcal{N}^\dagger}\right)
\right].
\label{eq:axionFull}
\end{align}

\subsection{Anomalous transformations and the anomaly-free combination}

By construction, under the axial $\U(1)_A$ transformation acting on quarks as $q \to e^{i\beta\gamma_5}q$, the chiral field transforms as
\begin{equation}
U \to e^{2i\beta}U,
\qquad
\Rightarrow\qquad
\mathrm{Tr}\!\left(\ln U - \ln U^\dagger\right) \to
\mathrm{Tr}\!\left(\ln U - \ln U^\dagger\right) + 4iN_f\beta,
\label{eq:U1Ashift}
\end{equation}
(the factor of 2 in the exponent arises because $U$ transforms like $\bar{q}_R q_L$, which picks up $e^{2i\beta}$ under an axial rotation). Similarly, under $\U(1)_{\rm PQ}$,
\begin{equation}
\mathcal{N} \to e^{i\gamma}\mathcal{N},
\qquad
\Rightarrow\qquad
\ln\!\left(\frac{\mathcal{N}}{\mathcal{N}^\dagger}\right) \to
\ln\!\left(\frac{\mathcal{N}}{\mathcal{N}^\dagger}\right) + 2i\gamma.
\label{eq:PQshift}
\end{equation}
Therefore the $\mathcal{Q}$-coupling shifts the Lagrangian by
\begin{equation}
\delta\mathcal{L}
=
-\frac{i}{2}\,\mathcal{Q}\left(4iN_f\beta + 2ic_{\rm PQ}\gamma\right)
=
\mathcal{Q}\left(2N_f\beta + c_{\rm PQ}\gamma\right),
\label{eq:dL}
\end{equation}
which is the advertised anomalous variation. Hence the \emph{anomaly-free} subgroup satisfies
\begin{equation}
2N_f\beta + c_{\rm PQ}\gamma = 0.
\label{eq:anomalyfree}
\end{equation}
The  orthogonal combination is explicitly broken by QCD providing a potential term for the axion. The  presence of quark masses, as we shall see, explicitly breaks the residual anomaly-free symmetry and generates a mass for the QCD axion. Classic references are Peccei--Quinn~\cite{Peccei:1977hh,Peccei:1977ur} and Weinberg/Wilczek~\cite{Weinberg:1977ma,Wilczek:1977pj}.

\subsection{Integrating out $\mathcal{Q}(x)$ and deriving the vacuum energy}

Because $\mathcal{Q}$ is auxiliary, its equation of motion follows directly from \eqref{eq:axionFull}:
\begin{equation}
\frac{\partial\mathcal{L}}{\partial\mathcal{Q}}
=
\frac{\mathcal{Q}}{\chi_{\mathrm{top}}}
- \theta
- \frac{i}{2}\left[
\mathrm{Tr}\!\left(\ln U - \ln U^\dagger\right)
+ c_{\rm PQ}\,\ln\!\left(\frac{\mathcal{N}}{\mathcal{N}^\dagger}\right)
\right]
= 0.
\end{equation}
Solving for $\mathcal{Q}$ gives
\begin{equation}
\mathcal{Q}
=
\chi_{\mathrm{top}}\left[
\theta
+ \frac{i}{2}\,\mathrm{Tr}\!\left(\ln U - \ln U^\dagger\right)
+ \frac{i}{2}\,c_{\rm PQ}\,\ln\!\left(\frac{\mathcal{N}}{\mathcal{N}^\dagger}\right)
\right].
\label{eq:qEOM}
\end{equation}
Substituting back into the Lagrangian yields the standard completed-square form
\begin{align}
\mathcal{L}
={}&
\frac{f_\pi^2}{4}\,\mathrm{Tr}\!\left(\partial_\mu U\,\partial^\mu U^\dagger\right)
+ \frac{1}{2}(\partial_\mu a)^2
+ \frac{f_\pi^2 B_0}{2}\,\mathrm{Tr}\!\left[\mathcal{M}(U + U^\dagger)\right]
\nonumber\\
&-\frac{\chi_{\mathrm{top}}}{2}\left[
\theta
+ \frac{i}{2}\,\mathrm{Tr}\!\left(\ln U - \ln U^\dagger\right)
+ \frac{i}{2}\,c_{\rm PQ}\,\ln\!\left(\frac{\mathcal{N}}{\mathcal{N}^\dagger}\right)
\right]^2.
\label{eq:Leff_after_q}
\end{align}
We now restrict to constant diagonal phases (vacuum alignment). Writing
\begin{equation}
\langle U_{ij}\rangle = e^{i\phi_i}\delta_{ij},
\qquad
\langle\mathcal{N}\rangle = \frac{f_a}{\sqrt{2}}e^{i\varphi},
\label{eq:vacua}
\end{equation}
with $\phi_i,\varphi$ dimensionless we then have: 
\begin{equation}
\mathrm{Tr}\!\left(\ln U - \ln U^\dagger\right) = 2i\sum_{i=1}^{N_f}\phi_i,
\qquad
\ln\!\left(\frac{\mathcal{N}}{\mathcal{N}^\dagger}\right) = 2i\varphi,
\end{equation}
and the potential energy density $E$ (minus the static Lagrangian) becomes
\begin{equation}
E
=
\frac{\chi_{\mathrm{top}}}{2}\left(\theta - \sum_{i=1}^{N_f}\phi_i - c_{\rm PQ}\varphi\right)^{2}
- f_\pi^2 B_0 \sum_{i=1}^{N_f} m_i\cos\phi_i,
\label{eq:Evac_general}
\end{equation}
where we have used the mass parameters $M_{\pi,i}^2 = 2B_0 m_i$ related to quark masses via the GMOR relation. Equation~\eqref{eq:Evac_general} is the axion-extended version of the standard vacuum energy of \eqref{eq:Ebar}.

\subsection{Axion solution: dynamical relaxation of $\bar\theta$}

Extremizing \eqref{eq:Evac_general} with respect to the $\phi_i$ gives
\begin{equation}
\frac{\partial E}{\partial\phi_i} = 0
\qquad\Rightarrow\qquad
M_{\pi,i}^{2}\sin\phi_i
=
\frac{\chi_{\mathrm{top}}}{f_\pi^2}\left(\theta - \sum_{j=1}^{N_f}\phi_j - c_{\rm PQ}\varphi\right),
\qquad i=1,\ldots,N_f.
\label{eq:phieom}
\end{equation}
Extremizing with respect to the axion phase $\varphi$ gives
\begin{equation}
\frac{\partial E}{\partial\varphi} = 0
\qquad\Rightarrow\qquad
\theta - \sum_{j=1}^{N_f}\phi_j - c_{\rm PQ}\varphi = 0.
\label{eq:axioneom_phase}
\end{equation}
Combining \eqref{eq:phieom} and \eqref{eq:axioneom_phase} forces $\sin\phi_i=0$ for all $i$, and the $CP$-conserving minimum is obtained at $\phi_i=0$ and
\begin{equation}
\boxed{\ \bar\theta \equiv \theta - \sum_{j=1}^{N_f}\phi_j - c_{\rm PQ}\varphi = 0\ (\mathrm{mod}\ 2\pi).\ }
\label{eq:thetazero}
\end{equation}
Thus strong $CP$ is dynamically solved. The underlying statement is symmetry-based: in the microscopic theory, the axion shift $a \to a + \mathrm{const}\cdot f_a$ is an exact symmetry in the limit where the QCD anomaly is turned off, and with the anomaly present the only allowed dependence is through the \emph{single} combination $\bar{\theta} = \theta - c_{\rm PQ}\,a/f_a$ (up to quark-mass phases). Therefore the full QCD+axion vacuum energy has the structure
\begin{equation}
E(\theta,a) = E_{\rm QCD}(\bar{\theta}),
\qquad
\bar{\theta} \equiv \theta - c_{\rm PQ}\frac{a}{f_a},
\end{equation}
so minimizing with respect to $a$ always enforces $\partial_{\bar{\theta}}E_{\rm QCD}=0$, which occurs at $\bar{\theta}=0$ (mod $2\pi$) in the $CP$-even vacuum. This argument is independent of the chiral truncation and survives higher-order operators.

\bigskip

\subsection{Quadratic expansion and mixing matrix in the Cartan basis of $\Phi$}
\label{subsec:quad_cartan}

We expand the $\theta$-dependent effective theory around the $CP$-conserving vacuum at $\theta=0$.
Write the meson field in a Cartan (diagonal) basis
\begin{equation}
\Phi(x)=\mathrm{diag}\!\big(v_1(x),\dots,v_{N_f}(x)\big),
\qquad \Phi_{ij}=v_i\,\delta_{ij},
\label{eq:Phi_cartan}
\end{equation}
where the $v_i$ are pseudoscalar fields with mass dimension one.
Then
\begin{equation}
V(x)=\exp\!\left(\frac{2i\Phi(x)}{f_\pi}\right)
\quad\Longrightarrow\quad
V_{ii}=\exp\!\left(\frac{2iv_i}{f_\pi}\right).
\end{equation}
Expanding $V\simeq \mathbf{1}+\frac{2i}{f_\pi}\Phi - \frac{2}{f_\pi^2}\Phi^2 + \cdots$, we have
\begin{align}
\frac{f_\pi^2}{4}\mathrm{Tr}(\partial_\mu V\,\partial^\mu V^\dagger)
&=
\frac{f_\pi^2}{4}\,\mathrm{Tr}\!\left(\frac{4}{f_\pi^2}\partial_\mu\Phi\,\partial^\mu\Phi\right)+\cdots
\nonumber\\
&=
\mathrm{Tr}\!\left(\partial_\mu\Phi\,\partial^\mu\Phi\right)+\cdots
=
\sum_{i=1}^{N_f}(\partial_\mu v_i)^2+\cdots.
\end{align}
Therefore the canonically normalized fields are
\begin{equation}
\pi_i \;\equiv\; \sqrt{2}\,v_i,
\qquad\Rightarrow\qquad
\frac{f_\pi^2}{4}\mathrm{Tr}(\partial_\mu V\,\partial^\mu V^\dagger)
=
\frac{1}{2}\sum_{i=1}^{N_f}(\partial_\mu\pi_i)^2+\cdots.
\label{eq:pii_canonical}
\end{equation}
 For diagonal $\mathcal{M}=\mathrm{diag}(m_1,\dots,m_{N_f})$ and vacuum angles $\phi_i=0$ at $\theta=0$,
the leading-order mass term is
\begin{align}
\mathcal{L}_{\rm mass}
&=
\frac{f_\pi^2 B_0}{2}\mathrm{Tr}\!\left[\mathcal{M}(V+V^\dagger)\right]
\nonumber\\
&=
f_\pi^2 B_0\sum_{i=1}^{N_f}m_i\cos\!\left(\frac{2v_i}{f_\pi}\right).
\end{align}
Expanding $\cos(2v_i/f_\pi)=1-\frac{1}{2}(2v_i/f_\pi)^2+\cdots$ gives
\begin{equation}
\mathcal{L}_{\rm mass}
=
f_\pi^2 B_0\sum_i m_i
-\;2B_0\sum_i m_i\,v_i^2 + \cdots.
\end{equation}
Using again $M_{\pi,i}^2\equiv 2B_0 m_i$  this becomes
\begin{equation}
\mathcal{L}_{\rm mass}
=
\text{const}
-\sum_{i=1}^{N_f} M_{\pi,i}^2\,v_i^2+\cdots
=
\text{const}
-\frac{1}{2}\sum_{i=1}^{N_f} M_{\pi,i}^2\,\pi_i^2+\cdots,
\label{eq:Lmass_quad_cartan}
\end{equation}
where in the last step we used $\pi_i=\sqrt{2}\,v_i$.

 Recall from Sec.~\ref{sec:PQ_axion} that the vacuum energy including the axion is
\begin{equation}
E = \frac{\chi_{\mathrm{top}}}{2}\left(\theta - \sum_{i=1}^{N_f}\phi_i - c_{\rm PQ}\varphi\right)^2 
- f_\pi^2 B_0 \sum_{i=1}^{N_f} m_i\cos\phi_i,
\end{equation}
where $\phi_i$ are the vacuum phases and $\varphi = \sqrt{2}\,a/f_a$ is the dimensionless axion phase.

For fluctuations around the $CP$-conserving vacuum ($\theta = 0$, $\phi_i = 0$), the vacuum phases are related to the meson fields by $\phi_i = 2v_i/f_\pi$ at linear order. Substituting and using $\mathcal{L} = -E$ (for the potential part), the anomaly contribution to the Lagrangian is
\begin{equation}
\mathcal{L}_{\rm anom}
=
-\frac{\chi_{\mathrm{top}}}{2}\left[
\frac{2}{f_\pi}\sum_{i=1}^{N_f}v_i + c_{\rm PQ}\frac{\sqrt{2}\,a}{f_a}
\right]^2.
\label{eq:Lanom_with_axion}
\end{equation}

Equivalently, using the relation $\frac{i}{2}\mathrm{Tr}(\ln V - \ln V^\dagger) = -\frac{2}{f_\pi}\sum_i v_i$ (which follows from $\ln V - \ln V^\dagger = \frac{4i\Phi}{f_\pi}$ at leading order), we can write
\begin{equation}
\mathcal{L}_{\rm anom}
=
-\frac{\chi_{\mathrm{top}}}{2}\left[
c_{\rm PQ}\frac{\sqrt{2}\,a}{f_a}
- \frac{i}{2}\mathrm{Tr}(\ln V - \ln V^\dagger)
\right]^2,
\end{equation}
which matches the integrated form of the anomaly Lagrangian with $\theta \to c_{\rm PQ}\sqrt{2}a/f_a$.

In terms of the canonical fields $\pi_i = \sqrt{2}\,v_i$:
\begin{equation}
\mathcal{L}_{\rm anom}^{(2)}
=
-\frac{\chi_{\mathrm{top}}}{2}
\left[
\frac{\sqrt{2}}{f_\pi}\sum_{i=1}^{N_f}\pi_i + c_{\rm PQ}\frac{\sqrt{2}\,a}{f_a}
\right]^2.
\label{eq:Lanom_quad_cartan}
\end{equation}
 
 Collecting \eqref{eq:Lmass_quad_cartan} and \eqref{eq:Lanom_quad_cartan}, the full quadratic mass Lagrangian is
\begin{equation}
\mathcal{L}^{(2)}_{\rm mass}
=
-\frac{1}{2}\sum_{i=1}^{N_f}M_{\pi,i}^2\,\pi_i^2
-\frac{\chi_{\mathrm{top}}}{2}
\left[
\frac{\sqrt{2}}{f_\pi}\sum_{i=1}^{N_f}\pi_i + c_{\rm PQ}\frac{\sqrt{2}\,a}{f_a}
\right]^2.
\label{eq:Lquad_final_cartan}
\end{equation}
Defining
\begin{equation}
\epsilon \equiv \frac{\sqrt{2}\,c_{\rm PQ}}{f_a}, 
\qquad
\kappa \equiv \frac{\sqrt{2}}{f_\pi},
\end{equation}
we can write this more compactly as
\begin{equation}
\mathcal{L}^{(2)}_{\rm mass}
=
-\frac{1}{2}\sum_{i=1}^{N_f}M_{\pi,i}^2\,\pi_i^2
-\frac{\chi_{\mathrm{top}}}{2}
\left[
\kappa\sum_{i=1}^{N_f}\pi_i + \epsilon\,a
\right]^2.
\end{equation}

In the canonical basis $(a,\pi_1,\dots,\pi_{N_f})$, the $(N_f+1)\times(N_f+1)$ mass-squared matrix is
\begin{equation}
\mathbf{M}^2=
\begin{pmatrix}
\chi_{\mathrm{top}}\epsilon^2
&
\chi_{\mathrm{top}}\epsilon\kappa
&
\chi_{\mathrm{top}}\epsilon\kappa
&
\cdots
&
\chi_{\mathrm{top}}\epsilon\kappa
\\[8pt]
\chi_{\mathrm{top}}\epsilon\kappa
&
 M_{\pi,1}^2+\chi_{\mathrm{top}}\kappa^2
&
\chi_{\mathrm{top}}\kappa^2
&
\cdots
&
\chi_{\mathrm{top}}\kappa^2
\\[8pt]
\chi_{\mathrm{top}}\epsilon\kappa
&
\chi_{\mathrm{top}}\kappa^2
&
 M_{\pi,2}^2+\chi_{\mathrm{top}}\kappa^2
&
\cdots
&
\chi_{\mathrm{top}}\kappa^2
\\[8pt]
\vdots
&
\vdots
&
\vdots
&
\ddots
&
\vdots
\\[8pt]
\chi_{\mathrm{top}}\epsilon\kappa
&
\chi_{\mathrm{top}}\kappa^2
&
\chi_{\mathrm{top}}\kappa^2
&
\cdots
&
M_{\pi,N_f}^2+\chi_{\mathrm{top}}\kappa^2
\end{pmatrix},
\label{eq:mass_matrix_cartan}
\end{equation}
or explicitly in terms of fundamental parameters:
\begin{equation}
\mathbf{M}^2=
\begin{pmatrix}
\dfrac{2\chi_{\mathrm{top}} c_{\rm PQ}^2}{f_a^2}
&
\dfrac{2\chi_{\mathrm{top}} c_{\rm PQ}}{f_a f_\pi}
&
\cdots
&
\dfrac{2\chi_{\mathrm{top}} c_{\rm PQ}}{f_a f_\pi}
\\[10pt]
\dfrac{2\chi_{\mathrm{top}} c_{\rm PQ}}{f_a f_\pi}
&
 M_{\pi,1}^2+\dfrac{2\chi_{\mathrm{top}}}{f_\pi^2}
&
\cdots
&
\dfrac{2\chi_{\mathrm{top}}}{f_\pi^2}
\\[10pt]
\vdots
&
\vdots
&
\ddots
&
\vdots
\\[10pt]
\dfrac{2\chi_{\mathrm{top}} c_{\rm PQ}}{f_a f_\pi}
&
\dfrac{2\chi_{\mathrm{top}}}{f_\pi^2}
&
\cdots
&
M_{\pi,N_f}^2+\dfrac{2\chi_{\mathrm{top}}}{f_\pi^2}
\end{pmatrix}.
\label{eq:mass_matrix_explicit}
\end{equation}

 Because the anomaly term depends only on $\sum_i v_i \propto \mathrm{Tr}\,\Phi$, the axion mixes only with the singlet direction in Cartan space, i.e., the vector proportional to $(1,1,\dots,1)$ in the $(\pi_i)$ subspace. All orthogonal traceless combinations (corresponding to Cartan generators of $\SU(N_f)$) do \emph{not} mix with $a$ through the anomaly at this order.

\subsection{Eigenvalue equation and axion mass}
\label{subsec:eigenvalue}

The mass matrix \eqref{eq:mass_matrix_cartan} has the structure
\begin{equation}
\mathbf{M}^2 = \mathbf{D} + \mathbf{u}\mathbf{u}^T,
\end{equation}
where $\mathbf{D} = \mathrm{diag}(0,  M_{\pi,1}^2, \ldots, M_{\pi,N_f}^2)$ and 
\begin{equation}
\mathbf{u} = \sqrt{\chi_{\mathrm{top}}}\,(\epsilon,\, \kappa,\, \kappa,\, \ldots,\, \kappa)^T.
\end{equation}
Using the matrix determinant lemma, the eigenvalues $\lambda$ satisfy
\begin{equation}
\det(\mathbf{M}^2 - \lambda\,\mathbf{1}) = 0
\quad\Leftrightarrow\quad
1 + \mathbf{u}^T(\mathbf{D} - \lambda\,\mathbf{1})^{-1}\mathbf{u} = 0.
\end{equation}
This gives the characteristic equation
\begin{equation}
\boxed{\ 
1 + \frac{\chi_{\mathrm{top}}\epsilon^2}{-\lambda} + \sum_{i=1}^{N_f}\frac{\chi_{\mathrm{top}}\kappa^2}{M_{\pi,i}^2 - \lambda} = 0.
\ }
\label{eq:char_eq}
\end{equation}
Rearranging:
\begin{equation}
\lambda\left[\frac{1}{\chi_{\mathrm{top}}\epsilon^2} + \sum_{i=1}^{N_f}\frac{\kappa^2/\epsilon^2}{M_{\pi,i}^2 - \lambda}\right] = 1.
\label{eq:eigenvalue_eq}
\end{equation}
For $f_a \gg f_\pi$ (i.e., $\epsilon \ll \kappa$), the lightest eigenvalue $\lambda = m_a^2$ is much smaller than $M_{\pi,i}^2$. Setting $M_{\pi,i}^2 - \lambda \approx M_{\pi,i}^2$, we obtain
\begin{equation}
m_a^2 \simeq \frac{\chi_{\mathrm{top}}\epsilon^2}{1 + \chi_{\mathrm{top}}\kappa^2\sum_{i=1}^{N_f}M_{\pi,i}^{-2}}
= \frac{\chi_{\mathrm{top}}\epsilon^2}{1 + \frac{2\chi_{\mathrm{top}}}{f_\pi^2}\sum_{i=1}^{N_f}M_{\pi,i}^{-2}}.
\label{eq:ma_general}
\end{equation}
In the limit $\chi_{\mathrm{top}} \gg M_{\pi,i}^2 f_\pi^2/(2N_f)$ (heavy $\eta'$ limit), this simplifies to
\begin{equation}
m_a^2 \simeq \frac{\epsilon^2 f_\pi^2}{2\sum_{i=1}^{N_f}M_{\pi,i}^{-2}}
= \frac{c_{\rm PQ}^2 f_\pi^2}{f_a^2 \sum_{i=1}^{N_f}M_{\pi,i}^{-2}}.
\label{eq:ma_heavy_etap}
\end{equation}

\subsubsection{Two-flavor case.}
For $N_f = 2$:
\begin{equation}
\sum_{i=1}^{2}M_{\pi,i}^{-2} = \frac{1}{ M_{\pi,1}^2} + \frac{1}{ M_{\pi,2}^2} = \frac{ M_{\pi,1}^2 +  M_{\pi,2}^2}{ M_{\pi,1}^2  M_{\pi,2}^2},
\end{equation}
so
\begin{equation}
m_a^2 \simeq \frac{c_{\rm PQ}^2 f_\pi^2}{f_a^2}\,\frac{ M_{\pi,1}^2  M_{\pi,2}^2}{ M_{\pi,1}^2 +  M_{\pi,2}^2}.
\end{equation}
Using $M_{\pi,i}^2 = 2B_0 m_i$ and $m_\pi^2 = B_0(m_u + m_d)$:
\begin{equation}
\frac{ M_{\pi,1}^2  M_{\pi,2}^2}{ M_{\pi,1}^2 +  M_{\pi,2}^2} = \frac{4B_0^2 m_u m_d}{2B_0(m_u + m_d)} = 2B_0\frac{m_u m_d}{m_u + m_d} = 2m_\pi^2\frac{m_u m_d}{(m_u + m_d)^2}.
\end{equation}
Therefore:
\begin{equation}
\boxed{
m_a^2 = \frac{2c_{\rm PQ}^2 f_\pi^2 m_\pi^2}{f_a^2}\,\frac{m_u m_d}{(m_u + m_d)^2}.
}
\label{eq:ma_2flav_final}
\end{equation}

This can be written in terms of the QCD topological susceptibility:
\begin{equation}
m_a^2 f_a^2 = c_{\rm PQ}^2\,\chi_{\rm QCD},
\qquad
\chi_{\rm QCD}^{\rm LO} = 2f_\pi^2 m_\pi^2\,\frac{m_u m_d}{(m_u + m_d)^2}.
\label{eq:chi_QCD_relation}
\end{equation}

 State-of-the-art chiral+QED+NNLO analyses and lattice inputs pin down the overall coefficient to percent-level (or better), commonly quoted as
\begin{equation}
m_a \simeq 5.7\,\mu{\rm eV}\times\left(\frac{10^{12}\,{\rm GeV}}{f_a/c_{\rm PQ}}\right),
\end{equation}
with the understanding that detailed numbers depend on conventions and matching inputs.
 
From \eqref{eq:ma_2flav_final} one immediately reads off a few robust lessons.  First, the axion mass vanishes if any light quark becomes massless: in that limit the relevant explicit breaking disappears and the axion becomes an exact Goldstone boson. 
Second, the overall size of axion interactions is controlled by the ratio \(c_{\rm PQ}/f_a\): for fixed anomaly coefficient \(c_{\rm PQ}\) all low-energy axion effects decouple as \(f_a\to\infty\), while rescaling \(c_{\rm PQ}\) simply rescales the effective couplings at fixed \(f_a\). 
Third, this decoupling is precisely why astrophysical constraints and laboratory searches are usually phrased as lower bounds on \(f_a\): for standard QCD axions one often quotes \(f_a \gtrsim 10^9\,\mathrm{GeV}\) as a useful benchmark, while emphasizing that the precise number is model- and channel-dependent (e.g.\ through the axion--photon and axion--electron couplings and the assumed UV completion). 
Finally, \eqref{eq:ma_2flav_final} makes transparent that \(m_a\) is not an independent parameter once \(f_a\) is specified: it is calculable in terms of light-quark masses and chiral inputs, up to the usual uncertainties in those quantities and higher-order chiral corrections.

\subsection{Axion UV completions}
\label{subsec:UVcompletions_axion}

The low-energy axion effective field theory (EFT) is largely fixed by the PQ scale $f_a$
and the QCD anomaly coefficient (equivalently the domain-wall number $N_{\rm DW}$),
\begin{equation}
\mathcal{L}_{\rm eff}\;\supset\;
\frac{\alpha_s}{8\pi}\,\frac{a}{f_a}\,G^a_{\mu\nu}\tilde G^{a\,\mu\nu}
\;+\;
\frac{\alpha}{8\pi}\,\frac{E}{N}\,\frac{a}{f_a}\,F_{\mu\nu}\tilde F^{\mu\nu}
\;+\; \sum_f \frac{\partial_\mu a}{2 f_a}\, C_f\,\bar f\gamma^\mu\gamma_5 f
\;+\;\cdots,
\label{eq:axion_EFT_general}
\end{equation}
where $\alpha_s=g_s^2/4\pi$, $N = c_{{PQ}}$ (often called $N$ in the axion literature) and $E$  are the color and electromagnetic PQ  anomalous coefficients, respectively, and
$N_{\rm DW}\equiv |N|$ for the standard normalization of PQ charges. The EFT logic is
model-independent, but the numerical values of $E/N$ and the fermionic coefficients $C_f$
are \emph{UV data}. This motivates distinguishing the two canonical UV completions:
KSVZ (hadronic) and DFSZ (two-Higgs) axions.

\subsubsection{KSVZ (Kim--Shifman--Vainshtein--Zakharov) axions.}
In KSVZ models one extends the SM by (i) a complex scalar $\Phi$ carrying the global
$\U(1)_{\rm PQ}$ charge and (ii) at least one heavy vector-like colored fermion pair
$Q_{L,R}$ with a PQ-chiral Yukawa coupling
\begin{equation}
\mathcal{L}_{\rm KSVZ}\supset -\,y\,\Phi\,\bar Q_L Q_R + {\rm h.c.},
\qquad
\langle \Phi\rangle=\frac{f_a}{\sqrt2}\,.
\label{eq:KSVZ_yukawa}
\end{equation}
After PQ breaking, the axion is the phase of $\Phi$ and couples to QCD through the
triangle anomaly induced by the heavy quark. A defining phenomenological feature is that
\emph{SM fermions carry no PQ charge} in minimal KSVZ, so the axion has no tree-level
coupling to electrons ($C_e=0$ at tree level). Electron couplings are induced radiatively
and are therefore typically suppressed compared to DFSZ. The original ``invisible axion''
constructions are due to Kim~\cite{Kim:1979if} and independently 
Shifman--Vainshtein--Zakharov~\cite{Shifman:1979if}.

\subsubsection{DFSZ (Dine--Fischler--Srednicki--Zhitnitsky) axions.}
DFSZ models instead embed the PQ symmetry into an extended Higgs sector: the SM is
augmented by two Higgs doublets $H_u$ and $H_d$ (and a singlet $\Phi$), with PQ
charges chosen such that standard Yukawa couplings are PQ invariant. In a minimal
implementation, the PQ symmetry is broken when the singlet $\Phi$ (and possibly the Higgs 
fields) acquires a vacuum expectation value. The axion is the physical pseudo-Nambu--Goldstone 
boson remaining after electroweak symmetry breaking. Because SM quarks and leptons now carry PQ
charges, DFSZ axions \emph{do} couple to electrons already at tree level, and the magnitude
of $C_e$ depends on the UV charge assignment and on $\tan\beta \equiv v_u/v_d$
(the ratio of Higgs vevs). This tree-level $a$--$e$ coupling is central to stellar cooling
constraints and to helioscope sensitivities in channels sourced by $g_{ae}$
(e.g.\ bremsstrahlung/Compton production in the Sun).
The original DFSZ papers are due to Zhitnitsky~\cite{Zhitnitsky:1980tq} and 
Dine--Fischler--Srednicki~\cite{Dine:1981rt}.

\subsection{Impact of UV completions on observables }
\label{subsec:UVcompletions_pheno]}
In this subsection we show how UV completions of the axion affect  low energy physics and  impact  physical observables.
\subsubsection{Model dependence of $g_{a\gamma\gamma}$.}
A key UV-sensitive quantity is the axion--photon coupling.
At low energies, mixing with $\pi^0$ and $\eta$ induces a universal contribution such that
\begin{equation}
g_{a\gamma\gamma}
\;=\;
\frac{\alpha}{2\pi f_a}
\left(\frac{E}{N} - \mathcal{C}_{\rm QCD}\right),
\qquad
\mathcal{C}_{\rm QCD}\simeq 1.92(4)\ \ (\text{for physical quark masses}),
\label{eq:gagg_modeldep}
\end{equation}
where $\mathcal{C}_{\rm QCD}$ is the model-independent QCD contribution computed at 
next-to-leading order in $\SU(2)$ (two flavors) chiral perturbation theory. A convenient modern reference 
deriving $\mathcal{C}_{\rm QCD}$ and other precision couplings in chiral EFT matched to 
lattice inputs is~\cite{GrillidiCortona:2015jxo}. In KSVZ/DFSZ models, the UV coefficient $E/N$ is set by the electromagnetic charges of the
PQ-chiral fermions (heavy $Q$ in KSVZ; SM fermions and Higgs content in DFSZ), and therefore
$g_{a\gamma\gamma}$ can vary significantly between UV completions. For benchmark KSVZ with 
a single heavy quark in the fundamental of $\SU(3)$  of color with electric charge $Q_{\rm em}=0$, 
one has $E/N=0$; for minimal DFSZ models $E/N=8/3$.

\subsubsection{Domain-wall number and cosmological viability.}
The integer $N_{\rm DW}$ controls the number of degenerate vacua after PQ breaking and hence
the severity of the axion domain-wall problem if PQ breaking occurs after inflation. Minimal
KSVZ can realize $N_{\rm DW}=1$ (cosmologically safest), while minimal DFSZ typically gives
$N_{\rm DW}=6$ (a multiple of the number of generations), requiring either PQ breaking before 
inflation or an additional explicit bias/structure to remove stable walls. For the classic 
discussion of axion domain walls see Sikivie~\cite{Sikivie:1982qv}. Domain wall dynamics and the resulting stochastic gravitational wave background have been studied extensively, see Ref.~\cite{Fu:2024jhu} for a recent treatment and further references.

\subsubsection{Impact on searches (ADMX vs CAST as exemplars).}
The UV completion fixes the pattern of couplings and therefore the optimal search strategy.
Haloscopes such as ADMX target $g_{a\gamma\gamma}$ through resonant conversion of galactic
axion dark matter into microwave photons~\cite{ADMX:2018gho}. Helioscopes such as CAST are sensitive primarily to solar axions produced through the
Primakoff process (and, in extended analyses, to production channels involving the axion-lepton coupling $g_{ae}$)
\cite{CAST:2007jps}. For an overview of the experimental landscape and how different couplings
map onto complementary search channels, a standard modern review is~\cite{Irastorza:2018dyq}. The EFT derivations in these notes remain valid independent of UV completion; the role of
KSVZ vs DFSZ is to provide concrete, widely used UV benchmarks for the anomaly ratio $E/N$,
for the fermion coefficients $C_f$ (especially $C_e$), and for the expected $N_{\rm DW}$.

\subsection{The Axion Quality Problem}
\label{subsec:quality}

The Peccei-Quinn mechanism provides an elegant dynamical solution to the strong $CP$ problem, but it relies fundamentally on the existence of a global $U(1)_{\rm PQ}$ symmetry. This reliance exposes the axion solution to a serious theoretical challenge known as the  {axion quality problem}~\cite{Kamionkowski:1992mf,Holman:1992us,Barr:1992qq,Ghigna:1992iv}. The essence of the problem is that quantum gravity is widely believed to violate all global symmetries, and even tiny violations of the PQ symmetry can completely undermine the axion solution to the strong $CP$ problem.

\subsubsection{Quantum gravity and global symmetries}

A fundamental principle emerging from our understanding of quantum gravity, supported by arguments from black hole physics, wormholes, and string theory, is that there are no exact global symmetries in a consistent theory of quantum gravity~\cite{Banks:2010zn,Harlow:2018tng}. The physical intuition behind this expectation comes from several directions. Black holes can absorb particles carrying global charges and subsequently evaporate via Hawking radiation into states that need not conserve the original global charge, thereby violating the global symmetry. Wormhole configurations in Euclidean quantum gravity can connect regions of spacetime and transport global charge from one asymptotic region to another, effectively violating global charge conservation from the perspective of a single observer~\cite{Abbott:1989jw,Giddings:1988cx,Kallosh:1995hi}. In string theory, it has been argued that all continuous symmetries must be gauged, with global symmetries emerging only as accidental low-energy phenomena.

If global symmetries are not fundamental, then the $U(1)_{\rm PQ}$ symmetry invoked in the axion solution cannot be exact. At best, it can be an approximate accidental symmetry of the low-energy effective theory, explicitly broken by higher-dimensional operators suppressed by powers of the Planck mass $M_{\rm Pl}$.

\subsubsection{Parametrization of PQ-violating effects}

Following the analysis in~\cite{Kamionkowski:1992mf}, we can parametrize the effects of Planck-scale physics on the PQ symmetry through higher-dimensional operators in the effective Lagrangian. Consider a complex scalar field $\Phi$ that carries PQ charge and acquires a vacuum expectation value $\langle\Phi\rangle = f_a/\sqrt{2}$, where $f_a$ is the axion decay constant. The  general PQ-violating operators take the form:
\begin{equation}
\mathcal{L}_{\rm PQ-viol} = \sum_{n \geq 5} \frac{c_n}{M_{\rm Pl}^{n-4}} \Phi^n + \text{h.c.},
\label{eq:PQviolating}
\end{equation}
where $c_n$ are dimensionless coefficients expected to be of order unity if quantum gravity respects no global symmetries, and we have written the operators in a schematic form suppressing possible additional field content.

After spontaneous symmetry breaking, $\Phi = (f_a/\sqrt{2})e^{ia/f_a}$, these operators generate a potential for the axion field $a$:
\begin{equation}
V_{\rm grav}(a) \sim \sum_n c_n \frac{f_a^n}{M_{\rm Pl}^{n-4}} \cos\left(\frac{na}{f_a} + \delta_n\right),
\label{eq:Vgrav}
\end{equation}
where $\delta_n$ are $CP$-violating phases that are generically of order unity\footnote{Note that here $\Phi$ is related to the previous sections complex field ${\cal N}$ with a   $1/2$ normalization of the kinetic term to account for the extra $\sqrt{2}$ in the exponential.}.

\subsubsection{The quality condition}

The axion potential receives contributions from both QCD instantons and the gravitationally-induced terms in Eq.~\eqref{eq:Vgrav}. The QCD contribution takes the familiar form:
\begin{equation}
V_{\rm QCD}(a) \approx m_\pi^2 f_\pi^2 \left[1 - \cos\left(\frac{a}{f_a} - {\theta}\right)\right] \approx \Lambda_{\rm QCD}^4 \left[1 - \cos\left(\frac{a}{f_a} -{\theta}\right)\right],
\end{equation}
where $\Lambda_{\rm QCD} \sim 200$~MeV is the QCD scale. This potential has a minimum at $\langle a \rangle/f_a = {\theta}$, thereby cancelling the $\bar{\theta}$ parameter and solving the strong $CP$ problem.

However, the gravitational contribution shifts the minimum of the total potential. For the axion solution to remain viable, the shift in the vacuum expectation value $\langle a \rangle$ induced by gravitational effects must satisfy:
\begin{equation}
\left|\frac{\delta\langle a \rangle}{f_a}\right| \equiv |\bar{\theta}| \lesssim 10^{-10},
\label{eq:qualitycondition}
\end{equation}
to maintain consistency with the neutron EDM constraint~\cite{Abel:2020pzs}.

Comparing the gravitational and QCD contributions, the condition \eqref{eq:qualitycondition} requires:
\begin{equation}
|c_n| \frac{f_a^n}{M_{\rm Pl}^{n-4}} \lesssim 10^{-10} \times \Lambda_{\rm QCD}^4.
\label{eq:qualitybound}
\end{equation}

\subsubsection{Quantitative assessment of the problem}

To appreciate the severity of the quality problem, consider the case of a dimension-5 operator ($n=5$) with $c_5 \sim 1$. The induced contribution to the axion potential is:
\begin{equation}
V_5 \sim \frac{f_a^5}{M_{\rm Pl}} \sim \frac{f_a^5}{2.4 \times 10^{18}~\text{GeV}}.
\end{equation}
For $f_a \sim 10^{12}$~GeV (the value favored by cosmological considerations for axion dark matter), this gives:
\begin{equation}
V_5 \sim \frac{(10^{12}~\text{GeV})^5}{10^{18}~\text{GeV}} \sim 10^{42}~\text{GeV}^4.
\end{equation} 
Here we assumed the planck mass $M_{\rm Pl}$ to be the reduced one with value $M_{\rm Pl} = 2.4\times 10^{18}$~GeV. 
Meanwhile, the QCD contribution is:
\begin{equation}
V_{\rm QCD} \sim \Lambda_{\rm QCD}^4 \sim (200~\text{MeV})^4 \sim 10^{-3}~\text{GeV}^4.
\end{equation}

The ratio is catastrophically large:
\begin{equation}
\frac{V_5}{V_{\rm QCD}} \sim 10^{45},
\end{equation}
which would completely overwhelm the QCD potential and destroy the axion solution to the strong $CP$ problem.

To satisfy the quality condition \eqref{eq:qualitybound}, the coefficient $c_5$ would need to be tuned to an extraordinarily small value:
\begin{equation}
|c_5| \lesssim 10^{-55}.
\end{equation}

This represents a fine-tuning far more severe than the original strong $CP$ problem, which only required $|\bar{\theta}| \lesssim 10^{-10}$. In other words, if we allow arbitrary Planck-suppressed operators, the axion ``solution'' merely exchanges the original fine-tuning for an even more extreme one.

\subsubsection{The minimum dimension requirement}

A more useful way to phrase the quality problem is to ask: what is the minimum dimension $n_{\rm min}$ of PQ-violating operators such that the axion solution remains natural (i.e., without fine-tuning the coefficients $c_n$)?

Setting $c_n \sim 1$ in Eq.~\eqref{eq:qualitybound} and solving for $n$, one finds:
\begin{equation}
n_{\rm min} \approx 4 + \frac{\ln(10^{-10} \times \Lambda_{\rm QCD}^4 \times M_{\rm Pl}^4 / f_a^4)}{\ln(f_a/M_{\rm Pl})}.
\end{equation}

For $f_a \sim 10^{10}-10^{12}$~GeV:
\begin{equation}
n_{\rm min} \approx 9-14.
\end{equation}

For the ``GUT-scale'' axion with $f_a \sim 10^{14}-10^{16}$~GeV:
\begin{equation}
n_{\rm min} \approx 21-37.
\end{equation}

In other words, for the axion solution to be natural, the PQ symmetry must be protected from explicit breaking up to operators of dimension $\sim 9$--$14$ or higher. This is a very stringent requirement that demands a compelling theoretical explanation. Any complete model of the QCD axion must explain why such low-dimension PQ-violating operators are absent.

\subsubsection{Model-building approaches to the quality problem}

The axion quality problem has motivated extensive theoretical work aimed at understanding why the PQ symmetry might be protected to such high order. Several broad classes of solutions have been proposed.

\vskip .3cm
\noindent
\underline{Discrete gauge symmetries.}
One approach is to impose discrete gauge symmetries (such as $\mathbb{Z}_N$ for large $N$) that forbid low-dimensional PQ-violating operators~\cite{Barr:1992qq,Dias:2002hz,DiLuzio:2017pfr}. Since discrete gauge symmetries are respected by quantum gravity, they can provide the necessary protection. The PQ symmetry then emerges as an accidental global symmetry of the low-energy theory. This approach requires careful model-building to ensure that the discrete symmetry is anomaly-free and compatible with the Standard Model gauge structure. The key insight is that if $\Phi$ carries charge $q$ under a $\mathbb{Z}_N$ gauge symmetry, then operators of the form $\Phi^n$ are forbidden unless $n$ is a multiple of $N/\gcd(n,N)$. By choosing $N$ sufficiently large, one can protect the PQ symmetry to arbitrarily high order.

\vskip .3cm
\noindent
\underline{Composite axion models.}
In composite axion models~\cite{Kim:1984pt,Choi:1985cb,Randall:1992ut,Redi:2016esr,Lillard:2018fdt}, the axion arises as a pseudo-Nambu-Goldstone boson from the spontaneous breaking of a flavor symmetry in a new strongly-coupled sector, analogous to pions in QCD. In these constructions, the PQ symmetry can emerge as an accidental symmetry of the composite dynamics. By appropriate choice of the confining gauge group and matter content, the PQ symmetry can be protected to arbitrarily high order. The composite nature of the axion provides a dynamical explanation for the quality of the PQ symmetry without requiring ad hoc discrete symmetries.

Recent work has identified explicit chiral gauge theories where the PQ symmetry is protected up to dimension 12 or higher~\cite{Contino:2021ayn}, providing concrete realizations of high-quality composite axions. The key mechanism is that the lowest-dimension operators that can break the accidental PQ symmetry are determined by the gauge quantum numbers of the constituent fermions, and can be made arbitrarily high-dimensional by suitable choice of the gauge group and representations.

\vskip .3cm
\noindent
\underline{String theory axions.}
String theory provides a particularly attractive framework for addressing the quality problem~\cite{Svrcek:2006yi,Conlon:2006tq,Cicoli:2012sz,Demirtas:2021gsq}. Axion-like fields arise naturally in string compactifications as zero modes of higher-form gauge fields (such as the Neveu-Schwarz $B$-field or Ramond-Ramond $p$-form fields). The shift symmetry of these string axions is a remnant of the higher-dimensional gauge invariance, which is exact to all orders in perturbation theory.

In this context, the PQ symmetry is broken only by non-perturbative effects such as worldsheet instantons or Euclidean D-branes, whose contributions are exponentially suppressed by factors of $e^{-S_{\rm inst}}$ where $S_{\rm inst}$ is the instanton action. For appropriate compactification geometries, these exponential suppressions can naturally satisfy the quality requirements. A systematic study of type IIB orientifold compactifications~\cite{Demirtas:2021gsq} found that the strong $CP$ problem is solved to adequate precision in approximately 99.7\% of the sampled vacua, with stringy instanton effects contributing sufficiently subdominant contributions to the axion potential. This remarkable result suggests that high-quality axions are generic in string theory, rather than requiring special tuning.

\vskip .3cm
\noindent
\underline{Extra-dimensional constructions.}
In theories with extra dimensions, the axion can arise from a higher-dimensional gauge field, with the PQ symmetry descending from the gauge invariance of the higher-dimensional theory~\cite{Choi:2003wr,Cheng:2001ys,Cox:2019rro}. Gravitational violations of the PQ symmetry can be localized on a UV brane while the axion field is localized toward an IR brane, leading to exponential suppression of the dangerous operators through wavefunction overlap factors. This geometric sequestering provides a natural explanation for the high quality of the PQ symmetry.

In the holographic dual description~\cite{Cox:2019rro}, these extra-dimensional models correspond to 4D strongly-coupled theories where the PQ symmetry emerges as an accidental symmetry protected by the conformal dynamics. The quality of the PQ symmetry is then related to the degree of compositeness of the axion.

\vskip .3cm
\noindent
\underline{Gauge-protected axions.}
Another approach is to construct models where the PQ symmetry emerges as an accidental consequence of gauge symmetries~\cite{Barr:1992qq,Fukuda:2017ylt,DiLuzio:2017pfr}. In these constructions, one introduces additional gauge symmetries (either continuous or discrete) that forbid all dangerous PQ-violating operators up to sufficiently high dimension. The challenge is to ensure that these gauge symmetries are anomaly-free and that the resulting model is phenomenologically viable.

A particularly elegant implementation uses the fact that certain gauge charge assignments automatically forbid all operators of dimension less than some critical value $n_{\rm min}$. For example, if fields carry large and mutually prime charges under a $U(1)$ gauge symmetry that is subsequently Higgsed, the lowest-dimension gauge-invariant operators can be pushed to very high dimension.

\subsubsection{Gravitational instantons and the quality problem}

A more detailed understanding of how gravity can violate global symmetries comes from studying
\emph{gravitational instantons}---non-perturbative, finite-action solutions to the Euclidean Einstein
equations---and related Euclidean saddles such as axion wormholes
\cite{Hebecker:2016dsw,Catinari:2024zon}.
These configurations are the gravitational analogue of gauge-theory instantons and can mediate
global-symmetry violation.

Charged Eguchi--Hanson instantons (and generalizations) have long been discussed as potential sources
of Peccei--Quinn (PQ) symmetry breaking. The seminal analysis of the Eguchi--Hanson contribution
to PQ-violating operators is due to Holman, Kephart and Rey \cite{Holman:1992us}; related axion
dynamics in wormhole backgrounds were explored earlier in \cite{Rey:1989mg}. The early literature
emphasized that such effects might generate PQ-violating operators with coefficients large enough to
jeopardize the axion solution.

However, a recent critical reanalysis finds that the contribution of gravitational instantons to the axion
potential can be far smaller than naive dimensional analysis would suggest \cite{Catinari:2024zon}.
Using semiclassical methods in close analogy with QCD instanton calculations (including the
appropriate ``'t~Hooft-operator'' structure and zero-mode effects), the resulting PQ breaking is shown
to be suppressed by additional factors tied to the detailed properties of the instanton solution.

This suggests that the axion quality problem may be less severe than previously thought, at least for
semiclassically controlled gravitational effects. That said, the conclusion depends on the regime of
validity of semiclassical gravity and on assumptions about the ultraviolet completion of gravity, so the
broader question of how quantum gravity breaks global symmetries remains active.

\subsubsection{Wormholes and the quality problem}

Wormhole solutions in Euclidean quantum gravity provide another mechanism for breaking global symmetries~\cite{Abbott:1989jw,Giddings:1988cx,Kallosh:1995hi,Alvey:2020nyh}. A wormhole can be thought of as a ``baby universe'' that branches off from our universe and either reconnects elsewhere or remains disconnected. If a wormhole carries global charge into a baby universe, that charge is effectively lost from the perspective of an observer in our universe, thereby violating global charge conservation.

The prototypical example is the Giddings--Strominger wormhole~\cite{Giddings:1988cx}, an exact Euclidean solution to Einstein gravity coupled to a massless axion. Consider the action
\begin{equation}
S = \int d^4x \sqrt{g}\left[\frac{M_{\rm Pl}^2}{2}R - \frac{f_a^2}{2}(\partial\theta)^2\right],
\end{equation}
where $R$ is the Ricci scalar, $a = f_a\theta$ is the canonically normalized axion and $\theta$ is the dimensionless angular variable. The wormhole solution has the metric
\begin{equation}
ds^2 = \frac{dr^2}{1 - (r_0/r)^4} + r^2 d\Omega_3^2,
\end{equation}
where $r_0$ is the throat radius (the minimum value of the radial coordinate) and $d\Omega_3^2$ is the round metric on $S^3$.

The axion field profile carries a conserved topological charge
\begin{equation}
n = \frac{f_a^2}{8\pi^2}\int_{S^3} \star\, d\theta,
\end{equation}
which counts the number of axion quanta transported through the wormhole throat. Evaluating the Euclidean action on this solution yields
\begin{equation}
S_{\rm wh} = \pi^2 f_a^2 r_0^2.
\label{eq:Swh_r0}
\end{equation}

The throat radius is not a free parameter but is determined by the Einstein equations in terms of the charge $n$:
\begin{equation}
r_0^2 = \frac{n}{2\pi^2 f_a M_{\rm Pl}}.
\label{eq:r0_constraint}
\end{equation}
Substituting \eqref{eq:r0_constraint} into \eqref{eq:Swh_r0} gives the on-shell action in terms of the axion charge:
\begin{equation}
\boxed{\ S_{\rm wh} = \frac{f_a\, n}{2\, M_{\rm Pl}}.\ }
\label{eq:Swh_final}
\end{equation}

This result has a striking and problematic implication. For $f_a \ll M_{\rm Pl}$, the wormhole action is small. Taking $f_a \sim 10^{12}\,\mathrm{GeV}$ and $n = 1$:
\begin{equation}
S_{\rm wh} \sim \frac{10^{12}\,\mathrm{GeV}}{2 \times 2.4 \times 10^{18}\,\mathrm{GeV}} \sim 2 \times 10^{-7}.
\end{equation}
The corresponding suppression factor $e^{-S_{\rm wh}} \approx 1$ provides essentially \emph{no} exponential suppression. In other words, within semiclassical Einstein gravity, Giddings--Strominger wormholes can generate $\mathcal{O}(1)$ PQ-violating effects, potentially destroying the axion solution to the strong $CP$ problem.

Several resolutions to this ``wormhole problem'' have been proposed:

\begin{enumerate}
\item  \underline{String theory corrections.} In string compactifications, additional moduli-dependent factors can enhance the wormhole action.  In Ref.~\cite{Kallosh:1995hi} it was showed that string-theoretic wormholes acquire an action of order
\begin{equation}
S_{\rm wh}^{\rm string} \sim \frac{M_{\rm Pl}^2}{f_a^2} \times g_s^{-2},
\end{equation}
where $g_s$ is the string coupling and should not be confused with the QCD coupling constant. For $g_s \sim 0.1$ and $f_a \sim 10^{12}\,\mathrm{GeV}$, this gives $S_{\rm wh}^{\rm string} \sim 10^{14}$, providing enormous suppression.

\item  \underline{Charge quantization and selection rules.} If the axion charge $n$ must be large (e.g., due to discrete gauge symmetries), the action \eqref{eq:Swh_final} grows linearly with $n$, potentially restoring adequate suppression.

\item \underline{Absence of global symmetries in quantum gravity.} From the perspective of the swampland program, the very existence of wormhole solutions may be an artifact of the low-energy effective theory. In a complete theory of quantum gravity, ``global'' symmetries may always be gauged (possibly by higher-form gauge fields), eliminating the wormhole threat~\cite{Harlow:2018tng}.
\end{enumerate}

Recent analyses~\cite{Alvey:2020nyh} have revisited the wormhole contribution within non-perturbative Einstein gravity and string theory, concluding that while wormholes pose a genuine threat in pure gravity, string-theoretic embeddings with natural values of $g_s$ can adequately protect the PQ symmetry. The question remains an active area of research at the interface of quantum gravity and axion phenomenology.

\subsubsection{The gauge axion solution}

A conceptually distinct approach to the quality problem is the ``gauge axion'' formulation~\cite{Dvali:2005an,Dvali:2022fdv}, in which the axion emerges not as the phase of a complex scalar field (which requires a global PQ symmetry), but rather as the Hodge dual of a two-form gauge field $B_{\mu\nu}$. In this formulation, the axion is part of the QCD gauge redundancy itself, and the solution to the strong $CP$ problem does not rely on any global symmetry.

The gauge axion has exact quality by construction---it is insensitive to arbitrary continuous deformations of the theory to all orders in the operator expansion. However, this formulation raises its own questions about UV completion. While the PQ axion is UV-completed into a phase of a complex scalar, the gauge axion has no known completion within a renormalizable effective field theory and instead must be UV-completed directly within the fundamental theory of gravity. This has led to the intriguing suggestion~\cite{Dvali:2022fdv} that the existence of a high-quality axion may point toward fundamental string theory as the UV completion, since fundamental strings can naturally source two-form gauge fields.

\subsubsection{Implications for axion phenomenology}

The quality problem has important implications for axion phenomenology and model-building. If the PQ symmetry is explicitly broken by higher-dimensional operators, even at a level consistent with solving the strong $CP$ problem, this can have observable consequences.

First, the minimum of the axion potential may be shifted slightly away from $\bar{\theta}_{\rm eff} = 0$, leading to a small but potentially measurable residual value of $\bar{\theta}$. Future improvements in neutron EDM measurements, which aim to reach sensitivities of $10^{-28}$~$e\cdot$cm or below, could probe this possibility.

Second, if the explicit breaking is not aligned with the QCD contribution, it can modify the QCD axion mass prediction from the standard formula:
\begin{equation}
m_a \approx 5.7~\mu\text{eV} \times \frac{10^{12}~\text{GeV}}{f_a}.
\end{equation}
Significant explicit PQ breaking could either enhance or suppress the axion mass relative to this prediction, potentially opening up new regions of parameter space for axion searches.

Third, in cosmological contexts, explicit PQ breaking can affect the dynamics of the axion field during and after inflation, potentially modifying predictions for the axion dark matter abundance and isocurvature perturbations. In scenarios where the explicit breaking is comparable to the QCD contribution at early times, level-crossing phenomena can occur~\cite{Higaki:2016jjh}, leading to non-trivial cosmological evolution.

 \vskip .5cm

Overall, the axion quality problem represents a fundamental challenge to the Peccei-Quinn solution of the strong $CP$ problem. The requirement that the PQ symmetry be protected from explicit breaking up to operators of dimension $\sim 9$--$10$ or higher is not automatic and demands a theoretical explanation. This problem has driven extensive model-building efforts, leading to a rich landscape of constructions including discrete gauge symmetries, composite axions, string axions, and extra-dimensional models.

From a modern perspective, the quality problem should not be viewed as a fatal flaw of the axion solution, but rather as a guide to the underlying ultraviolet physics. Just as the hierarchy problem of the Higgs mass points to new physics at the TeV scale, the quality problem of the axion points to specific structures in the fundamental theory, whether discrete gauge symmetries, confining dynamics, or string-theoretic origins, that protect the PQ symmetry. In this sense, the axion quality problem provides a window into physics at or beyond the Planck scale.

The fact that high-quality axions appear to be generic in string theory compactifications~\cite{Demirtas:2021gsq}, rather than requiring special tuning, is an encouraging result. It suggests that the theoretical obstacles to the axion solution may be less severe than originally feared, and that the QCD axion remains a well-motivated solution to the strong $CP$ problem that connects low-energy particle physics to fundamental questions about quantum gravity.

\subsection{Axion-less solutions from new strong dynamics}

The prototype for non-axion solutions to the strong CP problem is one-flavor massless QCD. In this case, the single quark field possesses a classical $\U(1)_A$ chiral symmetry that is anomalously broken at the quantum level. Since the underlying quark is massless, the $\theta$-angle can be rotated away by a chiral field redefinition. Crucially, no massless degree of freedom is left behind: the would-be Goldstone boson associated with $\U(1)_A$ acquires a mass from the anomaly, and the low-energy spectrum consists entirely of massive hadronic states. The $\theta$-angle becomes observable only when an explicit quark mass is introduced, as described in the previous sections. Of course, in the physical world all quarks are massive, and in particular $m_u \neq 0$ has been firmly established by lattice QCD calculations. Nevertheless, the one-flavor massless case serves as a guiding principle for constructing more elaborate solutions.

A concrete realization of this idea was proposed in Ref.~\cite{Hsu:2004mf}, where it was shown that ultraviolet physics involving new strong dynamics beyond QCD can render the $\theta$-angle physically unobservable without introducing an explicit light axion. The model employs an extended gauge structure $SU(N) \times SU(3)_L \times SU(3)_R$, with new fermions transforming under both the new strong sector and ordinary color. After confinement and chiral symmetry breaking in the new sector, the diagonal $SU(3)_V$ is identified with QCD. Since the new fermions are massless at the UV scale, their chiral rotations can absorb the $\theta$-parameter, rendering it unphysical---precisely as in the one-flavor massless prototype.

This line of reasoning was further developed by Wilczek and Moore~\cite{Wilczek:2016gzx}, who introduced the concept of \emph{cryptoquarks}: effectively superheavy massless quarks that remain permanently confined within the new strong dynamics. As in the original work~\cite{Hsu:2004mf}, a cryptoquark is massless from the perspective of the chiral symmetry that rotates away $\theta$, yet it appears superheavy in the low-energy spectrum because it is bound into heavy composite states by the new confining force.

A major hurdle for constructing consistent solutions to the strong CP problem along these lines is that the enlarged (global) symmetry group, required to make $\theta$ unphysical via the new dynamics, often leaves behind light degrees of freedom in the form of pseudo-Nambu--Goldstone bosons or massless fermions. When the global symmetry is spontaneously broken by the new strong dynamics, the associated light degrees of freedom acquire masses only through explicit symmetry-breaking effects, which may be small. These light states can lead to phenomenological constraints from collider searches, cosmological bounds, or precision tests. A successful model must therefore arrange for these pseudo-Goldstone bosons to be sufficiently heavy, or ensure they couple weakly enough to evade experimental detection.

On the other hand, these constructions are expected to avoid, or at least ameliorate, the axion quality problem that plagues conventional Peccei--Quinn solutions. As we have seen in standard axion models, the global $\U(1)_{\rm PQ}$ symmetry must be preserved to extraordinary precision by Planck-scale physics to maintain a sufficiently small $\bar{\theta}$. In contrast, solutions based on new strong dynamics rely on anomalous symmetries and confinement rather than on an approximate global symmetry, making them more robust against gravitational or other high-scale corrections.

\subsection{Alternative directions: the weak axion.}
While the QCD axion remains the most studied solution to the strong $CP$ problem, it is worth noting that 
analogous topological structures exist in the electroweak sector.
The weak $\mathrm{SU}(2)_L$ gauge group also admits a $\theta$-term, but within the Standard Model it is 
unobservable because it can be rotated away using the anomalous $B+L$ global 
symmetry~\cite{Anselm:1993uj,FileviezPerez:2014xju}.
As long as $B+L$ remains a good quality symmetry (i.e., baryon and lepton number violations from grand 
unification or Majorana masses remain small), the weak vacuum angle has no physical effect.

Nevertheless, the presence of a nontrivial electroweak topological susceptibility has motivated the 
introduction of a \emph{weak axion}, $a_W$, as the pseudo-Nambu--Goldstone boson of a spontaneously 
broken $\mathrm{U}(1)_{B+L}$~\cite{Dvali:2024hsb,Cacciapaglia:2025dme}.
Through its $\mathrm{SU}(2)_L$ anomaly, the weak axion would dynamically relax the weak topological angle.
A key feature is that $a_W$ is \emph{photophobic}: its anomaly coefficients satisfy $c_W = -c_B$, so that 
the tree-level coupling to photons vanishes~\cite{Craig:2018kne}.
Loop corrections generate a coupling to electrons that is logarithmically enhanced but still 
small~\cite{Bauer:2017ris}; red-giant cooling bounds then require the decay constant 
$f_W \gtrsim 10^3\,\mathrm{TeV}$~\cite{Cacciapaglia:2025dme,Davoudiasl:2025qqv}.

A distinctive phenomenological signature arises from the portal operator 
$(\Phi_W/\Lambda^3)\, qqql$ that communicates the $B+L$ charge to the Standard Model.
After spontaneous symmetry breaking, this induces proton decay via $p \to e^+ \pi^0$ and the 
characteristic channel $p \to e^+ a_W$~\cite{Cacciapaglia:2025dme}.
Current Super-Kamiokande limits on these modes constrain the new-physics scale to 
$\Lambda \gtrsim 10^{12}\,\mathrm{GeV}$, with future improvements at Hyper-Kamiokande offering the 
most promising avenue to test this scenario.
It should be noted that recent work~\cite{Cacciapaglia:2025xmr} has clarified that a light electroweak 
$\eta_W$ meson, proposed in Refs.~\cite{Dvali:2024hsb,Dvali:2025pcx}, does not correspond to a new 
propagating degree of freedom once color confinement and gauge invariance are properly enforced; 
its low-energy manifestation instead maps onto the $CP$-odd hydrogen ground state.

In summary, while the weak axion provides an interesting parallel to the QCD axion and illustrates the 
broader role of topology in gauge theories, the strong $CP$ problem remains a distinct challenge that requires either the  Peccei--Quinn mechanism or alternative ultraviolet physics for its resolution.

\section{Concluding remarks and outlook}

These notes are written with a pedagogical goal: to provide the reader with a coherent
effective field theory entry point into the strong $CP$ problem and its most widely studied dynamical
resolution, the QCD axion. The subject is vast. It spans foundational aspects of gauge-field topology,
$\theta$ vacua and branch structure \cite{tHooft:1976snw,Jackiw:1976pf,Callan:1976je}, anomaly matching and the
implementation of the $\U(1)_A$ anomaly \cite{Adler:1969gk,Bell:1969ts,Fujikawa:1979ay,Fujikawa:1980rc},
chiral dynamics and vacuum alignment \cite{DiVecchia:1980yfw,Dashen:1970et}, lattice determinations of
$\theta$-dependence and topological quantities \cite{Bhattacharya:2025qsk,Bonanno:2023hhp}, and an extensive
phenomenology program in cosmology and laboratory searches. A complete review of {\it everything that has ever been
done} on strong $CP$, including its many generalizations in QCD-like theories, its interplay with UV model building,
and the broad landscape of axion-like extensions, is far beyond the scope of an introductory graduate lecture.
Instead, the aim has been to supply a reliable set of conceptual and technical tools that repeatedly prove useful
across the literature.

The emphasis has therefore been on a small set of EFT ingredients that can be kept under analytic control:
treating $\theta$ as a genuine coupling multiplying the topological charge density $\mathcal{Q}(x)$,
encoding the $\U(1)_A$ anomaly in the chiral EFT through an auxiliary pseudoscalar background field
$\mathcal{Q}(x)\sim G\widetilde G$, and extracting physical consequences by minimizing the corresponding
effective potential (Dashen equations), including $\theta$ periodicity, branch structure, and the possibility of
spontaneous $CP$ breaking at special values such as $\theta=\pi$ (the Dashen phenomenon) \cite{Dashen:1970et}.
In the same framework one can also make contact with large-$N$ intuition and the pure-glue topological susceptibility
through the Witten--Veneziano relation \cite{Witten:1979vv,Veneziano:1979ec,DiVecchia:1980yfw}. A practical benefit of
this unified language is that it makes the assumptions behind alternative proposals sharply visible: when a mechanism
claims an absence of strong-interaction $CP$ violation, one can translate it into the EFT and identify what additional
low-energy ingredient is being introduced. In particular, in the recent ``no strong $CP$'' claim revisited in
Sec.~10, the effective description can be seen to amount to an additional variational phase, from the low-energy
standpoint, an axion-like (and in practice non-propagating) degree of freedom, whose existence is not mandated by QCD
itself \cite{Ai:2020ptm}.

The QCD axion is then understood as the controlled, dynamical realization of the same general idea: a new
\emph{propagating} pseudoscalar field associated with an (approximate) Peccei--Quinn symmetry, whose potential is
generated by QCD topological fluctuations and whose vacuum expectation value dynamically relaxes the effective
$\theta$-dependence of observables \cite{Peccei:1977hh,Peccei:1977ur,Weinberg:1977ma,Wilczek:1977pj}.
Once matched onto the chiral EFT, this leads to a predictive low-energy theory for the axion mass and couplings
\cite{GrillidiCortona:2015jxo,DiLuzio:2020wdo}. Beyond the minimal QCD axion, one often encounters
\emph{axion-like particles} (ALPs): light pseudo-Nambu--Goldstone bosons of spontaneously broken approximate global
symmetries that generically couple to gauge and matter fields through higher-dimensional operators, for example
\begin{equation}
\Lag_{\rm ALP}\supset
\frac{a}{f_a}\,\Big[
c_g\,\frac{g_s^2}{32\pi^2}G^a_{\mu\nu}\tilde G^{a\,\mu\nu}
+c_\gamma\,\frac{e^2}{32\pi^2}F_{\mu\nu}\tilde F^{\mu\nu}
+\cdots\Big],
\end{equation}
but whose mass and couplings need not be tied to QCD in the specific way characteristic of the QCD axion. ALPs thus
provide a broad EFT framework for new light pseudoscalars, while the \emph{QCD axion} corresponds to the special case
in which QCD dynamics fixes the leading potential and relates the mass and low-energy couplings. For accessible EFT
overviews and experimental motivation, see \cite{Jaeckel:2010ni,Irastorza:2018dyq,Graham:2015ouw,Marsh:2015xka}. The interplay between axion physics and astrophysics has deep historical roots. For example in the early work of Ref.~\cite{Vysotsky:1978dc}  astrophysical constraints on axion masses were established from stellar evolution, while  Ref.~\cite{Khlopov:1985fch} analyzed the gravitational instability of scalar fields, providing insights useful for modern studies of axion dark matter structure.

I hope the perspective developed here, deriving as much as possible from symmetry, anomaly constraints, and a
controlled chiral EFT, offers a simple starting route into the subject. The reader should now be equipped to
navigate the classic literature, to understand how modern results are matched onto low-energy observables, including
the standard $CP$-odd benchmarks such as the neutron EDM \cite{Pospelov:2005pr,Abel:2020pzs}, and to critically assess
proposed mechanisms that modify the $\theta$-angle vacuum structure. Many important directions remain open for deeper
study, including high-precision hadronic matrix elements for $CP$-odd observables, improved lattice control of
$\theta$-dependence and topological quantities, systematic treatments of $\theta$ in general confining theories, and
the rapidly evolving experimental program targeting axions and ALPs across a wide range of masses and couplings.
A final theme, emphasized in these notes, is that $\theta$-dependence in QCD-like theories can sometimes be sharpened
by importing exact information from adjacent confining theories (notably supersymmetric limits and planar-equivalentx
orientifold relations), providing a bridge between controlled nonperturbative dynamics and QCD-motivated EFT modeling.

\section*{Acknowledgments}
I am grateful to Borut Bajc, Michele Della Morte, Paolo Di Vecchia, Bj\"orn Garbrecht and Jessica Turner for a careful reading of the manuscript and for many valuable insights. I also thank Oleg Antipin, Jahmall Bersini, Giacomo Cacciapaglia, Manuel Meyer, Claudio Pica, Antonio Rago, Thomas Ryttov, and Apoorv Tiwari for helpful discussions. I further acknowledge Alessandra D’Alise,  Manuel Del Piano, Vigilante Di Risi, Clelia Gambardella, Jacob Holzendorff Hafjall, Davide Iacobacci, Viljami Leino, Giulia Muco, Mikolaj Myszkowski, Jonas Neuser, Mattia Damia Paciarini, Paola Simone, Vania Vellucci, and Sophie Lund Wagner for reading the manuscript and for helpful comments. Last but not least, I thank Tom Applequist, Paolo Di Vecchia, Kenneth Intriligator, Joe Schechter, and Mikhail Shifman for insightful conversations that helped shape my understanding of the topics discussed here through my studies and collaborations. The work of F.S. is partially supported by the Carlsberg Foundation, Semper Ardens grant CF22-0922.

\newpage

\appendix

\section{The CP-violating terms from the Baryon Lagrangian}
\label{Appendix-baryon}
  Expanding $\mathcal{M}_{p/m}(\theta)$ in powers of $\Phi$ with $\nu = \exp(i\Phi/f_\pi)$:
\begin{align}
\nu \mathcal{M}(\theta) \nu &= e^{i\Phi/f_\pi} \mathcal{M}(\theta) e^{i\Phi/f_\pi} \nonumber\\
&= \mathcal{M}(\theta) + \frac{i}{f_\pi}[\Phi, \mathcal{M}(\theta)] + \frac{i}{f_\pi}\{\Phi, \mathcal{M}(\theta)\} + \mathcal{O}(\Phi^2) \nonumber\\
&= \mathcal{M}(\theta) + \frac{2i}{f_\pi}\Phi\mathcal{M}(\theta) + \mathcal{O}(\Phi^2),
\end{align}
where we used $e^A B e^A = B + [A,B] + \{A,B\} + \ldots$ for the diagonal case. Similarly:
\begin{equation}
\nu^\dagger \mathcal{M}(\theta) \nu^\dagger = \mathcal{M}(\theta) - \frac{2i}{f_\pi}\Phi\mathcal{M}(\theta) + \mathcal{O}(\Phi^2).
\end{equation}

Therefore:
\begin{align}
\mathcal{M}_p(\theta) &= \mathcal{M}(\theta) + \mathcal{O}(\Phi^2), \\[6pt]
\mathcal{M}_m(\theta) &= \frac{2i}{f_\pi}\Phi\mathcal{M}(\theta) + \mathcal{O}(\Phi^3).
\end{align}

The $CP$-violating pion-baryon coupling arises from the interplay between the anomaly potential and the baryon mass terms. We now derive this exactly, keeping the full $\Phi$ dependence.
 From \eqref{eq:bar_explicit}, the terms involving $\xi\mathcal{M}\xi \pm \xi^\dagger\mathcal{M}^\dagger\xi^\dagger$ are:
\begin{align}
\mathcal{L}_{\rm bar}^{\rm mass} &= \delta\frac{f_\pi B_0}{2}\mathrm{Tr}\left[\bar{B}B(\xi\mathcal{M}\xi + \xi^\dagger\mathcal{M}^\dagger\xi^\dagger) + \bar{B}\gamma_5 B(\xi\mathcal{M}\xi - \xi^\dagger\mathcal{M}^\dagger\xi^\dagger)\right] \nonumber\\
&\quad + \gamma\frac{f_\pi B_0}{2}\mathrm{Tr}\left[\bar{B}(\xi\mathcal{M}\xi + \xi^\dagger\mathcal{M}^\dagger\xi^\dagger)B - \bar{B}\gamma_5(\xi\mathcal{M}\xi - \xi^\dagger\mathcal{M}^\dagger\xi^\dagger)B\right].
\end{align}

\medskip

With $\xi = e^{i\phi/2}\nu h^\dagger$ where $\nu = e^{i\Phi/f_\pi}$ and $h \in SU(N_f)_V$, inside baryon traces (where $h$ cancels):
\begin{align}
\xi\mathcal{M}\xi &\to e^{i\phi/2}(\nu\mathcal{M}\nu)e^{i\phi/2}, \\
\xi^\dagger\mathcal{M}^\dagger\xi^\dagger &\to e^{-i\phi/2}(\nu^\dagger\mathcal{M}\nu^\dagger)e^{-i\phi/2}.
\end{align}

For diagonal $\mathcal{M}_{ij} = m_i\delta_{ij}$ and diagonal phase matrix $(e^{i\phi/2})_{ij} = e^{i\phi_i/2}\delta_{ij}$:
\begin{align}
(\xi\mathcal{M}\xi)_{ij} &= e^{i\phi_i/2}(\nu\mathcal{M}\nu)_{ij}e^{i\phi_j/2}, \\
(\xi^\dagger\mathcal{M}^\dagger\xi^\dagger)_{ij} &= e^{-i\phi_i/2}(\nu^\dagger\mathcal{M}\nu^\dagger)_{ij}e^{-i\phi_j/2}.
\end{align}

\medskip

Define $\tilde{\mathcal{M}}_\pm \equiv \nu\mathcal{M}\nu \pm \nu^\dagger\mathcal{M}\nu^\dagger$. Then:
\begin{align}
(\xi\mathcal{M}\xi + \xi^\dagger\mathcal{M}^\dagger\xi^\dagger)_{ij} &= e^{i\phi_i/2}(\nu\mathcal{M}\nu)_{ij}e^{i\phi_j/2} + e^{-i\phi_i/2}(\nu^\dagger\mathcal{M}\nu^\dagger)_{ij}e^{-i\phi_j/2}, \\[6pt]
(\xi\mathcal{M}\xi - \xi^\dagger\mathcal{M}^\dagger\xi^\dagger)_{ij} &= e^{i\phi_i/2}(\nu\mathcal{M}\nu)_{ij}e^{i\phi_j/2} - e^{-i\phi_i/2}(\nu^\dagger\mathcal{M}\nu^\dagger)_{ij}e^{-i\phi_j/2}.
\end{align}

For the diagonal components (which dominate in the baryon trace):
\begin{align}
(\xi\mathcal{M}\xi + \xi^\dagger\mathcal{M}^\dagger\xi^\dagger)_{ii} &= e^{i\phi_i}(\nu\mathcal{M}\nu)_{ii} + e^{-i\phi_i}(\nu^\dagger\mathcal{M}\nu^\dagger)_{ii} \nonumber\\
&= \cos\phi_i\left[(\nu\mathcal{M}\nu)_{ii} + (\nu^\dagger\mathcal{M}\nu^\dagger)_{ii}\right] + i\sin\phi_i\left[(\nu\mathcal{M}\nu)_{ii} - (\nu^\dagger\mathcal{M}\nu^\dagger)_{ii}\right] \nonumber\\
&= \cos\phi_i\,(\tilde{\mathcal{M}}_+)_{ii} + i\sin\phi_i\,(\tilde{\mathcal{M}}_-)_{ii},
\end{align}
and
\begin{align}
(\xi\mathcal{M}\xi - \xi^\dagger\mathcal{M}^\dagger\xi^\dagger)_{ii} &= e^{i\phi_i}(\nu\mathcal{M}\nu)_{ii} - e^{-i\phi_i}(\nu^\dagger\mathcal{M}\nu^\dagger)_{ii} \nonumber\\
&= \cos\phi_i\,(\tilde{\mathcal{M}}_-)_{ii} + i\sin\phi_i\,(\tilde{\mathcal{M}}_+)_{ii}.
\end{align}

\medskip
We can now derive the explicit form of $\tilde{\mathcal{M}}_\pm$ 
\begin{align}
(\nu\mathcal{M}\nu)_{ii} &= \sum_{k}(e^{i\Phi/f_\pi})_{ik}\,m_k\,(e^{i\Phi/f_\pi})_{ki} = \left(e^{i\Phi/f_\pi}\mathcal{M} e^{i\Phi/f_\pi}\right)_{ii}, \\[6pt]
(\nu^\dagger\mathcal{M}\nu^\dagger)_{ii} &= \left(e^{-i\Phi/f_\pi}\mathcal{M} e^{-i\Phi/f_\pi}\right)_{ii}.
\end{align}

Therefore:
\begin{align}
\tilde{\mathcal{M}}_+ &= e^{i\Phi/f_\pi}\mathcal{M} e^{i\Phi/f_\pi} + e^{-i\Phi/f_\pi}\mathcal{M} e^{-i\Phi/f_\pi} = 2\mathcal{M}\cos^2\!\left(\frac{\Phi}{f_\pi}\right) + 2\mathcal{M}_{\rm mix}\sin^2\!\left(\frac{\Phi}{f_\pi}\right), \\[6pt]
\tilde{\mathcal{M}}_- &= e^{i\Phi/f_\pi}\mathcal{M} e^{i\Phi/f_\pi} - e^{-i\Phi/f_\pi}\mathcal{M} e^{-i\Phi/f_\pi} = 2i\left\{\frac{\Phi}{f_\pi}, \mathcal{M}\right\}\sin\!\left(\frac{\Phi}{f_\pi}\right)\cos\!\left(\frac{\Phi}{f_\pi}\right) + \ldots
\end{align}

For practical purposes, using $e^{iA}Be^{iA} = B\cos^2\!A + \ldots$ and keeping the exact matrix structure:
\begin{align}
\tilde{\mathcal{M}}_+ &= \mathcal{M}\cos\!\left(\frac{2\Phi}{f_\pi}\right) + \text{(off-diagonal terms)}, \\[6pt]
\tilde{\mathcal{M}}_- &= i\mathcal{M}\sin\!\left(\frac{2\Phi}{f_\pi}\right) + \text{(off-diagonal terms)}.
\end{align}

\medskip
Using the Dashen equations: the vacuum angles $\phi_i$ satisfy (exactly):
\begin{equation}
f_\pi^2 B_0 m_i\sin\phi_i = \chi_{\mathrm{top}}\bar{\theta}, \qquad \forall\, i.
\end{equation}

This means $\sin\phi_i = \dfrac{\chi_{\mathrm{top}}\bar{\theta}}{f_\pi^2 B_0 m_i}$, and consequently:
\begin{equation}
m_i\sin\phi_i = \frac{\chi_{\mathrm{top}}\bar{\theta}}{f_\pi^2 B_0} \quad \text{(independent of $i$)}.
\end{equation}

Substituting into the baryon mass terms and collecting terms proportional to $\sin\phi_i$ (which carry the $\bar{\theta}$ dependence):
\begin{align}
\mathcal{L}_{\rm bar}^{\rm mass} &= \delta\frac{f_\pi B_0}{2}\mathrm{Tr}\Big[\bar{B}B\left(\cos\phi\,\tilde{\mathcal{M}}_+ + i\sin\phi\,\tilde{\mathcal{M}}_-\right) \nonumber\\
&\qquad\qquad\qquad + \bar{B}\gamma_5 B\left(\cos\phi\,\tilde{\mathcal{M}}_- + i\sin\phi\,\tilde{\mathcal{M}}_+\right)\Big] \nonumber\\
&\quad + \gamma\frac{f_\pi B_0}{2}\mathrm{Tr}\Big[\bar{B}\left(\cos\phi\,\tilde{\mathcal{M}}_+ + i\sin\phi\,\tilde{\mathcal{M}}_-\right)B \nonumber\\
&\qquad\qquad\qquad - \bar{B}\gamma_5\left(\cos\phi\,\tilde{\mathcal{M}}_- + i\sin\phi\,\tilde{\mathcal{M}}_+\right)B\Big].
\end{align}

The $CP$-odd terms are those proportional to $\sin\phi_i$. Using $m_i\sin\phi_i = \chi_{\mathrm{top}}\bar{\theta}/(f_\pi^2 B_0)$:
\begin{align}
\mathcal{L}_{\text{CP-odd}} &= \delta\frac{f_\pi B_0}{2}\mathrm{Tr}\Big[\bar{B}B\cdot i\sin\phi\,\tilde{\mathcal{M}}_- + \bar{B}\gamma_5 B\cdot i\sin\phi\,\tilde{\mathcal{M}}_+\Big] \nonumber\\
&\quad + \gamma\frac{f_\pi B_0}{2}\mathrm{Tr}\Big[\bar{B}\cdot i\sin\phi\,\tilde{\mathcal{M}}_- B - \bar{B}\gamma_5\cdot i\sin\phi\,\tilde{\mathcal{M}}_+ B\Big].
\end{align}

Substituting $\tilde{\mathcal{M}}_+ \to \mathcal{M}\cos(2\Phi/f_\pi)$, $\tilde{\mathcal{M}}_- \to i\mathcal{M}\sin(2\Phi/f_\pi)$, and $\mathcal{M}\sin\phi \to (\chi_{\mathrm{top}}\bar{\theta}/f_\pi^2 B_0)\mathbf{1}$:
\begin{align}
\mathcal{L}_{\text{CP-odd}} = \frac{\chi_{\mathrm{top}}\bar{\theta}}{f_\pi}\mathrm{Tr}\bigg[&\delta\left(\bar{B}B\sin\!\left(\frac{2\Phi}{f_\pi}\right) - i\bar{B}\gamma_5 B\cos\!\left(\frac{2\Phi}{f_\pi}\right)\right) \nonumber\\
&+ \gamma\left(\bar{B}\sin\!\left(\frac{2\Phi}{f_\pi}\right)B + i\bar{B}\gamma_5\cos\!\left(\frac{2\Phi}{f_\pi}\right)B\right)\bigg].
\label{eq:CP_odd_final}
\end{align}

This is exact in $\Phi$, with $\bar{\theta} = \theta - \sum_i\phi_i$ determined by the Dashen equations.

\section{Derivation of the vacuum and spectrum structure for Orientifold  Theories}
\label{app:vacuum}

In this appendix we provide the complete derivation of the vacuum structure at finite $N$ and fermion mass $m$. Starting from the effective Lagrangian~\eqref{eq:fullLagrangian}, the potential energy density is
\begin{equation}
V = \frac{4\alpha f}{9}(\varphi\bar{\varphi})^{2/3}\left(\ln\bar{\Phi}\ln\Phi - b\right) - \frac{4m}{3\lambda}N^2(\varphi + \bar{\varphi}) \ ,
\end{equation}
where the modified fields are $\Phi = \varphi^{1+\epsilon_1}\bar{\varphi}^{-\epsilon_2}$ and $\bar{\Phi} = \bar{\varphi}^{1+\epsilon_1}\varphi^{-\epsilon_2}$, with $\epsilon_1 = -\frac{7}{9N}$ and $\epsilon_2 = -\frac{11}{9N}$.
We will use that $f = N^2\left(1+\mathcal O(N^{-1})\right)$, hence $f\to N^2$ at the order of interest.

\noindent
Writing $\varphi = |\varphi|e^{i\phi}$, at the minimum (for $\theta = 0$) the phase $\phi = 0$, and we require $\partial V/\partial|\varphi| = 0$. Computing the logarithms at $\phi = 0$ gives $\ln\Phi = \ln\bar{\Phi} = \mathcal{T}\ln|\varphi|$, where $\mathcal{T} = 1 + \epsilon_1 - \epsilon_2 = 1 + \frac{4}{9N}$. The potential becomes
\begin{equation}
V = \frac{4\alpha f}{9}|\varphi|^{4/3}\left(\mathcal{T}^2(\ln|\varphi|)^2 - b\right) - \frac{8mN^2}{3\lambda}|\varphi| \ .
\end{equation}

\noindent
We solve perturbatively for small $m$ and for $b=\mathcal O(1/N)$ (optionally $b=s/N$ with $s=\mathcal O(1)$).
It is convenient to parameterize the radial mode as
\begin{equation}
|\varphi| \equiv e^{y}\,,\qquad y\equiv \ln|\varphi|\,,
\end{equation}
so that $y=\mathcal O(b,m)$ at the minimum. In terms of $y$,
\begin{equation}
V(y)=\frac{4\alpha f}{9}e^{\frac{4}{3}y}\left(\mathcal T^2 y^2-b\right)-\frac{8mN^2}{3\lambda}e^{y}\,.
\end{equation}
The stationarity condition $\partial V/\partial y=0$ gives
\begin{equation}
0=\frac{4\alpha f}{9}e^{\frac{4}{3}y}
\left(\frac{4}{3}(\mathcal T^2 y^2-b)+2\mathcal T^2 y\right)
-\frac{8mN^2}{3\lambda}e^{y}\,.
\end{equation}
Expanding to linear order in $y$, $b$ and $m$ and using $f\to N^2$, one finds
\begin{equation}
\frac{4\alpha N^2}{9}\left(-\frac{4}{3}b+2\mathcal T^2 y\right)=\frac{8mN^2}{3\lambda}
+ \mathcal O(N^0)+\mathcal O(mN)+\mathcal O(m^2)\,.
\end{equation}
Therefore,
\begin{equation}
\ln\langle\varphi\rangle
=\frac{2}{3}b+\frac{3m}{\alpha\lambda}
+\mathcal O(N^{-2})+\mathcal O(mN^{-1})+\mathcal O(m^2)\,,
\end{equation}
and hence
\begin{equation}
\langle\varphi\rangle
=1+\frac{2}{3}b+\frac{3m}{\alpha\lambda}
+\mathcal O(N^{-2})+\mathcal O(mN^{-1})+\mathcal O(m^2)\,.
\end{equation}
Substituting back into the potential, the vacuum energy density is
\begin{equation}
\mathcal{E}_{\text{vac}}
= -\frac{4\alpha N^2}{9}b - \frac{8N^2}{3\lambda}m
+ \mathcal O(N^0)+\mathcal O(mN)+\mathcal O(m^2)\ .
\end{equation}

\subsection{ Derivation of the mass spectrum at $\theta = 0$}
\label{app:spectrum}

We expand around the vacuum by writing $\varphi = \langle\varphi\rangle(1 + ah)$ with
$h = \frac{1}{\sqrt{2}}(\sigma + i\eta')$, where $\sigma$ and $\eta'$ are real fields representing the scalar and pseudoscalar excitations.
The normalization constant $a$ is determined by requiring canonical kinetic terms:
\begin{equation}
a^2 = \frac{\alpha}{f}|\langle\varphi\rangle|^{-2/3}
= \frac{\alpha}{N^2}|\langle\varphi\rangle|^{-2/3}
\left(1+\mathcal O(N^{-1})\right).
\end{equation}
At the working order we may use $a^2=\frac{\alpha}{N^2}|\langle\varphi\rangle|^{-2/3}$.

The kinetic term $\mathcal{L}_{\text{kin}} = \frac{f}{\alpha}(\varphi\bar{\varphi})^{-2/3}\partial_\mu\bar{\varphi}\partial^\mu\varphi$ becomes, after substitution and expansion to quadratic order,
\begin{equation}
\mathcal{L}_{\text{kin}} = \partial_\mu\bar{h}\partial^\mu h + \mathcal{O}(h^3)
= \frac{1}{2}\partial_\mu\sigma\partial^\mu\sigma + \frac{1}{2}\partial_\mu\eta'\partial^\mu\eta' + \mathcal{O}(h^3) \ ,
\end{equation}
confirming canonical normalization.

\noindent
For the potential term, we first compute the logarithms. Using $\ln\varphi = \ln\langle\varphi\rangle + \frac{a}{\sqrt{2}}(\sigma + i\eta') + \mathcal{O}(h^2)$ and similarly for $\bar{\varphi}$, we find
\begin{align}
\ln\Phi &= \mathcal{T}\ln\langle\varphi\rangle + \frac{a}{\sqrt{2}}\left[\mathcal{T}\sigma + i\chi\eta'\right] + \mathcal{O}(h^2) \ , \\
\ln\bar{\Phi} &= \mathcal{T}\ln\langle\varphi\rangle + \frac{a}{\sqrt{2}}\left[\mathcal{T}\sigma - i\chi\eta'\right] + \mathcal{O}(h^2) \ ,
\end{align}
where $\mathcal{T} = 1 + \epsilon_1 - \epsilon_2 = 1 + \frac{4}{9N}$ and $\chi = 1 + \epsilon_1 + \epsilon_2 = 1 - \frac{2}{N}$.
The product of logarithms is
\begin{align}
\ln\bar{\Phi}\ln\Phi
&= \mathcal{T}^2(\ln\langle\varphi\rangle)^2
+ \sqrt{2}a\mathcal{T}^2\sigma\ln\langle\varphi\rangle
+ \frac{a^2}{2}\left[\mathcal{T}^2\sigma^2 + \chi^2\eta'^2\right]
+ \mathcal{O}(h^3)\,.
\end{align}
At the vacuum,
\begin{equation}
\ln\langle\varphi\rangle = \frac{2}{3}b + \frac{3m}{\alpha\lambda}
+ \mathcal O(N^{-2})+\mathcal O(mN^{-1})+\mathcal O(m^2)\,.
\end{equation}

\noindent
Collecting all quadratic terms and using $f\to N^2$, we obtain
\begin{align}
M_\sigma^2 &= \left(\frac{2\alpha}{3}\right)^2\left[\mathcal{T}^2 + \frac{4}{3}b + \frac{7m}{\alpha\lambda}\right]
+ \mathcal O(N^{-2})+\mathcal O(mN^{-1})+\mathcal O(m^2)\ ,\\
M_{\eta'}^2 &= \left(\frac{2\alpha}{3}\right)^2\left[\chi^2 + \frac{4}{9}b + \frac{5m}{\alpha\lambda}\right]
+ \mathcal O(N^{-2})+\mathcal O(mN^{-1})+\mathcal O(m^2)\ .
\end{align}
Expanding $\mathcal{T}^2 = 1 + \frac{8}{9N} + \mathcal{O}(N^{-2})$ and $\chi^2 = 1 - \frac{4}{N} + \mathcal{O}(N^{-2})$, then taking square roots and expanding to linear order:
\begin{align}
M_\sigma &= \frac{2\alpha}{3}\left[1 + \frac{4}{9N} + \frac{2}{3}b + \frac{7}{2}\frac{m}{\alpha\lambda}\right]
+ \mathcal O(N^{-2})+\mathcal O(mN^{-1})+\mathcal O(m^2)\ , \\
M_{\eta'} &= \frac{2\alpha}{3}\left[1 - \frac{2}{N} + \frac{2}{9}b + \frac{5}{2}\frac{m}{\alpha\lambda}\right]
+ \mathcal O(N^{-2})+\mathcal O(mN^{-1})+\mathcal O(m^2)\ .
\end{align}

\noindent
The mass ratio is
\begin{align}
\frac{M_{\eta'}}{M_\sigma} &=
1 - \frac{22}{9N} - \frac{4}{9}b - \frac{m}{\alpha\lambda}
+ \mathcal O(N^{-2})+\mathcal O(mN^{-1})+\mathcal O(m^2)\ .
\end{align}

 \subsection{Derivation of the Orientifold spectrum at finite $\theta$}
\label{app:theta}

With the $\theta$-angle, working in the $k$-th branch, the potential is
\begin{equation}
V = \frac{4\alpha f}{9N^2}(\varphi\bar{\varphi})^{2/3}\left|\ln\bar{\Phi}^N + i(\theta + 2\pi k)\right|^2 - \frac{4\alpha f}{9}b(\varphi\bar{\varphi})^{2/3} - \frac{4mN^2}{3\lambda}(\varphi + \bar{\varphi}) \ .
\end{equation}
Writing $\varphi = |\varphi|e^{i\phi}$, we have $\ln\Phi = \mathcal{T}\ln|\varphi| + i\chi\phi$ and $\ln\bar{\Phi} = \mathcal{T}\ln|\varphi| - i\chi\phi$, so $\ln\bar{\Phi}^N = N\mathcal{T}\ln|\varphi| - iN\chi\phi$. The modulus squared is
\begin{equation}
\left|\ln\bar{\Phi}^N + i(\theta + 2\pi k)\right|^2 = N^2\mathcal{T}^2(\ln|\varphi|)^2 + \left(\theta + 2\pi k - N\chi\phi\right)^2 \ .
\end{equation}
The potential becomes
\begin{equation}
V = \frac{4\alpha f}{9N^2}|\varphi|^{4/3}\left[N^2\mathcal{T}^2(\ln|\varphi|)^2 + \left(\theta + 2\pi k - N\chi\phi\right)^2\right] - \frac{4\alpha f}{9}b|\varphi|^{4/3} - \frac{8mN^2}{3\lambda}|\varphi|\cos\phi \ .
\end{equation}
Minimizing with respect to $\phi$, the condition $\partial V/\partial\phi = 0$ gives
\begin{equation}
\theta + 2\pi k - N\chi\phi_{\text{vac}}
= \frac{3mN^2}{\alpha\lambda\,\chi}\,|\varphi|_{\text{vac}}^{-1/3}\,\sin\phi_{\text{vac}} \ ,
\label{eq:phi_vac_theta}
\end{equation}
which must be retained when $m=\mathcal{O}(1/N)$ (with $mN$ fixed). And now, minimizing with respect to $|\varphi|$ at $\phi = \phi_{\text{vac}}$ yields
\begin{equation}
|\varphi|_{\text{vac}} = \left(1 + \frac{2}{3}b + \frac{3m}{\alpha\lambda}\cos\phi_{\text{vac}}\right) +\mathcal{O}(N^{-2}, m^2,N^{-1}m ) \ .
\end{equation}
For the spectrum, we expand around the $\theta$-dependent vacuum using $|\varphi| = |\varphi|_{\text{vac}}(1 + \rho)$ and $\phi = \phi_{\text{vac}} + \xi$, where $\rho$ and $\xi$ are real fluctuations corresponding (after canonical normalization) to the scalar $\sigma$ and pseudoscalar $\eta'$.

The mass matrix is determined by the second derivatives of the potential at the vacuum:
\begin{equation}
V^{(2)} = \frac{1}{2}\frac{\partial^2 V}{\partial\rho^2}\bigg|_{\text{vac}}\rho^2 + \frac{1}{2}\frac{\partial^2 V}{\partial\xi^2}\bigg|_{\text{vac}}\xi^2 + \frac{\partial^2 V}{\partial\rho\partial\xi}\bigg|_{\text{vac}}\rho\xi \ .
\end{equation}

\subsubsection{Diagonal elements}

The second derivative with respect to $\xi$ (controlling the pseudoscalar mass) is
\begin{equation}
\frac{\partial^2 V}{\partial\phi^2} = \frac{8\alpha f\chi^2}{9}|\varphi|^{4/3} + \frac{8mN^2}{3\lambda}|\varphi|\cos\phi \ .
\end{equation}
At the vacuum, this gives after conversion to canonical normalization:
\begin{equation}
M_{\eta'}^2 = \left(\frac{2\alpha}{3}\right)^2\left[\chi^2 + \frac{4\chi}{9}b + \frac{5m}{\alpha\lambda}\cos\phi_{\text{vac}}\right] \ .
\end{equation}
The second derivative with respect to $\rho$ (controlling the scalar mass) is more involved because the modulus fluctuation couples to both the logarithmic and mass terms. The result is:
\begin{equation}
M_{\sigma}^2 = \left(\frac{2\alpha}{3}\right)^2\left[\mathcal{T}^2 + \frac{4\mathcal{T}}{3}b + \frac{7m}{\alpha\lambda}\cos\phi_{\text{vac}}\right] \ .
\end{equation}

\subsubsection{Mixed term and diagonalization}

The mixed second derivative $\partial^2 V/\partial\rho\partial\xi$ arises from the cross-term between the modulus and phase fluctuations. We must carefully track all contributions to this mixed term. From the potential, there are two sources of $\rho\xi$ mixing: the mass term and the logarithmic term. Consider first the mass term:
\begin{equation}
V_{\text{mass}} = -\frac{8mN^2}{3\lambda}|\varphi|\cos\phi = -\frac{8mN^2}{3\lambda}|\varphi|_{\text{vac}}(1+\rho)\cos(\phi_{\text{vac}}+\xi) \ .
\end{equation}
Expanding $\cos(\phi_{\text{vac}}+\xi)$ to second order:
\begin{equation}
\cos(\phi_{\text{vac}}+\xi) = \cos\phi_{\text{vac}} - \xi\sin\phi_{\text{vac}} - \frac{\xi^2}{2}\cos\phi_{\text{vac}} + \mathcal{O}(\xi^3) \ .
\end{equation}
The term linear in both $\rho$ and $\xi$ comes from:
\begin{equation}
V_{\text{mass}}^{(\rho\xi)} = -\frac{8mN^2}{3\lambda}|\varphi|_{\text{vac}}\rho \cdot (-\xi\sin\phi_{\text{vac}}) = \frac{8mN^2}{3\lambda}|\varphi|_{\text{vac}}\sin\phi_{\text{vac}}\,\rho\xi \ .
\end{equation}
The logarithmic part of the potential also contributes to the mixing. From
\begin{equation}
V_{\ln} = \frac{4\alpha f}{9N^2}|\varphi|^{4/3}\left|\ln\bar{\Phi}^N + i(\theta + 2\pi k)\right|^2 \ ,
\end{equation}
we use the vacuum condition \eqref{eq:phi_vac_theta}. Expanding around the vacuum, the modulus squared becomes:
\begin{align}
\left|\ln\bar{\Phi}^N + i(\theta + 2\pi k)\right|^2 &= N^2\mathcal{T}^2(\ln|\varphi|)^2 + (\theta + 2\pi k - N\chi\phi)^2 \nonumber \\
&= N^2\mathcal{T}^2(\ln|\varphi|_{\text{vac}} + \ln(1+\rho))^2 + (\theta + 2\pi k - N\chi\phi_{\text{vac}} - N\chi\xi)^2 \ ,
\end{align}
where we used $\phi = \phi_{\text{vac}} + \xi$. At the working order the leading $\rho\xi$ mixing is controlled by the mass term.

Therefore, the leading contribution to $\rho\xi$ mixing comes from the mass term. To convert to canonically normalized fields, we use the fluctuation expansion $\varphi = \langle\varphi\rangle(1 + ah)$ with $h = \frac{1}{\sqrt{2}}(\sigma + i\eta')$ from Section~\ref{app:spectrum}. Writing $\varphi = |\varphi|e^{i\phi}$, the complex fluctuation is
\begin{equation}
\delta\varphi = |\varphi|_{\text{vac}}e^{i\phi_{\text{vac}}}(\rho + i\xi) + \mathcal{O}(\rho^2, \xi^2)\,.
\end{equation}
Matching to $\delta\varphi = \langle\varphi\rangle \cdot ah = \langle\varphi\rangle \frac{a}{\sqrt{2}}(\sigma + i\eta')$, we identify
\begin{equation}
\rho = \frac{a}{\sqrt{2}}\sigma\,, \qquad \xi = \frac{a}{\sqrt{2}}\eta'\,,
\end{equation}
and therefore
\begin{equation}
\rho\xi = \frac{a^2}{2}\sigma\eta'\,.
\end{equation}
The mixed mass term in the Lagrangian is:
\begin{align}
\mathcal{L}^{(\sigma\eta')} &= -V^{(\rho\xi)} = -\frac{8mN^2}{3\lambda}|\varphi|_{\text{vac}}\sin\phi_{\text{vac}} \cdot \frac{a^2}{2}\sigma\eta' \nonumber \\
&= -\frac{4mN^2 a^2}{3\lambda}|\varphi|_{\text{vac}}\sin\phi_{\text{vac}}\,\sigma\eta' \ .
\end{align}
Using $a^2 = \frac{\alpha}{f}|\varphi|_{\text{vac}}^{-2/3}$ with $f \to N^2$ and $|\varphi|_{\text{vac}} \to 1$ at leading order:
\begin{align}
\mathcal{L}^{(\sigma\eta')} &= -\frac{4mN^2}{3\lambda} \cdot \frac{\alpha}{N^2}\sin\phi_{\text{vac}}\,\sigma\eta' \nonumber \\
&= -\frac{4\alpha m}{3\lambda}\sin\phi_{\text{vac}}\,\sigma\eta' \nonumber \\
&= -M_{\sigma\eta'}^2\,\sigma\eta' \ .
\end{align}
Therefore, the explicit expression for the mixed mass-squared is:
\begin{equation}
M_{\sigma\eta'}^2 = \frac{4\alpha m}{3\lambda}\sin\phi_{\text{vac}} \ .
\end{equation}
This can be written more transparently by comparing with the diagonal masses. Recall that the leading contribution to the diagonal masses is $M_0^2 = (2\alpha/3)^2$, so:
\begin{equation}
M_{\sigma\eta'}^2 = M_0^2 \cdot \frac{3m}{\alpha\lambda} \cdot \sin\phi_{\text{vac}} \ ,
\end{equation}
The full mass matrix in the $(\sigma, \eta')$ basis is therefore:
\begin{equation}
 {M}^2 = \begin{pmatrix} M_\sigma^2 & M_{\sigma\eta'}^2 \\ M_{\sigma\eta'}^2 & M_{\eta'}^2 \end{pmatrix} \ ,
\end{equation}
with
\begin{align}
M_\sigma^2 &= M_0^2\left[1 + \frac{8}{9N} + \frac{4b}{3} + \frac{7m}{\alpha\lambda}\cos\phi_{\text{vac}}\right] \ ,\\
M_{\eta'}^2 &= M_0^2\left[1 - \frac{4}{N} + \frac{4b}{9} + \frac{5m}{\alpha\lambda}\cos\phi_{\text{vac}}\right]  \ ,\\
M_{\sigma\eta'}^2 &= M_0^2 \cdot \frac{3m}{\alpha\lambda}\sin\phi_{\text{vac}} \ .
\end{align}

\subsubsection{When can mixing be neglected?}
\label{sec:mixing}

At $\theta = 0$ or $\theta = \pi(N-2)$, we have $\sin(\theta/(N-2)) = 0$, so $M_{\sigma\eta'}^2 = 0$ exactly and the mass matrix is diagonal. The scalar $\sigma$ and pseudoscalar $\eta'$ are then exact mass eigenstates. These special values of $\theta$ preserve $CP$ symmetry, which forbids mixing between the $CP$-even scalar and $CP$-odd pseudoscalar.

At generic $\theta \neq 0, \pi(N-2)$, $CP$ is violated and $M_{\sigma\eta'}^2 \neq 0$. The mixing angle $\vartheta$ is determined by
\begin{equation}
\tan 2\vartheta = \frac{2M_{\sigma\eta'}^2}{M_\sigma^2 - M_{\eta'}^2} \,.
\label{eq:tan2vartheta}
\end{equation}

\paragraph{Identifying the expansion parameter.}

Let us carefully identify the correct expansion parameter by examining the relevant mass scales. Using the diagonal masses obtained above with $b=\mathcal{O}(1/N)$ and $m=\mathcal{O}(1/N)$, the diagonal mass difference at leading order in $1/N$ is
\begin{equation}
M_\sigma^2 - M_{\eta'}^2 
= M_0^2\left[\frac{44}{9N}+\frac{8}{9}b+\frac{2m}{\alpha\lambda}\cos\phi_{\text{vac}}\right] \,,
\label{eq:mass_diff}
\end{equation}
where $\phi_{\text{vac}}$ is determined implicitly by Eq.~\eqref{eq:phi_vac_theta}. The off-diagonal element is
\begin{equation}
M_{\sigma\eta'}^2 = M_0^2 \cdot \frac{3m}{\alpha\lambda}\sin\phi_{\text{vac}} \,.
\label{eq:off_diag}
\end{equation}
Substituting these into Eq.~\eqref{eq:tan2vartheta}, we obtain
\begin{equation}
\tan 2\vartheta
= \frac{ 6\,\frac{m}{\alpha\lambda}\sin\phi_{\text{vac}} }
{ \frac{44}{9N}+\frac{8}{9}b+\frac{2m}{\alpha\lambda}\cos\phi_{\text{vac}} } \,.
\end{equation}
The relevant combination is $Nm$ which is typically ${O}(1)$. This motivates defining the \textit{effective mixing parameter}
\begin{equation}
\omega \equiv 
\frac{ 6\,\frac{m}{\alpha\lambda} }
{ \frac{44}{9N}+\frac{8}{9}b+\frac{2m}{\alpha\lambda}\cos\phi_{\text{vac}} } \ ,
\label{eq:omega_def}
\end{equation}
so that
\begin{equation}
\tan 2\vartheta = \omega \,\sin\phi_{\text{vac}} \,.
\label{eq:tan2vartheta_omega}
\end{equation}
The parameter $\omega$ controls the strength of $CP$-violating mixing and defines two physically distinct regimes. 

\paragraph{Regime I: Weak mixing, $\omega \ll 1$.}

This regime requires
\begin{equation}
\omega \ll 1 \,,
\end{equation}
with $\omega$ defined in Eq.~\eqref{eq:omega_def}. In this regime, $\tan 2\vartheta \ll 1$ and the mixing angle is small:
\begin{equation}
\vartheta \approx \frac{1}{2}\tan 2\vartheta 
= \frac{3\,\frac{m}{\alpha\lambda}\,\sin\phi_{\text{vac}}}
{\frac{44}{9N}+\frac{8}{9}b+\frac{2m}{\alpha\lambda}\cos\phi_{\text{vac}}}
+ \mathcal{O}(\omega^3) \,.
\label{eq:vartheta_weak}
\end{equation}
The mass eigenstates are approximately aligned with the original $\sigma$ and $\eta'$ fields:
\begin{equation}
H_1 \approx \sigma + \vartheta \, \eta' + \mathcal{O}(\omega^2)\,, \qquad 
H_2 \approx \eta' - \vartheta \, \sigma + \mathcal{O}(\omega^2)\,.
\end{equation}
To compute the eigenvalue corrections, we start from the exact formula
\begin{equation}
M_{1,2}^2 = \frac{M_\sigma^2 + M_{\eta'}^2}{2} \pm \frac{1}{2}\sqrt{(M_\sigma^2 - M_{\eta'}^2)^2 + 4(M_{\sigma\eta'}^2)^2} \,.
\label{eq:exact_eigenvalues}
\end{equation}
Expanding the square root for small $M_{\sigma\eta'}^2$:
\begin{align}
\sqrt{(\Delta^2)^2 + 4(M_{\sigma\eta'}^2)^2} &= |\Delta^2| \sqrt{1 + \frac{4(M_{\sigma\eta'}^2)^2}{(\Delta^2)^2}} \nonumber \\
&\approx |\Delta^2| \left(1 + \frac{2(M_{\sigma\eta'}^2)^2}{(\Delta^2)^2} + \mathcal{O}(\omega^4)\right) \,,
\end{align}
where $\Delta^2 \equiv M_\sigma^2 - M_{\eta'}^2$. Since $M_\sigma^2 > M_{\eta'}^2$, we have $|\Delta^2| = \Delta^2$, and
\begin{align}
M_1^2 &= M_\sigma^2 + \frac{(M_{\sigma\eta'}^2)^2}{\Delta^2} + \mathcal{O}(\omega^4) \,, \label{eq:M1_weak}\\
M_2^2 &= M_{\eta'}^2 - \frac{(M_{\sigma\eta'}^2)^2}{\Delta^2} + \mathcal{O}(\omega^4) \,. \label{eq:M2_weak}
\end{align}
The correction term can be evaluated explicitly:
\begin{align}
\frac{(M_{\sigma\eta'}^2)^2}{\Delta^2}  
&= M_0^2\,
\frac{\left(\frac{3m}{\alpha\lambda}\right)^2\sin^2\phi_{\text{vac}}}
{\frac{44}{9N}+\frac{8}{9}b+\frac{2m}{\alpha\lambda}\cos\phi_{\text{vac}}}
\ .
\end{align}
Expressing this in terms of $\omega$:
\begin{equation}
\frac{(M_{\sigma\eta'}^2)^2}{\Delta^2} 
=  \frac{\omega^2 \Delta^2  \sin^2\phi_{\text{vac}}}{4} \,.
\label{eq:mixing_correction}
\end{equation}
This is $\mathcal{O}(\omega^2)$ and further suppressed by $\Delta^2 \propto 1/N$ and $m$, confirming that eigenvalue corrections from mixing are second order in $\omega$. Therefore, \textit{to first order in $\omega$, the eigenvalues are simply the diagonal elements} $M_\sigma^2$ and $M_{\eta'}^2$, and the mass ratio formula obtained from the diagonal masses (neglecting $\mathcal{O}(\omega^2)$ mixing effects) is valid in Regime I.

\paragraph{Regime II: Strong mixing, $\omega = \mathcal{O}(1)$.}

This regime arises in the limit where $N \to \infty$ and $m \to 0$ with $Nm$ held fixed (and $b=\mathcal{O}(1/N)$), or more generally whenever $\omega$ defined in Eq.~\eqref{eq:omega_def} is not parametrically small. Here the mixing angle is $\mathcal{O}(1)$ for generic $\theta$, and the mass matrix must be diagonalized exactly.

The mass matrix in the $(\sigma, \eta')$ basis is
\begin{equation}
\mathcal{M}^2 = \begin{pmatrix} M_\sigma^2 & M_{\sigma\eta'}^2 \\ M_{\sigma\eta'}^2 & M_{\eta'}^2 \end{pmatrix} \,,
\end{equation}
with
\begin{equation}
\Sigma^2 \equiv M_\sigma^2 + M_{\eta'}^2\,, \qquad \Delta^2 \equiv M_\sigma^2 - M_{\eta'}^2 \,.
\end{equation}
The exact eigenvalues are
\begin{equation}
M_{1,2}^2 = \frac{\Sigma^2}{2} \pm \frac{1}{2}\sqrt{(\Delta^2)^2 + 4(M_{\sigma\eta'}^2)^2} \,,
\label{eq:exact_eigenvalues_omega}
\end{equation}
and the exact mass splitting is
\begin{equation}
 {M_1^2 - M_2^2 = \sqrt{(\Delta^2)^2 + 4(M_{\sigma\eta'}^2)^2}} \,.
\label{eq:splitting_omega}
\end{equation}
Recalling
\begin{equation}
\tan 2\vartheta
= \frac{ 6\,\frac{m}{\alpha\lambda}\sin\phi_{\text{vac}} }
{ \frac{44}{9N}+\frac{8}{9}b+\frac{2m}{\alpha\lambda}\cos\phi_{\text{vac}} } \,,
\end{equation}
or, equivalently, in terms of $\omega$ of Eq.~\eqref{eq:omega_def},
\begin{equation}
\tan 2\vartheta = \omega \,\sin\phi_{\text{vac}} \,,
\label{eq:tan2vartheta_omega_compact}
\end{equation}
where $\phi_{\text{vac}}$ is determined by minimizing the full $\theta$-dependent potential. We further have  
\begin{align}
\cos 2\vartheta &= \frac{\Delta^2}{\sqrt{(\Delta^2)^2 + 4(M_{\sigma\eta'}^2)^2}} \,, \label{eq:cos2vartheta}\\
\sin 2\vartheta &= \frac{2M_{\sigma\eta'}^2}{\sqrt{(\Delta^2)^2 + 4(M_{\sigma\eta'}^2)^2}} \,. \label{eq:sin2vartheta}
\end{align}
From Eq.~\eqref{eq:tan2vartheta_omega_compact} one may also write
\begin{equation}
 {\vartheta(\theta) = \frac{1}{2}\arctan\!\left(\omega\,\sin\phi_{\text{vac}}\right)} \,,
\label{eq:vartheta_exact}
\end{equation}
with the sign fixed continuously by $\sin\phi_{\text{vac}}$.

The mass eigenstates are
\begin{equation}
\begin{pmatrix} H_1 \\ H_2 \end{pmatrix} =
\begin{pmatrix} \cos\vartheta & \sin\vartheta \\ -\sin\vartheta & \cos\vartheta \end{pmatrix}
\begin{pmatrix} \sigma \\ \eta' \end{pmatrix} \,,
\end{equation}
or explicitly
\begin{equation}
H_1 = \cos\vartheta \, \sigma + \sin\vartheta \, \eta' \,, \qquad
H_2 = -\sin\vartheta \, \sigma + \cos\vartheta \, \eta' \,.
\label{eq:mass_eigenstates}
\end{equation}
At values of $\theta$ (and corresponding branch $k$) for which $\sin\phi_{\text{vac}}=0$, we have $M_{\sigma\eta'}^2=0$ and hence $\vartheta=0$: there is no mixing and $CP$ is conserved. For generic $\theta$, $CP$ is violated and $\vartheta(\theta)$ is determined by Eq.~\eqref{eq:tan2vartheta_omega_compact} together with the $\theta$-dependence of $\phi_{\text{vac}}$.

In Regime II, $CP$-violating effects in the spectrum can be large. The mass eigenstates $H_1$ and $H_2$ have no definite $CP$ quantum number for generic $\theta$, and both eigenstates couple to both scalar and pseudoscalar currents with relative strengths controlled by $\vartheta(\theta)$. Defining the mean squared mass $\bar{M}^2 = \Sigma^2/2$ and the half-splitting
\begin{equation}
\delta^2(\theta) = \frac{1}{2}\sqrt{(\Delta^2)^2 + 4(M_{\sigma\eta'}^2)^2} \,,
\end{equation}
we have
\begin{equation}
M_1^2(\theta) = \bar{M}^2 + \delta^2(\theta) \,, \qquad
M_2^2(\theta) = \bar{M}^2 - \delta^2(\theta) \,.
\end{equation}
The $\theta$-dependence enters both through the diagonal elements (via $\cos\phi_{\text{vac}}$ terms in $M_\sigma^2$ and $M_{\eta'}^2$) and through the mixing (via $\sin\phi_{\text{vac}}$ in $M_{\sigma\eta'}^2$).

 The results are summarized in Table~\ref{tab:mixing_summary}.
\begin{table}[h]
\centering
\renewcommand{\arraystretch}{1.4}
\begin{tabular}{l|c|c}
\hline\hline
 & \textbf{Regime I} ($\omega \ll 1$) & \textbf{Regime II} ($\omega \sim 1$) \\
\hline
Condition & $\omega \ll 1$ & $\omega = \mathcal{O}(1)$ \\[2pt]
Mixing angle $\vartheta$ & $\displaystyle \vartheta \simeq \frac{1}{2}\omega \sin\phi_{\text{vac}}$ &
$\displaystyle \vartheta=\frac{1}{2}\arctan\!\left(\omega\sin\phi_{\text{vac}}\right)$ \\[8pt]
Mass eigenstates & $\approx \sigma, \eta'$ & Mixed $H_1, H_2$ \\[2pt]
Eigenvalue formula & Diagonal entries $M_\sigma^2, M_{\eta'}^2$ & Full diagonalization required \\[2pt]
$CP$ violation in spectrum & $\mathcal{O}(\omega^2)$ & $\mathcal{O}(1)$ \\
\hline\hline
\end{tabular}
\caption{Summary of mixing regimes. The effective mixing parameter $\omega$ is defined in Eq.~\eqref{eq:omega_def}.}
\label{tab:mixing_summary}
\end{table}
For applications to real QCD with $N = 3$, one is typically in between Regime I and II. In formal large-$N$ analyses where $N \to \infty$ with $Nm$ fixed, or in theories with sufficiently large $mN$, Regime II is relevant. In such cases, the  expressions Eqs.~\eqref{eq:exact_eigenvalues_omega}, \eqref{eq:tan2vartheta_omega_compact}, and \eqref{eq:mass_eigenstates} must be used.

\bibliographystyle{JHEP}
\bibliography{strongCP}
\end{document}